%% file: ms_astroph.tex
\documentclass[12pt,preprint]{emulateapj}

\begin{document}

\shortauthors{Metchev et al.}
\shorttitle{New L and T Dwarfs from a Cross-Match of 2MASS and SDSS}

\title{A Cross-Match of 2MASS and SDSS: Newly-Found L and T Dwarfs
and an Estimate of the Space Densitfy of T Dwarfs}
\author{Stanimir A.\ Metchev}
\affil{Department of Physics and Astronomy, University of California, Los 
Angeles, California 90095}
\email{metchev@astro.ucla.edu}
\author{J.\ Davy Kirkpatrick, G.\ Bruce Berriman}
\affil{Infrared Processing and Analysis Center, MC 100--22, California Institute of Technology, 
Pasadena, California 91125}
\author{Dagny Looper}
\affil{Institute for Astronomy, University of Hawai'i, 2680 Woodlawn Drive, Honolulu, Hawai'i 96822}

\begin{abstract}
We report new L and T dwarfs found in a cross-match of the SDSS Data Release 1 and 2MASS.   Our 
simultaneous search of the two databases effectively allows us to relax the criteria for object detection 
in either survey and to explore the combined databases to a greater completeness level. We find two 
new T dwarfs in addition to the 13 already known in the SDSS DR1 footprint.  We also identify 22 new 
candidate and bona-fide L dwarfs, including a new young L2 dwarf and a peculiar L2 dwarf with 
unusually blue near-IR colors---potentially the result of mildly sub-solar metallicity.  These discoveries 
underscore the utility of simultaneous database cross-correlation in searching for rare objects.  Our 
cross-match completes the census of T dwarfs within the joint SDSS and 2MASS flux limits to the $
\approx$97\% level.  Hence, we are able to accurately infer the space density of T dwarfs.  We employ 
Monte Carlo tools to simulate the observed population of SDSS DR1 T dwarfs with 2MASS 
counterparts and find that the space density of T0--T8 dwarf systems is $0.0070_{-0.0030}^{+0.0032}
$~pc$^{-3}$ (95\% confidence interval), i.e., about one per 140~pc$^3$.  Compared to predictions for 
the T dwarf space density that depend on various assumptions for the sub-stellar mass function, this 
result is most consistent with models that assume a flat sub-stellar mass function $d N/d M \propto M^
{0.0}$.   No $>$T8 dwarfs were discovered in the present cross-match, though less than one was 
expected in the limited area (2099~deg$^2$) of SDSS DR1.  
\end{abstract}



\section{INTRODUCTION}

Our knowledge of the properties of ultra-cool L and T dwarfs has increased dramatically over the past 
decade as a result of the completion of several large-area optical and near-IR imaging surveys, and 
the implementation of fast computerized access to survey databases.  L and T dwarfs are readily 
identified in imaging surveys by their characteristic red optical minus near-infrared (near-IR) colors.  
There are now hundreds of L dwarfs and over 100 T dwarfs known\footnote{A database of known L 
and T dwarfs is maintained at http://DwarfArchives.org \citep{kirkpatrick03, gelino_etal04}.}, the vast 
majority of which have been found in the Two-Micron All-Sky Survey \citep[2MASS;][]{skrutskie_etal06} 
and in the Sloan Digital Sky Survey \citep[SDSS;][]{stoughton_etal02}.  The large number of L and T 
dwarfs identified in these two uniform and well-characterized data sets allows detailed investigations 
of the population properties of sub-stellar objects, namely their mass and luminosity functions and 
their multiplicity.  A detailed investigation focusing on a flux-limited sample of field L dwarfs has 
already been presented in \citet{cruz_etal07}.  However, a similarly comprehensive empirical 
investigation of field T dwarfs has not been performed yet.  The most detailed study of T dwarfs to date 
is the 2MASS T5--T8 dwarf survey of \citet{burgasser02}.  \citeauthor{burgasser02}'s focus on the T5--
T8 sub-range was driven by their characteristic {\sl blue} near-IR colors ($J-K_S\sim0$~mag) that set 
them apart from the majority of main-sequence stars in 2MASS.  T0--T4 dwarfs, on the other hand, 
have red to neutral near-IR colors (2.0~mag~$\gtrsim J-K_S \gtrsim0.5$~mag) and searches for them 
face a vast contamination by background low-mass stars.  As a result, our understanding of the field 
T0--T4 population has lagged.  Although a number of T0--T4 dwarfs have been identified in the optical 
in SDSS \citep[][and references therein]{geballe_etal02, knapp_etal04, chiu_etal06}, an adequate 
analysis of the population of early T dwarfs is still lacking.  Accurate knowledge of the number density 
of early T dwarfs relative to those of late L and mid T dwarfs is important for studies aimed at 
constraining the time scale of dust sedimentation and cloud formation in sub-stellar photospheres at 
the L/T transition.  Completing the census of known T dwarfs to allow such studies is the primary 
science motivation for the present study.

With hundreds of L and T dwarfs now known, a small number of peculiar L and T dwarfs have also 
emerged from the larger sample.  These unusual and rare objects are set apart from their counterparts 
either by having abnormal surface gravities \citep[e.g.,][]{kirkpatrick_etal06, burgasser_etal06, 
cruz_etal07} or lower metallicities \citep{burgasser_etal03b, burgasser04b}.  The recognition of such 
variety among the known L and T dwarfs has revealed a necessity for dimensional expansion of the 
present L and T dwarf classification schemes to include the effects of surface gravity and metallicity 
\citep{kirkpatrick05}.   However, the number of known peculiar objects is presently too small to enable 
their accurate characterization; a larger sample will be needed to adequately anchor an expanded 
classification scheme.  The defining photometric characteristics of peculiar ultra-cool dwarfs that set 
them apart from the normal population, e.g., redder near-IR colors for young L dwarfs \citep
{kirkpatrick_etal06}, or redder optical and bluer near-IR colors for metal-poor ultra-cool sub-dwarfs 
\citep{lepine_etal03b, burgasser_etal03b, cruz_etal07}, are only now being recognized.  Targeted 
photometric searches for such objects in the existing databases may be more fruitful in the near future.  
As a by-product of the present study, we remark on the characteristics of two peculiar L dwarfs 
discovered in our search.

Finally, the analysis of the late-T dwarf population of \citeauthor{burgasser02} (\citeyear{burgasser02}; 
see also \citealt{burgasser04, burgasser07}, \citealt{allen_etal05}) has shown that the number density 
of sub-stellar objects monotonically increases until the cool end of the present spectral type sequence 
(at T8; $T_{\rm eff}\approx750$~K), and is expected to continue increasing for even cooler objects.  
That is, brown dwarfs with spectral types $>$T8 are likely numerous, but have eluded detection in 
present large-area surveys because of being intrinsically faint.  The photospheres of extremely cool 
brown dwarfs, with effective temperatures below 400~K, are expected to have undergone a chemical 
transformation that is similar to the one occurring at the transition between the L and T spectral types, 
with the dominant source of opacity in the near-IR becoming water clouds, as opposed to methane 
clouds \citep{burrows_etal03}.  Even cooler ($\lesssim$200~K) brown dwarfs may have ammonia-
dominated photospheres that are very similar to those of giant planets in the Solar System.  At very low 
effective temperatures, the emergent spectral energy distribution (SED) may be such that these 
objects may require a new spectral type (``Y'') for classification.  The discovery and characterization of 
such extremely cool brown dwarfs are among the primary science drivers for present and future deep 
large area surveys, e.g., with UKIRT \citep[The UKIRT Infrared Deep Sky Survey;][]{lawrence_etal06}, 
with the Panoramic Survey Telescope and Rapid Response System \citep[Pan-STARRS;][]
{kaiser_etal02}, or with the Wide-field Infrared Survey Explorer \citep[WISE;][]{mainzer_etal06}.   These 
large sensitive projects will undoubtedly dramatically expand our knowledge of sub-stellar objects at 
the bottom of the main sequence.  Nevertheless, it is possible that a small population of such 
extremely cool objects may already be present in the current generation of sky surveys.  Among the 
existing surveys, SDSS and 2MASS offer the best chance for finding brown dwarfs later than spectral 
type T8 because they cover the most volume.  Given their anticipated faintness, very red optical colors, 
and potentially blue near-IR colors, $>$T8 dwarfs may be present only as low signal-to-noise ($S/N$) 
single-band detections in SDSS (at $z$) and 2MASS (at $J$).  As such, they are more likely to have 
been overlooked or flagged as artifacts in either survey.  A combined consideration of the optical and 
near-IR data from SDSS and 2MASS may improve the chance for their discovery.  That is, a cross-
correlation of the SDSS and 2MASS databases may allow us to not only probe deeper, but also 
cooler, than is possible in either survey alone.  Such a cross-correlation is the underlying approach of 
the present work.

The ability to cross-correlate large astronomical databases is one of the main technological goals of 
the National Virtual Observatory (NVO). In this paper we present results from a pilot project to test an 
implementation of this approach, focusing on the search of new brown dwarfs from a rapid cross-
match of the 2MASS All-Sky Point Source Catalog (PSC) and SDSS Data Release 1 (DR1). The 
project was selected by the NVO as one of three demonstration research projects that would inform of 
the long-term hardware and software technology needs of the NVO.  The brown dwarf project in 
particular was aimed at identifying the technologies that will be needed to cross-match source 
catalogs at scale.  In the present paper we describe the implementation of our cross-matching 
technique (\S~\ref{sec_xmatch}), and report first results from the project, including identifications of two 
previously overlooked T dwarfs, a new peculiar L dwarf, and a young L dwarf in SDSS DR1 and 
2MASS (\S~\ref{sec_results}).  We demonstrate that our dual-database cross-correlation search is 
more sensitive to T dwarfs than previous searches performed on SDSS or 2MASS alone, and take 
advantage of the high degree of completeness to T dwarfs attained in our search to estimate the T 
dwarf space density in the solar neighborhood (\S~\ref{sec_density}).  We discuss reasons for the 
omission of the newly identified T dwarfs in previous SDSS and 2MASS searches and draw lessons 
from our experience in cross-correlating large imaging databases  in \S~\ref{sec_discussion}. Finally, 
we outline the improved prospects for finding brown dwarfs cooler than spectral type T8 in a future 
iteration of the SDSS/2MASS cross-match using the much expanded Fifth Data Release (DR5) of the 
SDSS imaging survey (\S~\ref{sec_conclusion}).

\section{TARGET SELECTION: CROSS-MATCHING 2MASS AND SDSS} 
\label{sec_xmatch}

Our targets were selected from the 2099 deg$^2$ imaging footprint of SDSS DR1.  We used the 
combined optical (from SDSS) and near-IR (from 2MASS) characteristics of cataloged objects to 
identify suitable targets.  This Section details our cross-correlation approach and the target selection 
process.

\subsection{Cross-Correlation Approach \label{sec_approach}}

Rather than first identifying candidate brown dwarfs from one survey (e.g., SDSS) and subsequently 
investigating their parameters in the other (2MASS) to look for suitable ultra-cool dwarf candidates, our 
target selection was based on a simultaneous consideration of object parameters in both SDSS and 
2MASS.  This approach effectively allows us to decreas the number of requirements for object 
identification in either survey (e.g., minimum signal-to-noise ratio per band, number of bands in which 
the object is detected, number of error flag settings), and enables the identification of bona-fide objects 
at lower signal-to-noise ratios or with suspect error flags.  As a result, we can probe deeper and to a 
greater completeness level than can be reliably done in either survey alone.  

For the dual-database search we used a cross-comparison engine developed for this project at the 
NASA/IPAC Infrared Science Archive (IRSA) in collaboration with the National Partnership for 
Advanced Computational Infrastructure (NPACI) and the NVO.  The engine compared the positions of 
all sources contained in the 2MASS All-Sky PSC to those in the \verb|BestDR1| SDSS catalog and 
selected only those pairs of objects in the two databases that matched a pre-set $z-J$ color criterion 
(\S~\ref{sec_cull0}). The 2MASS PSC and SDSS DR1 catalogs were stored locally. Cross-comparison 
is input/output intensive, and was optimized by dividing the catalogs into declination strips that were 
sorted and cross-correlated in parallel. The comparison was executed on commodity hardware. A 
Web-based interface supported filtering of the resulting set of candidates by their attributes, e.g., by 
magnitudes or colors.

\subsection{Candidate Selection Based on Position and Color  \label{sec_cull0}}

We designed the cross-matching criteria with the properties of T dwarfs in mind.  
Our primary target selection procedure employed a $6\farcs$0 matching radius and a $z-J\geq2.75
$~mag color cut-off.  That is, we identified all sources in the SDSS DR1 catalog whose coordinates 
were within 6$\farcs$0 of the coordinates of a listed source in the 2MASS All-Sky PSC, and whose 
implied colors were redder than $z-J=2.75$~mag.  The color cut-off, with $z$ based on the SDSS AB 
sinh magnitude system \citep{fukugita_etal96, lupton_etal99} and $J$ on the 2MASS Vega magnitude 
system, ensured sensitivity to most T dwarfs, although also included objects with spectral types as 
early as L3. The matching radius was designed to be inclusive of objects with appreciable proper 
motions, while at the same time to avoid an unmanageable number of candidates.  As implemented, 
our cross-match is 100\% complete to objects with proper motions up to 1$\farcs$5~year$^{-1}$, 
based on the maximum difference between the observing epochs of 2MASS \citep[1997 June to 2001 
February;][]{skrutskie_etal06} and SDSS DR1 \citep[2000 April to 2001 June;][]{stoughton_etal02, 
abazajian_etal03}.  Thus designed, the 2MASS-PSC/SDSS-DR1 cross-match produced 860,040 ultra-
cool dwarf candidates fitting the initial color and position criteria over the 2099~deg$^2$ area of SDSS 
DR1.

\subsection{Further Selection Based on Color, Brightness, and Morphology \label{sec_cull1}}

Having completed the initial positional and color selection from the 2MASS All-Sky PSC and SDSS 
DR1 databases, we applied a secondary set of selection criteria to eliminate the majority of spurious 
candidates, as detailed below:
\begin{enumerate}
\item $z\leq21.0$ mag; \label{crit_z}
\item $i>21.3$ mag (SDSS 95\% completeness limit) or $i-z\geq3.0$ mag; \label{crit_i}
\item $g>22.2$ mag and $r>22.2$ mag (SDSS 95\% completeness limits); \label{crit_gr}
\item $J>14$ mag; \label{crit_J}
\item SDSS object flag setting \verb|type = 6| (i.e., SDSS point sources only); \label{crit_type}
\item 2MASS object flag setting \verb|ext_key = NULL| (i.e., not extended in 2MASS) and 
\verb|gal_contam = 0| (i.e., not contaminated by a nearby 2MASS extended source); \label{crit_ext}
\item 2MASS object flag setting \verb|mp_flg = 0| (i.e., not marked as a known minor planet). 
\label{crit_mp}
\end{enumerate}

The $z$-band limiting magnitude requirement (criterion~\ref{crit_z}) corresponds approximately to the 
level at which the completeness of the SDSS survey drops to zero \citep{stoughton_etal02}, and was 
set to weed out very low signal-to-noise ($S/N<3$) sources.  Criterion \ref{crit_i} effectively selects $i$-
band dropouts in SDSS: potential T dwarfs that are either undetected at $i$ or are very red in $i-z$.  
Similarly, criterion \ref{crit_gr} states that any candidate T dwarf should not be detected in either $g$ or 
$r$ bands.  Criterion \ref{crit_J} requires a more detailed explanation.  The $J>14$ magnitude cut-off 
was imposed to minimize the large number of candidates representing the cross-identification of a 
bright-star artifact in SDSS (e.g., a filter glint or a diffraction spike, especially near saturated stars) with 
the (unsaturated) image of the same star in 2MASS.    While this magnitude cut-off prevents us from 
potentially finding very bright nearby T dwarfs, in all likelihood all such $J\leq14$~mag ($z\leq17
$--18~mag) T dwarfs have already been found in SDSS, where they should be detectable at high 
signal to noise in both $z$ and $i$ bands.  Still, criterion \ref{crit_J} may also discard any objects 
redder than $z-J\sim4$~mag: potential $>$T8 dwarfs.  However, with expected absolute magnitudes 
$M_J\gtrsim17$, such very late dwarfs would have to be within $\sim$3~pc of the Sun to be detected 
at $J\leq14$~mag in 2MASS, and would likely have multi-arcsecond per year proper motions.  These 
proper motions would be much larger than our 1$\farcs$5~year$^{-1}$ proper motion completeness 
limit, and hence our 2MASS/SDSS-DR1 cross-match would be insensitive to them from the start.  
Therefore, criterion \ref{crit_J} incurs negligible penalty on our ability to recover T dwarfs, while it 
significantly decreases the number of artifacts posing as T dwarf candidates.  The remaining criteria 
ensure that the identified candidates are not known artifacts or flux measurements of the blank sky 
(see also discussion in \S~\ref{sec_lessons}) in SDSS (criterion \ref{crit_type}), that they are not 
extended or contaminated by nearby extended sources in either SDSS or 2MASS (criterions \ref
{crit_type} and \ref{crit_ext}; though see \S~\ref{sec_lessons}), and that the candidates are not known 
minor planets in 2MASS (criterion \ref{crit_mp}).  Application of the additional criteria limited the 
number of T dwarf candidates to 45,409, or 5.3\% of the initial number.

No other criteria based on 2MASS and SDSS object flags were applied.  In particular, we did not 
discriminate against candidates marked as single-band detections, potential cosmic rays, electronic 
ghosts, and other artifacts in either database.  The reasoning for this was that the optical/near-IR 
cross-match may recover low signal-to-noise objects mistakenly marked as artifacts in either 
database.

\subsection{Identification of Erroneous 2MASS/SDSS Matches \label{sec_cull2}}

The final round of automated candidate culling involved rejecting mis-associations among 2MASS 
and SDSS point sources.  It was our experience that, In most cases, a 2MASS star was erroneously 
associated with a fainter nearby SDSS star (not seen in 2MASS) rather than with its true SDSS 
counterpart.  Thus, although the actual 2MASS star was not redder than $z-J=2.75$~mag, a match 
with $z-J\geq2.75$~mag was reported.  In this simple scenario there are two SDSS objects in the 6$
\arcsec$-radius circle (i.e., two ``positional'' matches), and one of them appears to fit the imposed color 
criterion (one ``color'' match).  In reality, neither of the two SDSS objects has the colors of a T dwarf 
and both are bluer.  More generally, either single or multiple 2MASS objects may each have multiple 
``color'' and ``positional'' matches in SDSS, especially in denser stellar fields.  As in the simple 
example case, it remains true that each 2MASS candidate that has fewer ``color'' than ``positional'' 
matches in SDSS is most probably the result of a spurious alignment of different objects.  Such 
spurious alignments accounted for the overwhelming majority (97.4\%) of the T-dwarf candidates 
remaining after the previous cull (\S~\ref{sec_cull1}).  On the other hand, candidates for which the 
number of ``color'' and ``positional'' matches were equal remained potential bona-fide brown dwarfs.  
Our database cross-match (\S~\ref{sec_cull0}) produced the numbers of both ``color'' ($N_C$) and 
```positional'' ($N_P$) matches for all candidates.  Thus, we were easily able to screen against 
candidate T dwarfs in 2MASS that had fewer ``color'' than ``positional'' matches (i.e., $N_C<N_P$) in 
SDSS.

A possibility remains in the above scenarios that some bona fide T dwarf candidates may nevertheless 
get thrown out in the described procedure.  For example, in the case of thhe single 2MASS object 
matched to one of two SDSS objects, it is possible that the fainter of the two SDSS objects is indeed 
the one visible in 2MASS, in which case its $z-J$ color is red and the object is a probable T dwarf, 
whereas the brighter SDSS object is blue and is undetected in 2MASS.  This may occur because the 
$J$-band limiting magnitude of 2MASS ($J=16.1$~mag at the 99\% completeness level) is brighter 
than the $z$-band limiting magnitude of SDSS ($z=20.5$~mag at the 95\% completeness limit), so a 
main sequence star in SDSS with $z-J<3.4$~mag (but still sufficiently red not to be detected at $g$ 
and $r$) and $J>16.1$~mag could remain undetected in 2MASS.  We explored this possibility in each 
$N_C<N_P$ case by comparing the SDSS and 2MASS coordinates for each match.  If another bright 
($r<22.2$~mag and $i<21.3$~mag) and bluer ($z-J<2.75$~mag) SDSS star was found within 1$
\arcsec$ of the 2MASS source, the match was discarded.  This runs the risk of throwing out very close 
($\leq$1$\arcsec$) star---brown dwarf pairs (potential binaries).  However, given that $1\arcsec$ is the 
approximate seeing-limited resolution threshold of 2MASS and SDSS, a star---brown-dwarf binary 
with a smaller separation would have been unresolved anyway.  The above procedure was multiply 
checked to ensure that it did not miss any good candidates.  We note that a theoretical possibility still 
exists, in which a high-proper motion T dwarf passes during the 2MASS imaging epoch within 1$
\arcsec$ of a reddish star detected in SDSS (at $i$ and $z$ only), but not in 2MASS, and then moves 
to beyond 1$\arcsec$ from the star during the SDSS imaging epoch.  In this rare scenario the reported 
match would have $N_C=1$ and $N_P=2$, i.e., $N_C<N_P$, and would be discarded because the 
2MASS position of the T dwarf and the SDSS position of the infringing star would be within 1$\arcsec$ 
of each other.  We believe that such pathological cases are very rare, even in moderately dense stellar 
fields, and have chosen to disregard them to streamline our automated candidate selection.

Of the 45,409 candidates remaining after the cull described in \S~\ref{sec_cull1}, 654 (1.4\%) were 
such that $N_C=N_P$ (i.e., potential bona-fide brown dwarfs) and 44,755 (98.6\%) were such that 
$N_C<N_P$ (likely erroneous matches).  Of the latter, 506 survived the proximity criterion described in 
the preceding paragraph, and thus a total of $654+506=1160$ (0.13\% of all initial candidates) 
potential brown dwarfs remained.  

\subsection{Visual Selection of Candidates \label{sec_cull3}}

The 1160 candidates produced by the automated culling were examined through visual comparison of 
the 2MASS and SDSS images.  The examination confirmed that the majority were artifacts, such as 
cosmic rays in SDSS and persistence or line--128 artifacts in 2MASS\footnote{Various 2MASS artifacts 
are described at \url{http://www.ipac.caltech.edu/2mass/gallery/anomalies/ .}}, or faint background 
stars whose $r$- and $i$-band SDSS magnitudes were strongly affected by scattered light from the 
bright haloes of nearby saturated stars.  

The final inspection stage left us with 82 ``good'' ultra-cool dwarf candidates (0.0095\% of all initial 
candidates).  Sixty-three of these are new objects and nineteen were already known T (11) and L (8) 
dwarfs.  We describe our observational follow-up of the new candidates in \S~\ref{sec_followup} and 
present the results of our search in \S~\ref{sec_results}.  
The reliability of our cross-match in recovering the previously known T dwarfs is discussed in \S~\ref
{sec_completeness}.

\section{FOLLOW-UP OF BONA-FIDE CANDIDATES}
 \label{sec_followup}

The 63 new candidate ultra-cool dwarfs were the subject of an imaging and spectroscopic follow-up 
campaign.  To confirm the existence of the candidates, we imaged them with the Palomar 1.5~meter 
telescope, the Shane 3~meter telescope, and the University of Hawaii 2.2~meter telescope.  Further 
characterization of the most promising and/or confirmed candidates was obtained spectroscopically 
with Keck/LRIS in the optical or with IRTF/SpeX in the IR, or through 3--8~$\micron$ imaging with {\sl 
Spitzer}/IRAC.  Twenty-eight of the 63 new candidates were potential T dwarfs and all were followed 
up.  Twenty of the remaining candidates are likely L dwarfs based on their optical and near-IR colors, 
and did not require further imaging confirmation because they were detected in multiple bands in 
2MASS and SDSS at relatively high signal to noise ratios (S/N$>$10).  Finally, a set of 15 candidates 
near bright stars were followed up only with imaging, but not with spectroscopy, to confirm their 
existence and their red optical minus near-IR colors.  The spectroscopic characterization of the new 
candidate L dwarfs and of the candidates near bright stars is still on-going.  

\subsection{Ground-Based Imaging Follow-Up \label{sec_imaging_followup}}

Thirty of the candidate brown dwarfs were imaged in the Gunn $i$ and $z$ bands in queue-scheduled 
mode with the Palomar 1.5 meter automated telescope between 2004 March and 2005 December.  
The telescope operation and data acquisition have been described in detail in \citet{cenko_etal06}.  
Total integrations were 60~min at $i$ and 30~min at $z$, taken in series of 2~min-long exposures.  
The telescope pointing was dithered in a non-redundant circular disk pattern by up to 3$\arcmin$ 
between exposures in right ascension and in declination to allow the simultaneous reconstruction of a 
sky image.  The attained limiting (Vega) magnitudes were $i\approx25$~mag and $z\approx21$~mag.  
The $z$-band imaging depth approximately matched the depth of the SDSS $z$-band images, while 
our $i$-band exposures were somewhat deeper and allowed us to measure $i-z$ colors of the coolest 
and reddest ($i-z\sim4$~mag) potential T dwarfs that did not have $i$-band detections in SDSS.

Nine of the 15 candidates near bright stars were imaged in the \citet{bessel90} $I$-band filter with the 
Prime Focus Camera (PFCam) on the Shane 3~meter Lick Observatory telescope on 18 April 2007.  
Total integrations ranged between 8 and 120~min, taken in series of 1~min exposures, dithered along 
a box pattern on the array.  With the $I$-band imaging we tested whether the objects were red enough 
($I-z\gtrsim1.5$~mag) to be L or T dwarfs.  Such a check was necessary because in all cases the 
SDSS $i$-band data at these locations were contaminated by filter glints or saturation columns from 
the nearby bright star.

Another set of 24 candidates, some of which had already been imaged with the Palomar 1.5 meter 
telescope, were also imaged at $J$ band with the Ultra Low Background Camera (ULBCAM; Loose et 
al. 2007, in preparation) on the University of Hawaii 2.2 meter telescope.  Total integrations were 
90~s, consisting of two 45~s exposures dithered by 45$\arcsec$.  The attained imaging depth was 
$J\approx20$~mag, 4~magnitudes fainter than the 99\% completeness level of the 2MASS catalogue.

A final set of 11 candidates were imaged at $J$ band with the slit viewing cameras on the IRTF/SpeX 
and Keck/NIRSPEC instruments.  None of these were confirmed to be real.  These were likely the 
results  of alignments between asteroids and noise spikes or just between noise spikes in the two 
databases.

Altogether we identified 24 probable new L and T dwarfs through ground-based imaging and through 
inspection of the high signal-to-noise detections in 2MASS and SDSS.   We obtained further optical or 
near-IR spectroscopy (\S~\ref{sec_followup_spec}) and/or 3.6--8.0~$\micron$ {\sl Spitzer} photometry 
(\S~\ref{sec_followup_spitzer}) for 6 of the 24 new probable candidates.  We list optical and near-IR 
photometry for the 24 new L and T candidates and the 19 known L and T dwarfs in Table~\ref
{tab_izJHK}.  Seven of the 15 candidates near bright stars also remain as possible L dwarfs.  Although 
confirmed as real objects, the optical photometry of these seven candidates remains unreliable, and 
they require spectroscopy to check whether they are ultra-cool.   Because of their less likely 
confirmation as L or T dwarfs, we have listed these separately (Table~\ref
{tab_bright_star_candidates}) and have not counted them toward the 24 probable and bona-fide L 
and T candidates.  The remainder of the 82 candidates were discarded as being background M stars, 
bright star artifacts, or other 2MASS and SDSS artifacts.  These are listed in Table~\ref
{tab_junk_candidates}.

\subsection{Ground-Based Spectroscopic Follow-up \label{sec_followup_spec}}

The epochs and instrumental set-ups for the various spectroscopic observations are detailed in 
Table~\ref{tab_spectr_obs}.

\subsubsection{Optical Spectroscopy with Keck/LRIS}

We used the Keck Low Resolution Spectrograph \citep{oke_etal95} to obtain an optical spectrum of 
our first confirmed candidate, 2MASS~J01040750--0053283, on 03 Jan 2003 UT. A 400 line/mm 
grating blazed at 8500~\AA\ was used with a 1$\arcsec$ slit, a 2048$\times$2048 CCD, and the 
OG570 order blocking filter to block flux shortward of 5700~\AA. This produced 7-\AA\ resolution ($R
\approx900$) spectra covering the range 6300-Ð10100~\AA.  To minimize slit losses we oriented the 
slit along the parallactic angle.  Two separate exposures were obtained, with a 2$\arcsec$ dither 
along the slit between the two to mitigate the effect of bad pixels. A 1200-s exposure was taken at the 
first position and a 300-s exposure at the second.

Data were reduced and calibrated using standard {\sc IRAF} routines. As the array was read out in 
dual-amplifier mode, we subtracted off the bias for the separate halves of the array using the overscan 
applicable to each amplifier then stitched the two halves back together. Quartz-lamp flat-field 
exposures taken of the inside of the telescope dome were used to normalize the response of the 
detector. Individual spectra were traced and extracted using the {\sc apextract} routine and a sky 
background subtracted. Wavelength calibration was achieved using neon and argon arc lamp 
exposures taken immediately after the program object, and then the two separate spectra summed. 
Finally, the summed spectrum of the science target was flux calibrated using observations of the 
standard Hiltner~600 \citep{hamuy_etal94} taken the previous night and with the same setup. The 
data have not been corrected for telluric absorption, so atmospheric O$_2$ bands near 6850--6900~
\AA\ and 7600--7700~\AA, and H$_2$O bands near 7150--7300~\AA, 8150--8350~\AA, and 
8950--9650~\AA\ are still present in the spectrum (see \S~\ref{sec_2mass0104}).

\subsubsection{Near-IR Spectroscopy with IRTF/SpeX}

Four other candidates, 2MASS~J00521232+0012172, 2MASS~J01075242+0041563, \\
2MASS~J01262109+1428057, and 2MASS~J15461461+4932114, were observed spectroscopically 
at IRTF with SpeX \citep{rayner_etal03} between 2005 September and 2006 December.  All 
observations were taken in prism mode with the 0$\farcs$5 slit, resulting in a resolution of $R\sim150
$.  The slit was rotated to the parallactic angle for all targets.  We employed a standard A--B--B--A 
nodding sequence along the slit to record object and sky spectra.  Flat-field and argon lamps were 
observed immediately after each set of target and standard star observations for use in instrumental 
calibrations.  Standard stars were used for flux calibration and telluric correction.  All reductions were 
carried out in standard fashion using the SpeXtool package version 3.2 \citep{cushing_etal04, 
vacca_etal03}.

\subsection{Imaging Follow-Up with {\sl Spitzer}/IRAC \label{sec_followup_spitzer}}

For three of the candidates (one of which, 2MASS~J12144089+6316434, was subsequently 
independently discovered by \citealt{chiu_etal06}) we obtained 3.6--8.0~$\micron$ imaging 
observations with all four channels of the IRAC camera \citep{fazio_etal04} on the {\sl Spitzer} Space 
Telescope.  The data were acquired between June and November 2005 as part of {\sl Spitzer} 
program 244.  All observations shared a common {\sl Spitzer} Astronomical Observation Request 
(AOR) design.  Each target was observed with the same 5-position Gaussian dither pattern in all 4 
channels.  The dither pattern started with the target near the center of the array, and with subsequent 
relative offsets distributed within a radius of $\approx$100$\arcsec$ of the initial position.  The frame 
times were 12 or 30~s, which yielded net exposure times of 10.4 or 26.8~s per pointing, respectively.  
The total exposure time per target per filter was 52 or 134~s (Table~\ref{tab_spitzer_phot}).

The data were reduced with the S14.0.0 version of the {\sl Spitzer} data processing pipeline at the {\sl 
Spitzer} Science Center (SSC).  For each raw frame the IRAC pipeline software removes electronic 
bias, subtracts a dark sky image generated from observations of low stellar density regions near the 
ecliptic pole, flat-fields the data using a flat field generated from high-background observations near 
the ecliptic plane, and then linearizes the data using laboratory pixel-response  measurements
\footnote{The data reduction pipeline is described in greater detail in the IRAC Data Handbook 
(Version 3.0) at \url{http://ssc.spitzer.caltech.edu/irac/dh/iracdatahandbook3.0.pdf .}}.  For each 
science exposure the data reduction pipeline produces a Basic Calibrated Data (BCD) frame: an 
image reduced in the above manner and flux-calibrated with respect to photometric standard stars.  
For each AOR the pipeline also produces a high signal to noise median-combined ``post-BCD'' image 
from all individual dithered exposures of the science target.  For our purposes we used the BCD 
frames to measure object photometry because all targets were bright enough to be detected in the 
separate BCD frames and the individual measurements could be averaged for an empirical 
determination of magnitude errors.  The flux in the BCD and post-BCD frames is in units of MJy ster$^
{-1}$, which was converted back to DN, Janskys, and then to magnitudes using the values of {\sl 
FLUXCONV}, {\sl Calfac}, and the zero magnitude fluxes for each IRAC channel listed in Table~5.1 of 
the IRAC Data Handbook.

We measured target fluxes in each of the four camera channels in 3 pixel-radius apertures.  A 
measure of the local sky background was obtained from annuli with inner radii of 10 pixels and outer 
radii of 20 pixels.  This combination of  target aperture and background annulus radii represents one 
of the standard constructs for aperture photometry with IRAC (Table~5.7 of the IRAC Data Handbook) 
for which aperture corrections have been determined from bright standard stars to better than 2\% 
accuracy.   The flux for each target was obtained as the average of the aperture-corrected 
measurements from all five individual dithers.  The standard deviation of the mean of the 
measurements was used as an estimate of the flux error, to which we added in quadrature the 2\% 
uncertainty in the aperture correction.  When converting to magnitudes on the Vega system, the 
uncertainty in the zero magnitude flux (Table 5.1 of the IRAC Data Handbook) was also added in 
quadrature to the flux error.  Table~\ref{tab_spitzer_phot} lists the log of IRAC observations and the 
photometry for the three objects confirmed with {\sl Spitzer}.

\section{RESULTS}
 \label{sec_results}

Of the 24 likely L and T dwarfs identified during our visual inspection and observational follow-up (\S~
\ref{sec_followup}), two were confirmed as new T dwarfs (\S\S~\ref{sec_2mass1546}--\ref
{sec_2mass1324}), and four were found to be L dwarfs (\S\S~\ref{sec_2mass0052}--\ref
{sec_2mass0917}).  The remaining 20 are likely to be L dwarfs based on their optical/near-IR colors 
(\S~\ref{sec_remaining_candidates}).  We list optical/near-IR colors for all candidates and spectral 
types for the six confirmed new ultra-cool dwarfs in Table~\ref{tab_izJHK_colors}.
We also show optical ($z$-band) finding charts for the six new bona fide dwarfs in Figure~\ref
{fig_findcharts}.  Color-color diagrams of $z-J$ versus $i-z$ and of $z-J$ versus $J-K_S$ colors for all 
candidates are shown in Figures~\ref{fig_zJiz} and \ref{fig_zJJK}.    Figure~\ref{fig_irac_lt} overlays the 
mid-IR colors of three objects observed with {\sl Spitzer}/IRAC on an IRAC color-color diagram of ultra-
cool dwarfs from \citet{patten_etal06}.

In this section we discuss the six new bona-fide ultra-cool dwarfs confirmed with ground-based 
spectroscopy or {\sl Spitzer} mid-IR imaging.  We also present mid-IR photometry of the T dwarf 
2MASS~J12144089+6316434, independently discovered by \citet{chiu_etal06}.

\subsection{New and Confirmed T Dwarfs \label{sec_new_Ts}}

\subsubsection{2MASS~J15461461+4932114: a New T2.5 Dwarf \label{sec_2mass1546}}

Having identified 2MASS~J15461461+4932114 as a candidate T dwarf in the cross-match, we 
obtained a 0.8--2.5~$\micron$ $R\sim$150 prism spectrum of the object with IRTF/SpeX (\S~\ref
{sec_followup_spec}).  The spectrum is shown in Figure~\ref{fig_2mass1546_spec} alongside SpeX 
spectra of T1--T4 standards from \citet{burgasser_etal06b}.  We determined the spectral type of 
2MASS~J15461461+4932114 using both visual inspection and calibrated spectral type indices 
following the unified T dwarf spectral classification scheme of \citet{burgasser_etal06b}.  The 0.8--2.5~
$\micron$ spectrum of 2MASS~J15461461+4932114 is visually best matched by the SpeX prism 
spectrum of the T3 standard 2MASS~J12095613--1004008.  Use of the five primary water and 
methane indices of \citet{burgasser_etal06b} yielded spectral types in the T1--T3 range with a formal 
mean and standard deviation of T2.0$\pm$0.7.  Combining the two classification approaches, we 
adopt a final spectral type of T2.5$\pm$1.0 for 2MASS~J15461461+4932114.

2MASS~J15461461+4932114 has not been previously identified neither in 2MASS nor in SDSS.  
While early T dwarfs do not stand out from main-sequence stars in 2MASS because of their 
unremarkable ($0.5\lesssim J-K_S\lesssim1.5$~mag) near-IR colors, they are readily identifiable in 
SDSS because of their very red far-optical colors ($i-z>3$~mag).  At a $z$-band AB magnitude of 
19.06, 2MASS~J15461461+4932114 is brighter than the majority of the known T dwarfs in SDSS 
DR1.  Therefore, the omission of this T dwarf in compilations of ultra-cool dwarfs from SDSS \citep
{knapp_etal04, chiu_etal06} is intriguing.  A reason for its omission may be its proximity ($\approx$2$
\arcsec$) to another point source of comparable brightness (Fig.~\ref{fig_findcharts}).  Other 
possibilities are discussed in \S~\ref{sec_missed_Ts}.  

A comparison of the 2MASS and SDSS data, obtained 2.9 years apart, show that the proper motion of 
2MASS~J15461461+4932114 ($0\farcs57\pm0\farcs14$ year$^{-1}$) differs from that ($\approx0
\farcs0$~year$^{-1}$) of the nearby source, hence the two are unrelated.


\subsubsection{2MASS~J13243553+6358281: a New Early T Dwarf \label{sec_2mass1324}}

We followed up 2MASS~J13243553+6358281 through imaging with {\sl Spitzer}/IRAC.  This object 
was independently discovered by co-authors D.L.\ and J.D.K.\ in a separate survey of high proper 
motion objects in 2MASS.  A near-IR spectrum of 2MASS~J13243553+6358281 
is reported in \citet{looper_etal07}. 
Here we present only the {\sl Spitzer} data (Table~\ref
{tab_spitzer_phot}).  We use these data together with the optical and near-IR photometry of 
2MASS~J13243553+6358281 from SDSS and 2MASS to obtain a photometric estimate of its spectral 
type.

The 3.6--8.0~$\micron$ IRAC colors of ultra-cool dwarfs were recently characterized by \citet
{patten_etal06}.  A comparison of the IRAC colors of 2MASS~J13243553+6358281 with those of 
known L and T dwarfs (Fig.~\ref{fig_irac_lt}) illustrates that 2MASS~J13243553+6358281 is redder 
than the latest L dwarfs and is comparable in [3.6$\micron$]--[8.0$\micron$] color to T3--T6 dwarfs.  
The [4.5$\micron$]--[5.8$\micron$] color of 2MASS~J13243553+6358281 is marginally redder than 
those of any of the known T dwarfs.  Overall, the location of 2MASS~J13243553+6358281 on the 
IRAC color-color diagram in Figure~\ref{fig_irac_lt} is closest to the locus of T3--T5 dwarfs.

The optical and near-IR photometry of 2MASS~J13243553+6358281 point to it being an early T dwarf.  
This is apparent from Figure~\ref{fig_zJJK}, where the $z-J$ and $J-K_S$ colors of 
2MASS~J13243553+6358281 are near the boundary between the T dwarf and the L dwarf loci.  A 
more detailed comparison with the $z-J$ and $J-K$ colors of ultra-cool dwarfs from \citet
{knapp_etal04} and \citet{chiu_etal06} reveals that 2MASS~J13243553+6358281 has optical/near-IR 
colors most consistent with those of T0--T1 dwarfs.  Considering both the IRAC and the optical/near-IR 
data, we conclude that the spectral type of 2MASS~J13243553+6358281 is between T0 and T5 and 
assign it as T2.5:.   The discrepancy between the spectral types inferred from the mid-IR and the 
optical/near-IR data may indicate binarity, as is relatively common among early-T dwarfs \citep
{liu_etal06, burgasser_etal06}.  Unlike the new T2.5 dwarf 2MASS~J15461461+4932114 (\S~\ref
{sec_2mass1546}), which lies close to a field object, an obvious reason for the omission of 
2MASS~J13243553+6358281 from previous compilation of ultra-cool dwarfs from SDSS does not 
present itself immediately.  Possibilities are discussed alongside 2MASS~J15461461+4932114 in \S~
\ref{sec_missed_Ts}.

\subsubsection{2MASS~J12144089+6316434: a Confirmed Mid-T Dwarf \label{sec_2mass1214}}

The identification of 2MASS~J12144089+6316434 as a T dwarf was unknown at the time when it 
surfaced as a candidate in our cross-match.  It was subsequently announced in the recent update on 
ultra-cool dwarfs in SDSS by \citet{chiu_etal06}, where it is classified as a T3.5$\pm$1.0 based on 
spectroscopy with IRTF/SpeX.  We have not obtained spectroscopic observations of 
2MASS~J12144089+6316434.  However, our {\sl Spitzer}/IRAC photometry (Table~\ref
{tab_spitzer_phot}; Fig.~\ref{fig_irac_lt}) for this object is in agreement with the classification of 
\citeauthor{chiu_etal06}

\subsection{New and Confirmed L dwarfs \label{sec_Ls}}

Although our cross-matching criteria were not designed with L dwarfs in mind, eight known L3.5--L8 
dwarfs were recovered and 24 more L dwarf candidates were found.  These were allowed by the $i-z$ 
color cut because their $i$-band magnitudes were fainter than the $i=21.3$~mag 95\% completeness 
limit of SDSS (see criterion \ref{crit_i} in \S~\ref{sec_cull1}).  However, most of the L dwarfs were still 
sufficiently bright (above the $i\approx23.0$~mag 3$\sigma$ detection limit) to be detected at $i$.

\subsubsection{2MASS~J00521232+0012172: an Unusually Blue L2 Dwarf \label{sec_2mass0052}}

This object has a rather red $z-J=3.14\pm0.15$~mag color and a blue $J-K_S=0.90\pm0.19$~mag 
color compared to other L dwarfs, which set it in the T dwarf locus on a $z-J$ vs.\ $J-K_S$ diagram 
(Fig.~\ref{fig_zJJK}).  The $H-K_S=0.10\pm0.21$~mag color of 2MASS~J00521232+0012172 is also 
unusually blue and T dwarf-like.  However, its $i-z=1.96\pm0.16$~mag is quite ordinary for an L dwarf, 
much lower than the $i-z\gtrsim3.0$~mag typical of T dwarfs (Fig.~\ref{fig_zJiz}).  Therefore, 
2MASS~J00521232+0012172 is probably an L dwarf.   A comparison of the 0.8--1.3$\micron$ section 
of our 0.8--2.5~$\micron$ $R\sim150$ IRTF/SpeX spectrum (Fig.~\ref{fig_2mass0052_spec_Ls}a) to 
SpeX spectra of L dwarf standards from \citet{cushing_etal05} narrows down the spectral type range of 
2MASS~J00521232+0012172 to L2$\pm$1.  

As indicated by its blue near-IR colors, the SED of 2MASS~J00521232+0012172 differs from those of 
the standard L dwarfs in several important ways.  For one, 2MASS~J00521232+0012172 has an 
unusually pronounced 1.3~$\micron$ peak, a deep H$_2$O absorption band between 1.35--1.5~$
\micron$, and somewhat depressed $K$-band continuum (all of which explain its T-dwarf-like $z-J$ 
and $J-K_S$ colors).  In addition, the spectrum of 2MASS~J00521232+0012172 exhibits weaker than 
usual 
vanadium oxide absorption at 1.05~$\micron$, in line with the weaker metal oxide features in sub-
solar metallicity ultra-cool dwarfs \citep{gizis97, lepine_etal03b}.
Notably, however, the spectrum of 2MASS~J00521232+0012172 does not show unusually strong 
FeH absorption bands (at 0.99 and 1.09~$\micron$), as expected of metal-poor ultra-cool dwarfs.  
Furthermore, although its $J-K_S$ color is blue compared to other L dwarfs, it is still much redder than 
the $J-K_S<0.3$~mag colors of other known L sub-dwarfs \citep{burgasser_etal03b, burgasser04b}.  
Hence, 
while 2MASS~J00521232+0012172 exhibits some features of a metal poor early L dwarf, it
does not fit well into the current extent of our knowledge of sub-stellar effective temperature and 
metallicity.

Several other candidate mildly metal-deficient L dwarfs with comparably blue $J-K_S$ colors are 
discussed in \citet{knapp_etal04}, \citet{chiu_etal06}, and \citet{cruz_etal07}.  \citeauthor{cruz_etal07} 
point to the high inferred tangential velocities ($\sim$100~km s${^-1}$) of their two blue L dwarfs as an 
indication that they belong to the galactic thick disk population, and are therefore at least partially 
metal-deficient.  However, based on an inferred spectrophotometric distance of 65$\pm$7~pc and a 
measured proper motion of $0\farcs15\pm0\farcs12$~year$^{-1}$, the tangential velocity of 
2MASS~J00521232+0012172 ($46\pm37$~km~s$^{-1}$) is poorly constrained and fully consistent 
with the 1$\sigma$ range ($\sim$15--55~km~s$^{-1}$) of tangential velocities of L2 dwarfs in the solar 
neighborhood \citep[see][Fig.~3]{schmidt_etal07}.  Therefore, the metal-poor nature of 
2MASS~J00521232+0012172 is uncertain.

As an alternative to sub-solar metallicity, \citet{knapp_etal04}, \citet{chiu_etal06}, and \citet
{cruz_etal07} point out that the blue $J-K_S$ colors of some L dwarfs may be caused by a reduction in 
cloud condensate opacity.  Models of sub-stellar photospheres that incorporate more efficient dust 
sedimentation tend to produce bluer near-IR colors \citep[e.g.,][]{marley_etal02}.  The SED of 
2MASS~J00521232+0012172 may well be affected by both factors: mild metal deficiency and a 
marginally reduced condensate opacity.  The photospheres of metal-poor L dwarfs are indeed thought 
to have a reduced condensate formation efficiency because of the observed persistence of TiO bands 
and \ion{Ti}{1} and \ion{Ca}{1} lines in their optical spectra \citep{burgasser_etal03b, 
burgasser_etal07b}---features that normally weaken and disappear at the M/L transition \citep
{kirkpatrick_etal99}.  Identifying and studying L dwarfs with similarly blue near-IR colors will produce 
adequate anchor points to establish a sub-stellar metallicity scale, and will provide important empirical 
constraints for future theoretical efforts to model sub-stellar photospheres.

\subsubsection{2MASS~J0104075--005328: a New L5 Dwarf in the SDSS Early Data Release \label
{sec_2mass0104}}

The object 2MASS~J0104075--005328 was identified as a candidate ultra-cool dwarf in a preliminary 
run of our cross-matching algorithm on small subsets of the 2MASS and SDSS databases, namely the 
2MASS Second Incremental Data Release (IDR2) and the SDSS Early Data Release \citep[EDR;][]
{stoughton_etal02}.  The candidate was spectroscopically confirmed as an L5 dwarf with Keck/LRIS 
(\S~\ref{sec_followup_spec}) and first announced in \citet{berriman_etal03}.
An optical $R\approx900$ spectrum of 2MASS~J0104075--005328 is shown in Figure~\ref
{fig_2mass0104_spec} alongside spectra of L3--L7.5 dwarfs from \citet{kirkpatrick_etal99, 
kirkpatrick_etal00}.  The spectral type of 2MASS~J0104075--005328 was assigned following the 
guidelines in \citet{kirkpatrick_etal99}.  In particular, we used the CrH-$a$, Rb-$b$/TiO-$b$, Cs-$a$/
VO-$b$, and Color$-d$ ratios defined in \citet{kirkpatrick_etal99}, which measure the strengths of 
metal hydride, metal oxide, and alkali absorption, and the redness of the spectrum.  From these 
spectral ratios we infer a spectral type of L5$\pm$0.5 for 2MASS~J0104075--005328, in agreement 
with its bye-eye placement in the L3--L7.5 sequence in Figure~\ref{fig_2mass0104_spec}.

\subsubsection{2MASS~J01262109+1428057: a Young L2 Dwarf \label{sec_2mass0126}}

This object was given priority for spectroscopic follow-up because it is $\approx$0.3~mag redder in 
$z-J$ than the locus of known L dwarfs at a comparable $J-K_S$ color (Fig.~\ref{fig_zJJK}), i.e., 
probably ultra-cool and intriguingly distinct from known L and T dwarfs.  Our IRTF/SpeX spectrum 
contains the typical features of an L dwarf but does not fit well into the optically-anchored L dwarf 
spectral sequence (Fig.~\ref{fig_2mass0126_spec}a) because of its unusually bright $H$-band peak 
and relatively bright $K$-band continuum.  At first glance, such an inconsistency might not be 
unexpected, since it is known that the optical L dwarf spectral type sequence does not trace a 
continuous spectroscopic progression of in the near-IR \citep[e.g.,][]{mclean_etal03}.  This is because 
the optical and the near-IR regions of the spectrum sample different physical conditions in the L dwarf 
photosphere.  However, the discrepancy between the near-IR SED of 2MASS~J01262109+1428057 
and the SEDs of other L dwarfs with similar spectral types is much larger in this case, with the $H$-
band peak of 2MASS~J01262109+1428057, in particular, being much brighter.  Closer scrutiny of the 
spectrum of 2MASS~J01262109+1428057 reveals further differences from the spectra of other L 
dwarfs.  For example, 2MASS~J01262109+1428057 lacks the strong \ion{Na}{1} and \ion{K}{1} 
doublets between 1.14~$\micron$ and 1.26~$\micron$, indicating that it has lower surface gravity than 
field L dwarfs.  Such an interpretation is also supported by the more peaked shape of the $H$-band 
continuum of 2MASS~J01262109+1428057 compared to that of the other L3--L8 dwarfs.  Similarly 
peaked $H$-band continua are characteristic of young ultra-cool dwarfs \citep{lucas_etal01, 
luhman_etal04, kirkpatrick_etal06, allers_etal07}, where the effect is thought to be caused either by 
enhanced water vapor absorption on either side of the $H$ band \citep{luhman_etal04, allers_etal07} 
or by a decrease in the strength of CIA $H_2$ \citep{kirkpatrick_etal06} at low surface gravity.

In Figure~\ref{fig_2mass0126_spec}b we compare the spectrum of 2MASS~J01262109+1428057 to 
the spectra of known low surface gravity objects: a late M giant (IY~Pup) and two young early L dwarfs, 
2MASS~J01415823--4633574 \citep[L0; 1--50~Myr;][]{kirkpatrick_etal06} and G~196--3B \citep[L2; 
60--300~Myr;][]{rebolo_etal98, kirkpatrick_etal01}.  Similarly to 2MASS~J01262109+1428057, the 
spectra of the comparison low surface gravity objects also display peaked $H$-band continua, to 
varying extents.  The spectrum of G~196--3B provides the closest match to the spectrum of 
2MASS~J01262109+1428057.   Therefore, we conclude that 2MASS~J01262109+1428057 is also a 
young early-L dwarf.  

A precise spectroscopic classification of 2MASS~J01262109+1428057 is challenging.  Young L 
dwarfs have yet to be incorporated into the spectral classification schemes for $>$1~Gyr-old field L 
dwarfs.  Many of the spectroscopic indices currently used for L dwarf classification are based on the 
strengths of alkali, metal-oxide, or water absorption signatures in the optical and the near-IR, and are 
gravity sensitive.  Hence, they are inadequate indicators of effective temperature for the lower surface 
gravity photospheres of young L dwarfs.  A large sample of young L dwarfs that will allow detailed 
characterization of gravity- and temperature-sensitive features is still to be presented (K.~Cruz et al.\ 
2007, in preparation).  Nevertheless, a recent spectroscopic study of young, mostly late-M dwarfs by 
\citet{allers_etal07} includes two early L dwarfs, and can serve as a reference.  In particular, \citet
{allers_etal07} observe that the strength of water absorption in the blue end (1.49--1.56~$\micron$) of 
the $H$-band spectra of late M and early L dwarfs is approximately independent of surface gravity, 
and hence may be an adequate proxy for effective temperature.  We apply the $H$-band water 
continuum index defined by \citeauthor{allers_etal07} to both 2MASS~J01262109+1428057 and 
G~196--3B and find that the two objects have identical index values.  Therefore, we adopt a spectral 
type of L2$\pm$2 for 2MASS~J01262109+1428057, where the uncertainty includes the error ($\pm
$1.0 sub-type) in the spectral classification of G~196--3B, the scatter ($\pm$1.0 sub-type) in the index 
relation of \citeauthor{allers_etal07}, and the error of our index measurement (corresponding to $\pm
$1.5 sub-type).  

Finally, we observe that the depth of the 1.7--2.1~$\micron$ water absorption band in the spectrum of 
2MASS~J01262109+1428057 is somewhat shallower than in G~196--3B.  To the extent to which the 
continuum in this wavelength range may be gravity-sensitive \citep[e.g., potentially due to the 
diminishing strength of CIA H$_2$ absorption with decreasing surface gravity;][]{borysow_etal97, 
kirkpatrick_etal06}, this may indicate that 2MASS~J01262109+1428057 has comparable or slightly 
higher surface gravity, and hence may be marginally older than G~196--3B (0.06-0.3~Gyr).  Such a 
conclusion is backed by the slightly weaker VO absorption bands at 1.05~$\micron$ and 1.18~$
\micron$ in 2MASS~J01262109+1428057 than in G~196--3B.  Therefore, 2MASS~J01262109
+1428057 is probably 0.1--0.3~Gyr old.

We note that while we classify 2MASS~J01262109+1428057 alongside G~196--3B as an L2 dwarf, 
the 0.8--2.5~$\micron$ continua of both of these young L2 dwarfs are redder than those of early L 
dwarfs in the field (e.g., see Fig.~\ref{fig_2mass0052_spec_Ls}a).  Unusually red near-IR colors are a 
recurrent property of young L dwarfs \citep[e.g.,][]{kirkpatrick_etal06}, probably caused either by slower 
sedimentation of dust grains or by relative weakness of CIA H$_2$ in their lower surface gravity 
photospheres.  Unusually red far-optical and near-IR colors are therefore promising criteria for 
discovering more young ultra-cool dwarfs in the future.  


\subsubsection{2MASS~J09175418+6028065: a Probable Mid-L Dwarf \label{sec_2mass0917}}

2MASS~J09175418+6028065 has a similar $J-K_S$ color to other L dwarfs, but sits redder of the L 
dwarf locus in $z-J$ (Fig.~\ref{fig_zJJK}).  This photometric behavior is the same as observed for 
2MASS~J01262109+1428057 (\S~\ref{sec_2mass0126}), which we established to be a young ($
\gtrsim$0.1~Gyr) L dwarf.  Therefore, 2MASS~J09175418+6028065 could also be a young L dwarf.

Alternatively, 2MASS~J09175418+6028065 could also be a late M giant or a carbon star.  Very late 
($>$M7) giants often display similar far-optical and near-IR colors. In principle, we could use the 
proper motion of 2MASS~J09175418+6028065 between the SDSS and 2MASS imaging epochs to 
discern whether it is a nearby L dwarf with a high proper motion, or a distant, nearly stationary M giant 
or carbon star.  However, the proper motion of 2MASS~J09175418+6028065 ($0\farcs16\pm0\farcs18
$~year$^{-1}$) does not weigh more on either of these possibilities.  Judging merely by its apparent 
magnitude ($J=17.16\pm0.27$~mag), if 2MASS~J09175418+6028065 were a very late M giant, it 
would have to be at a distance of $\gtrsim$100~kpc, well into the halo of our Galaxy.

We followed up 2MASS~J09175418+6028065 as part of our {\sl Spitzer}/IRAC imaging of identified 
candidates.  The 3.6--8.0$\micron$ colors of 2MASS~J09175418+6028065 lie well within the L dwarf 
locus (Fig.~\ref{fig_irac_lt}), and not far from the locus of the early T dwarfs.  We therefore tentatively 
conclude that 2MASS~J09175418+6028065 is a mid-L dwarf.  Future near-IR spectroscopy of this 
object will establish its spectral type and address the possibility of it being another low surface gravity 
young L dwarf.


\subsubsection{Remaining Candidates: Probable L Dwarfs \label{sec_remaining_candidates}}

Twenty candidate ultra-cool dwarfs still await spectroscopic follow-up.  These are marked with crosses 
in the color-color diagrams in Figures~\ref{fig_zJiz} and \ref{fig_zJJK}.  The relatively red $z-J$ and 
blue $J-K_S$ colors of a handful of these appear very similar to those of T dwarfs (Fig.~\ref{fig_zJJK}).  
However, their $i-z$ colors are too blue ($i-z<3.0$~mag) for T dwarfs (Fig.~\ref{fig_zJiz}), indicating 
that their spectral types are earlier than T.  Their T dwarf-like $z-J$ and $J-K_S$ colors may thus be an 
indication of mild metal deficiencies, as we hypothesized for 2MASS~J00521232+0012172 (\S~\ref
{sec_2mass0052}).   

An alternative reason for the very red $z-J$ colors in the few cases above may be sought in the non-
logarithmic behavior of the inverse hyperbolic sine (asinh) magnitude system \citep{lupton_etal99} of 
SDSS.  For very faint objects, detected at signal-to-noise ($S/N$) levels less than 5, the deviation of 
the asinh magnitude system from a logarithmic one with the same flux zero point becomes $>
$0.05~mag, and quickly rises to 0.6~mag at $S/N\approx1$.  The $S/N=5$ level in SDSS $z$ 
corresponds to $z\approx20.8$~mag \citep{york_etal00}, with some minor variations among the 
individual CCDs due to their quantum efficiencies.  Hence, fainter objects will have $z-J$ colors 
(where $z$ is on the asinh scale and $J$ is on the logarithmic scale) that would be $\geq$0.05~mag 
too red compared to what identical, but apparently brighter, objects would have.  However, all of our 
candidates are brighter than $z=20.8$~mag (Table~\ref{tab_izJHK}).  Therefore their $z-J$ colors are 
not subject to such artificial reddening. 

The remaining candidates fall well into the L dwarf loci on the near-IR color-color diagrams in 
Figures~\ref{fig_zJiz} and \ref{fig_zJJK} and are thus probably L dwarfs.  Occasional late M giants or 
carbon stars among these are also possible.  Future spectroscopic observations of all of the remaining 
candidates promise to uncover several more metal-poor or young objects.

\section{THE SURFACE AND SPACE DENSITY OF T DWARFS}
 \label{sec_density}

The results from the present experiment may not be ideal for an analysis of the surface and space 
density of T dwarfs.  Our sensitivity to T dwarfs is limited by the relatively small cross-matching radius 
(6$\arcsec$), which may have excluded some nearby objects with very high proper motions ($>$1.5$
\arcsec$~year$^{-1}$), and the total number of T dwarfs known in the 2099~deg$^2$ area probed by 
the cross-match is small.  Both the cross-match radius and the area over which the cross-correlation 
was performed were chosen conservatively to limit the candidate identifications to a number 
manageable for a pilot project.  Nevertheless, the discovery of two new T dwarfs (\S~\ref
{sec_new_Ts}) and a very high recovery fraction of known SDSS DR1 T dwarfs (see \S~\ref
{sec_completeness}) point to a high degree of completeness to objects that match our cross-match 
criteria.  That is, albeit imprecise, an estimate of the T dwarf density from this sample would be 
accurate.  We take advantage of this opportunity and address the issue of the surface and space 
density of T dwarfs in \S~\ref{sec_surface_dens} and ~\S~\ref{sec_space_dens}, respectively.  To 
overcome the limitations of our sample arising from its limited sensitivity to high proper motion T 
dwarfs, we combine our data set with the results from previous searches for T dwarfs in SDSS DR1 
\citep{knapp_etal04, chiu_etal06}.

\subsection{Completeness of the 2MASS/SDSS Cross-Match to T Dwarfs  \label{sec_completeness}}

To estimate the completeness of our cross-match, we test if it successfully recovered all known T 
dwarfs in the overlap area between 2MASS and SDSS DR1.  We identified 11 known T dwarfs from 
the cross-match, whereas a total of 13 were known in SDSS DR1: 11 from \citet[][and references 
therein]{knapp_etal04}\footnote{In their work published shortly after the release of SDSS~DR1, \citet
{knapp_etal04} include one additional T dwarf, SDSS~J042348.57$-$041403.5, also observed with 
the SDSS telescope.  However, this T dwarf is outside the official SDSS footprint, and therefore does 
not contribute to the T dwarf statistics in the 2099~deg$^2$ area of SDSS DR1 (J.\ Knapp 2007, 
private communication).}, one from \citet{chiu_etal06}, and one from \citet{burgasser_etal99}\footnote
{2MASS~J121711.19$-$031113.3 (T7.5) was discovered in 2MASS and coincidentally resides in the 
SDSS DR1 footprint.  However, it has not been included in any compilations of T dwarfs in SDSS until 
now.}. One of the two T dwarfs that we did not recover, SDSS~J151603.03+025928.9 is a T0$\pm$1.5 
dwarf \citep{knapp_etal04} with a color $z-J=2.49$~mag: too blue for our $z-J\geq2.75$~mag cut-off.  
This indicates that our cross-match is not 100\% sensitive to brown dwarfs of spectral type $\lesssim
$T1.5.  The other overlooked T dwarf, SDSS~J020742.83+000056.2 \citep[T4.5;][]{geballe_etal02}, is 
classified as a galaxy (object \verb|type=3|) at both $i$ and $z$ bands in SDSS DR1, and hence was 
missed by our search which focused only on point sources (\verb|type=6|; criterion \ref{crit_type} in \S~
\ref{sec_cull1}).  With the re-reduction of the SDSS data for Data Release 2 \citep{abazajian_etal04}, 
SDSS~J020742.83+000056.2 has been re-classified as a point source, and we expect that it would 
have been successfully recovered by us, had we applied the cross-match to SDSS~DR2.   All of the 
remaining known SDSS DR1 T dwarfs were recovered.  Therefore, barring other misclassifications of 
brown dwarfs as extended sources in either SDSS DR1 or 2MASS, we expect the cross-match to be 
sensitive to 100\% of brown dwarfs with spectral types $\geq$T2 within the combined flux limits of the 
two surveys and within the employed 6$\arcsec$ matching radius.  

The total number of T dwarfs in the 2099~deg$^2$ footprint of SDSS~DR1 that are visible both in 
2MASS and SDSS is thus 15: 13 known previously (12 of which were recovered in prior searches in 
SDSS DR1) and 2 presented here.  These are listed in Table~\ref{tab_sdss_tdwarfs}.  We only 
consider T dwarfs that are detected both in SDSS and in 2MASS, in agreement with the construction of 
our cross-match and with the adopted confirmation procedure for T dwarf candidates found in SDSS 
data \citep{knapp_etal04, chiu_etal06}.

Throughout the remainder of this analysis we shall assume that the two new T dwarfs discovered in 
our 2MASS/SDSS-DR1 cross-match complete the census of T dwarfs down to the combined sensitivity 
limits of the two databases in the region of the SDSS DR1 footprint.  This assumption is stronger than 
what can be justified based solely on the present cross-match because it incorporates T dwarfs with 
higher proper motions and bluer $z-J$ colors than allowed by our cross-match criteria (\S~\ref
{sec_cull0}).  The premise is based on the combined sensitivity of the 2MASS/SDSS-DR1 cross-match 
and of previous searches for T dwarfs in SDSS DR1 that have not imposed such color and proper 
motion cut-offs \citep[][and references therein]{knapp_etal04, chiu_etal06}.  We can not empirically 
verify the robustness of this assumption because we have not tested the completeness of SDSS-only T 
dwarf identifications for objects with $z-J<2.75$~mag and proper motions $>1\farcs5$~year$^{-1}$, 
i.e., T dwarfs to which our approach was not 100\% sensitive.  However, given the overall success rate 
($12/15 = 80\%$) of SDSS-only T dwarf identifications, and the small fraction of L and T dwarfs with 
proper motions $>1\farcs5$~year$^{-1}$ (15\%, based on the compilation of L and T dwarf proper 
motions and parallaxes at {\url DwarfArchives.org}), we expect that only $0.20\times0.15=3.0\%$ of T 
dwarfs would be missed by a combination of the previous SDSS-only searches and our present 
2MASS/SDSS cross-match.  That is, the combined recovery rate for T dwarfs is $\approx$97\%.  Since 
the 3.0\% incompleteness correction amounts to a fraction (0.45) of a T dwarf, we will ignore it in the 
rest of our analysis.

\subsection{Surface Density \label{sec_surface_dens}}

Given 15 T known dwarfs in the 2099~deg$^2$ footprint of SDSS DR1, the surface density of T dwarfs 
in SDSS (that are also detectable in 2MASS) is $(7.1\pm1.8)\times10^{-3}$~deg$^{-2}$, or 1 per 
140~deg$^2$, in agreement with previous determinations (1 in 140~deg$^2$ [\citealt
{collinge_etal02}] or 1 in 100~deg$^2$[\citealt{knapp_etal04}]).  We use this surface density to 
estimate the completeness of the number of known T dwarfs in the latest SDSS Data Release 5 \citep
[DR5;][]{adelmanmccarthy_etal07} and in 2MASS.

Given the imaging surface area (8000~deg$^2$) of SDSS DR5, we would expect 57$\pm$15 T dwarfs 
that are detectable both in SDSS and in 2MASS.  Forty-six of all known and published T dwarfs 
(including the present two) have been detected in SDSS DR5.\footnote{Twelve more T dwarfs have 
been published based on SDSS data \citep{knapp_etal04, chiu_etal06}.  However, these are located 
in areas of SDSS that have not been publicly released, either because of not satisfying the image 
quality criteria or because of being part of SDSS-II (see {\url http://www.sdss.org}; G.\ Knapp 2007, 
private communication).}  Thus, we find that the current census of T dwarfs in SDSS DR5 is between 
60 and 100\% complete.  We expect that our 2MASS/SDSS cross-correlation technique will be highly 
instrumental in identifying any remaining T dwarfs in SDSS DR5.  With regard to the 2MASS T dwarf 
census, over the 4$\pi$ steradians (41253~deg$^2$) of the entire sky we would expect 294$\pm$76 T 
dwarfs in 2MASS.  Only 97, including the present 2, have 2MASS identifications.  That is, the 2MASS 
T-dwarf census is $\sim$33\% complete.  The majority of the ``missing'' 2MASS T dwarfs are likely 
outside of the SDSS footprint, and have not yet been identified either because they are early T's with 
near-IR colors that are indistinguishable from those of L dwarfs or earlier-type stars, or because they 
are projected along the galactic plane (at $\vert b\vert < 15\degr$), which has not yet been scrutinized 
for T dwarfs in detail.  Preliminary results from 2MASS-based searches for redder T dwarfs and 
extending to lower galactic latitudes 
are presented in \citet{looper_etal07}.

\subsection{Space Density \label{sec_space_dens}}

We estimate the space density of T dwarfs based on the $\approx$97\% complete (\S~\ref
{sec_completeness}) sample of 15 T dwarfs in the SDSS DR1 footprint.   In principle, the factor of 3 
larger population of known T dwarfs in the SDSS DR5 footprint can produce a more precise estimate 
of the T dwarf space density.  However, the unknown incompleteness of the SDSS DR5 T dwarf 
population means that such an estimate will be less reliable than one based on our DR1 sample.

We use a Monte Carlo approach to simulate the observed population of T dwarfs in SDSS DR1 and 
2MASS.  Given that the detectability of T dwarfs in SDSS and 2MASS is strongly dependent on their 
absolute magnitudes and colors, or equivalently, on their spectral types, we adopt the observed 
distribution of T dwarf spectral types in SDSS DR1 as an input to our Monte Carlo simulations.   We 
have updated the published spectral types of the previously known T dwarfs to conform with the 
unified T dwarf near-IR classification scheme of \citet{burgasser_etal06b}.  In addition, we use the 
known multiplicity properties of T dwarfs based on high-angular resolution imaging studies \citep[][and 
references therein]{burgasser07}.  We tabulate the observed spectral type distribution of the SDSS-
DR1/2MASS T dwarfs and the binary rate at each spectral type in Table~\ref{tab_absmags}.   Because 
of the small number of objects in our sample, we divide it in 3 bins, containing dwarfs with spectral 
types T0--T2.5, T3--T5.5, and T6--T8.  

We run independent Monte Carlo simulations for each of the three T spectral type bins and adjust the 
input volume density and binary fraction until we reproduce the observed data.  In deciding whether a 
simulated T dwarf is detected, we take into account its heliocentric distance, absolute magnitude, 
binarity, and the detection limits of SDSS and 2MASS.  We detail all of these considerations in \S~\ref
{sec_considerations}.  Unlike in the construction of our cross-match, we do not impose an upper limit 
on the proper motion of T dwarfs, or an explicit lower limit on their $z-J$ color.  This is because T-dwarf 
searches in SDSS that have employed the $i$-dropout technique \citep{fan_etal01, knapp_etal04, 
chiu_etal06} have not discriminated against proper motion or the $z-J$ colors, as long as the 
candidates were detected in 2MASS.  We describe the implementation of our considerations into the 
Monte Carlo analysis in \S~\ref{sec_montecarlo}.  In \S~\ref{sec_spacedens_anal} we summarize the 
result from the simulations and infer the space density of T dwarfs.

\subsubsection{Input Considerations for the Monte Carlo Analysis \label{sec_considerations}}

\paragraph{Heliocentric Distances and Simulation Volumes.}
The Monte Carlo simulations were performed by randomly generating T dwarfs in a spherical volume 
centered on the Sun, and by checking whether the simulated T dwarfs would be sufficiently bright to 
be detected in SDSS and 2MASS.  The radius of the spherical volume was chosen specifically for 
each bin of T sub-types so that it would be sufficiently large to include any {\sl binary} T dwarfs that are
4$\sigma$ outliers in the $z$ apparent magnitude.  The standard deviation $\sigma$ was obtained as 
the quadrature sum of the standard deviation of the $z$ absolute magnitude for the given T sub-type 
(as estimated from the $J$-band absolute magnitude and the $z-J$ color; Table~\ref{tab_absmags}) 
and the standard deviation of the survey limiting magnitude (see below).  We chose the SDSS $z$ 
apparent magnitude as the determining factor for the simulation volume because, as we shall see 
below, the SDSS $z$ images probe a similar heliocentric volume for T dwarfs as the 2MASS $J$, $H$, 
and $K_S$ images, and because a $z$-band detection is required for all T dwarfs discussed here.  
Our simulations thus account for 99.997\% of all observable binary T0--T8 dwarfs in SDSS and 
2MASS, and for virtually 100.000\% of all observable single T0--T8 dwarfs.

\paragraph{T-Dwarf Absolute Magnitudes and Colors.}
We estimated mean $J$-band absolute magnitudes for each spectral sub-type bin based on the 
trigonometric parallax studies of \citet{dahn_etal02}, \citet{tinney_etal03}, and \citet{vrba_etal04}.  
Known T-dwarf companions to stars with {\sl Hipparcos} parallaxes \citep{perryman_etal97} were also 
included.  Binarity, whenever known, was accounted for by assuming that each of the components in a 
binary system contributes equally to the combined flux.  Mean $z$, $H$, and $K_S$-band absolute 
magnitudes were estimated from the $J$-band absolute magnitudes and from the mean optical/near-
IR colors for each spectral type bin (Table~\ref{tab_absmags}).  We used the compilations of SDSS $z
$-band and near-IR (MKO) photometry of T dwarfs in \citet{knapp_etal04} and \citet{chiu_etal06}, and 
near-IR (2MASS) photometry compiled on the {\url DwarfArchives.org} web site.  Where necessary, we 
converted the MKO near-IR photometry to the 2MASS photometric system using the transformations 
for ultra-cool dwarfs from \citet{stephens_leggett04}.  Additional synthesized SDSS $z$-band 
photometry for three T dwarfs was taken from \citet{dahn_etal02}.  No $z$-band photometry has been 
published for T8 dwarfs (none are known in SDSS).  However, judging by the small range of the 
variation in $z-J$ color between spectral types T4 and T7.5, T8 dwarfs like havely similar $z-J$ colors 
(we assume $z-J\sim3.5$~mag).  

We draw the absolute magnitude of each T dwarf in our Monte Carlo simulations from a Gaussian 
distribution with mean and standard deviation equal to the mean and the standard deviation of the 
absolute magnitudes of T dwarfs in its spectral sub-type bin.  The scatter of absolute magnitudes in 
each bin includes three components: one due to the intrinsic luminosity scatter of T dwarfs at a given 
spectral type, another due to photon and detector noise, and a third due the accuracy of the 
photometric calibration.  The combined treatment of these three error terms in a single, empirically 
quantifiable parameter greatly simplifies our approach.  However, we note that the latter two terms, 
that describe the measurement errors, are only treated in an average sense.  An important systematic 
effect that arises when measurement errors dominate (at $S/N\lesssim5$) is flux over estimation of 
faint sources, a.k.a.\ \citet{malmquist1927} bias.  A source with an intrinsic brightness near the 
sensitivity limit of a measurement is more likely to be detected if noise drives up the measured 
brightness, as opposed to driving it down.  Simulations based on Gaussian noise statistics indicate 
that flux overestimation is $\sim$10\% at the $S/N=5$ level, and $\sim$5\% at $S/N=7$ \citep[see 
section V.3.\ of][]{cutri_etal03}.  Given that we have incorporated at least a zeroth-order treatment of 
the measurement errors in our selection of T dwarf absolute magnitudes, we expect the flux 
overestimation bias to be significantly smaller at these $S/N$ levels.  We will assume that the resultant 
over-abundance of T dwarfs detected in 2MASS and SDSS as a result of Malmquist bias is negligible 
compared to the errors due to small-number statistics.

\paragraph{Survey Limiting Magnitudes: 2MASS {\bf $JHK_S$}.}
We limit our analysis to objects detectable at a signal-to-noise ratio of $S/N\geq5$ in all three 2MASS 
bands or at $S/N\geq7$ in at least one 2MASS band.  These are the object detection requirements in 
the 2MASS All-Sky PSC \citep{skrutskie_etal06}.  We estimated the mean and the scatter of the 
2MASS limiting magnitudes at $S/N=5$ and $S/N=7$ at each of the $J$, $H$, and $K_S$ bands from 
the magnitudes of 2700 to 7800 point sources at moderate galactic latitudes ($30\degr < \vert b \vert < 
31\degr$) in the 2MASS Working Database.  The resulting mean $S/N=5$ and $S/N=7$ magnitudes 
and their standard deviations are tabulated in Table~\ref{tab_limmags}.

Figure~\ref{fig_limmags} shows that the distributions of the apparent magnitudes of objects detected at 
$S/N=5$ (solid lines) or at $S/N=7$ (thin lines) in any of the 2MASS filters are well approximated by 
Gaussians.  This is an important observation, as it underscores the fact that the limiting magnitude of 
2MASS, or any survey in general, is not a constant.  The variation of the limiting magnitude is an 
important factor to consider when estimating the completeness of a survey (e.g., through Monte Carlo 
simulations), as it can be used to reflect uncertainties caused by variations in the photometric 
conditions and detector performance during the survey.  Our present analysis demonstrates that a 
simple Gaussian parameterization of the limiting magnitude provides a realistic approximation to the 
complex set of variables that govern survey depth, at least in the case of 2MASS point sources.  We 
performed an independent check on this result by comparing our ensemble-averaged $S/N=5$ and 
$S/N=7$ limiting magnitudes to estimates that can be obtained from the predicted magnitudes of $S/
N=10$ point sources and from the photometric zero points in the various scans of the 2MASS survey.  
The latter parameters are contained in the 2MASS Scan Information Table. \footnote{See \S~IV.8 and 
\S~VI.2 of the 2MASS Explanatory Supplement \citep{cutri_etal03}: {\url http://www.ipac.caltech.edu/
2mass/releases/allsky/doc/explsup.html}.}  We found the two sets of limiting magnitudes to be largely 
indistinguishable.  We have given preference to our approach because of its universal applicability to 
imaging databases other than 2MASS.  Indeed, below we will assume that a similar parameterization 
also holds for SDSS point sources, although we will use a much more limited sample of objects to 
estimate the limiting magnitude and its standard deviation.

\paragraph{Survey Limiting Magnitudes: SDSS {\bf $z$}.}
SDSS object descriptors do not include the $S/N$ ratio of a detection.   However, the information can 
be gleaned from the \verb|psfCounts| and \verb|psfCountsErr| entries ($S/N=\verb|psfCounts|/
\verb|psfCountsErr|$) for each object in the object catalog (\verb|fpObjc|) file for each field. 
An important 
additional consideration in the case of SDSS $z$-band detections of T dwarfs is the large discrepancy 
between the slope of the red-optical continuum of T dwarfs and the throughput curve of SDSS at $z$-
band.  The SEDs of T dwarfs rise over an order of magnitude in strength between 0.8--1.0~$\micron$, 
whereas the throughput of the SDSS $z$ band steadily decreases between 0.85--1.0~$\micron$ due 
to the decreasing quantum efficiency of the optical CCD.  Because most of the red-optical photons of T 
dwarfs are emitted in a wavelength range in which their $z$-band detection is inefficient, the $z$-band 
sensitivity of the SDSS survey toward T dwarfs may be inferior compared to the one for objects with 
bluer, star-like colors.  Therefore, we use information only from the 48 known SDSS T dwarfs to 
estimate the appropriate $z$-band limiting magnitude for the survey.  This effect is not of concern in 
the 1.0--2.3~$\micron$ range probed by 2MASS, as T-dwarf colors are more similar to the colors of 
stars in the near-IR than in the optical.

We limit our analysis to objects detectable at $S/N\geq8.3$ in the SDSS $z$ band.  This limit was 
chosen to correspond to the 0.12~mag upper limit on the $z$-band magnitude error imposed in the 
most recent and broadest search for T dwarfs in SDSS by \citet{chiu_etal06}, which complements our 
cross-match for T dwarfs with spectral types earlier than T2 or with proper motions larger than 1$\farcs
$5~year$^{-1}$ (\S~\ref{sec_completeness}).  For each known SDSS T dwarf we estimate what its 
magnitude would have been if it were detected at the $S/N=8.3$ level using the information for the 
object (number of detected counts, error in the number of counts, background sky flux, effective area of 
the point-spread-function [PSF]), the detector (gain, dark current, read noise), and the observation 
(airmass, atmospheric extinction) available in the appropriate \verb|fpObjc| and \verb|tsField| files for 
each observation.  The mean and the standard deviation of the $S/N=8.3$ detection limit for T dwarfs 
is tabulated in Table~\ref{tab_limmags}, and the distribution of the $S/N=8.3$ magnitudes is plotted 
alongside the distributions of the 2MASS $S/N=5$ and $S.N=7$ magnitudes in Figure~\ref
{fig_limmags}.  
We have also computed the respective SDSS $z$-band $S/N=7$ and $S/N=5$ limiting magnitudes for 
reference (Table~\ref{tab_limmags}).  The mean survey depths to T dwarfs in SDSS (at $S/N=8.3$ at 
$z$ band) and 2MASS (at $S/N=5$) are shown in Table~\ref{tab_survey_depth}.

\paragraph{Binarity.}
All presently known T dwarf multiples are found in $<$1$\arcsec$ binaries that are unresolved in 
ground-based seeing-limited surveys.  Unresolved binarity leads to brighter apparent magnitudes for 
these T dwarfs, and to differences between the component and the observed (systemic) spectral types.  
To correctly account for these effects in our Monte Carlo analysis we assume that a certain fraction of 
the simulated T dwarfs are unresolved binaries.  The binary frequency among T dwarfs is known to be 
strongly dependent on the systemic T sub-type, with binaries among early T dwarfs being much more 
common than binaries among late T dwarfs \citep{liu_etal06, burgasser_etal06}.  This is likely the 
result of a blending of the spectroscopic features of the individual components in the unresolved 
systemic spectrum of a binary that produces a spectrum with intermediate characteristics; e.g., a binary 
comprised of an L/T transition dwarf and a mid-T dwarf will have an intermediate, early-T, systemic 
spectral type \citep{burgasser07}.  We incorporate this effect in our Monte Carlo simulations by 
adopting different binarity frequencies for T0--T2.5 dwarfs (50\%), T3--T5.5 dwarfs (21\%),
and T6--T8 dwarfs (13\%), based on the compilation of L and T dwarf multiplicity from high resolution 
imaging surveys in Table~1 of \citet{burgasser07}.  The actual binary frequency, including 
spectroscopic pairs unresolved in direct imaging may be up to a factor of 2 higher, as found for higher-
mass (spectral types M5--L5) binaries \citep{basri_reiners06}.  The results from our Monte Carlo 
experiments show that a factor of 2 increase in the frequency of T dwarf binaries leads to a $\lesssim
$20\% decrease in the mean space density of T dwarf systems, mostly among early T dwarfs (\S~\ref
{sec_spacedens_anal}).  We consider all binaries to have equal brightness components, in 
accordance with the strong peak near unity in the mass ratio distribution of L and T binaries \citep
{burgasser_etal06, reid_etal06}.

\subsubsection{Monte Carlo Simulations \label{sec_montecarlo}}

For each spectral sub-type bin between T0 and T8, we generated $N_{\rm sim}$ dwarfs within a 
spherical simulation volume of sufficiently large (\S~\ref{sec_considerations}) radius $r_{\rm sim}$.  
We drew their $z$, $J$, $H$, and $K_S$ absolute magnitudes from Gaussian distributions with the 
appropriate means and standard deviations adopted from optical/near-IR colors and $M_J$ absolute 
magnitudes listed in Table~\ref{tab_absmags}.  A fraction $f_{\rm bin, sim}$ of simulated dwarfs in 
each spectral type bin were set to be binaries, and their apparent fluxes were doubled.  This fraction 
was adjusted iteratively throughout each simulation to maintain a fixed fraction $f_{\rm bin, det}$ of 
detected binary systems equal to the fraction $f_{\rm bin, obs}$ of binaries observed in high-resolution 
imaging surveys (\S~\ref{sec_considerations}).  Because of the brighter systemic apparent magnitude 
of unresolved binaries, the fraction of simulated binaries $f_{\rm bin, sim}$ was lower than $f_{\rm bin, 
det}$.

Whether a T dwarf was detectable in SDSS and in 2MASS or not was decided by a comparison of its 
apparent magnitude to the limiting magnitudes of the two surveys at each band.  For each simulated 
observation of a T dwarf we assigned limiting magnitudes at $z$, $J$, $H$, and $K_S$ drawn from 
Gaussian distributions with the corresponding means and standard deviations discussed in \S~\ref
{sec_considerations} and listed in Table~\ref{tab_survey_depth}.  The detection limits in the three 
2MASS bands were assumed to be correlated because the data were taken contemporaneously.  All 
simulated T dwarfs with $z$-band magnitudes fainter than the $z$-band $S/N=8.3$ detection limit or 
than the $z=20.4$~mag threshold imposed by \citet{chiu_etal06} were ignored.  Simulated T dwarfs 
that were fainter than the $S/N=5$ detection limits in at least one 2MASS bands and fainter than the 
$S/N=7$ detection limit in the other two 2MASS bands, were also discarded.  Finally, from the sample 
of detectable T dwarfs we further selected only the fraction that would fall in any given 2099 deg$^2$ 
area of the sky, corresponding to the area of the SDSS DR1 footprint, and considered only these $N_
{\rm det}$ dwarfs as ``detected.''  This treatment correctly reproduces the stochastic errors in the 
number of expected T dwarfs in each spectral sub-type bin in our SDSS-DR1/2MASS cross-match.

The above simulations were repeated 10000 times for each spectral sub-type bin to drive down the 
stochastic errors associated with detecting few (4--6) T dwarfs per spectral type bin per simulation.  
The set of 10000 simulations was then repeated several more times for each spectral sub-type bin to 
iterate the number $N_{\rm sim}$ of simulated T dwarfs in the bin until the mean number of detected T 
dwarfs in the simulation $N_{\rm det}$ converged with the expect mean number of T dwarfs $N_{\rm 
exp}$ per 2099 deg$^2$ unit area in SDSS and 2MASS.  Using a Bayesian approach, as is 
appropriate for the small number statistics regime \citep[e.g.,][]{kraft_etal91}, we find that for any 
spectral type bin, the expectation value $N_{\rm exp}$ equals the number of detected T dwarfs in the 
bin plus one half (see Appendix \S~\ref{sec_appendix}).  The input parameters and details about the 
simulations are listed in Table~\ref{tab_montecarlo}.  Table~\ref{tab_space_dens} shows the results 
for the T dwarf space density at each spectral type, assuming different inputs for the observed 
frequency of T binaries: a binary frequency equal to the observed one \citep[23\% on average;][]
{burgasser07} in direct imaging, a binary frequency equal to twice the observed one (e.g., 
incorporating unresolved spectroscopic binaries), and a hypothetical binary frequency of 0.

We verified that the T dwarfs generated in the Monte Carlo simulations accurately represented the 
population of observed T dwarfs in SDSS by (1) comparing the apparent magnitude distributions of the 
simulated and of the observed T dwarfs, and (2) comparing the fraction of $K_S$- and $H+K_S$-band 
drop-outs between the simulated and the observed populations of T dwarfs.  Histograms of the 
apparent magnitude distributions at each of the $z$, $J$, $H$, and $K_S$ bands are shown in 
Figure~\ref{fig_mag_hist}.  Solid lines show the observed apparent magnitude distribution of the 15 
known T dwarfs in SDSS DR1 and 2MASS, while dashed lines show the apparent magnitude 
distribution of the combined set of $\approx$160,000 T0--T8 dwarfs detected in all of our simulations.  
Kolmogorov-Smirnov (K-S) tests on all histogram pairs show that the probabilities that the observed 
and the simulated distributions originate from the same parent distribution of T dwarf apparent 
magnitudes are 47\%, 57\%, 76\%, and 72\% at $z$, $J$, $H$, and $K_S$ bands, respectively.  That 
is, we find that the magnitude distribution of the simulated and of the observed T dwarf populations are 
in adequate agreement.  

The near-IR colors of T dwarfs are such that they are frequently not detected in all three 2MASS 
bands, dropping below the $S/N\approx3$ detection threshold most often at $K_S$ band, and 
sometimes at both $H$ and $K_S$ bands.  A correct model of the population of T dwarfs in the solar 
neighborhood and of their detectability in SDSS and in 2MASS should adequately predict the rates at 
which T dwarfs are missed at $H$ and $K_S$ bands.  We compare the $H$ and $K_S$ band drop-out 
rates for the known populations of T dwarfs in SDSS DR1 and DR5 to our simulations in Table~\ref
{tab_dropouts}.  The table lists the drop-out rates in two cases: for the entire T0--T8 population and for 
T6--T8 dwarfs only.  As we see, the drop-out rates of the simulated T dwarfs are in line with the 
observed ones within the statistical limitations.  Based on this and on the previous comparison, we 
conclude that the population of T dwarfs simulated in our Monte Carlo analysis provides an adequate 
representation of the observed one in SDSS DR1 and 2MASS.

\subsubsection{Inferred T0--T8 Dwarf Space Density \label{sec_spacedens_anal}}

Summing up the space densities in all spectral type bins, we find that the overall space density of T0--
T8 dwarf systems is $7.0_{-3.0}^{+3.2}\times10^{-3}$~pc$^{-3}$ (95\% confidence interval),
i.e., about one in 140~pc$^3$.  The space densities of early T0--T2.5 systems, the population that was 
not addressed in the 2MASS survey of \citet{burgasser02}, is $0.86_{-0.44}^{+0.48}\times10^{-3}$~pc
$^{-3}$, i.e., $\lesssim$1 in 1000~pc$^3$.  The error estimates on the space densities in the individual
spectral type bins are determined from the 95\% confidence limits on the number of observed T dwarfs 
per bin.  For the small number statistics case, these are described in the Appendix.  The error estimate 
on the overall space density is obtained from the convolution of the probability density distributions of 
all bins, under the assumption that the numbers of detected T dwarfs in all bins are independent of 
each other.  Because we concluded that the census of T dwarfs in SDSS DR1 was $\approx$97\% 
complete (\S~\ref{sec_space_dens}), we do not expect a significant systematic correction due to 
missed T dwarfs.  However, the errors do not include systematic effects that may result from our 
uncertain knowledge of the T dwarf binary fraction.  Our working assumption is that the fraction of 
known T dwarfs that are binaries equals the fraction of resolved systems in direct imaging.  Additional 
unresolved binaries likely exist among the known T dwarfs and may as much as double the T dwarf 
binary fraction (\S~\ref{sec_considerations}).  As seen from Table~\ref{tab_space_dens}, a factor of 2 
increase in the binary fraction leads to a $\approx$14\% decrease of the overall T dwarf space density, 
with the most significant (nearly two-fold) decrease being among early T dwarfs.  Conversely, in the 
hypothetical case in which all T dwarfs are single, the inferred space density of T0--T8 dwarfs is 14\% 
higher.  Therefore, we conclude that our result is not strongly dependent on systematic uncertainties in 
assumed the frequency of T dwarf binaries.

We point out that our space density estimate is true only for the systemic spectral types of T dwarfs.  In 
general, the individual components of binary T dwarfs have spectral types that differ from the 
composite spectral type of the binary.  The difference between the component and systemic spectral 
types can be up to 3--4 spectral sub-types, especially in binaries with early T systemic types \citep
{burgasser07}.  We have not considered the various combinations of component spectral types and 
their corresponding systemic spectral types in our Monte Carlo simulations because this would require 
an assumption for the binary mass ratio distribution of T dwarf binaries, whereas we have strived to 
keep our analysis purely empirical.  Therefore, although we find that the space density of T0--T2.5 
dwarfs is $0.9\times10^{-3}$~pc$^{-3}$ (Table~\ref{tab_space_dens}), if most of these are binaries 
consisting of late-L and mid-T dwarfs, the actual space density of individual objects of spectral type 
T0--T2.5 will be much lower.

Our estimate of the space density of T dwarfs is in good agreement with previous findings based either 
on less complete data from SDSS \citep[0.0068~pc$^{-3}$;][]{collinge_etal02} or on T5--T8 dwarfs 
from 2MASS \citep[$0.006\pm0.004$~pc$^{-3}$;][]{burgasser02}.   Comparing to the space density of 
L dwarfs and earlier-type main sequence stars, we find that T0--T8 dwarfs are a factor of $\approx$1.9 
more common than L dwarfs \citep[$\gtrsim$0.0038~pc$^{-3}$;][]{cruz_etal07}, a factor of 1.5 more 
common than ultra-cool M7--M9.5 dwarfs \citep[0.0049~pc$^{-3}$;][]{cruz_etal07}, and $\approx$8 
times {\sl less} common than 0.1--1.0~$M_\odot$ stars \citep[0.057~pc$^{-3}$;][]{reid_etal99}.  We also 
compare our result to previous semi-empirical analyses of the field brown dwarf population by \citet
{burgasser04, burgasser07} and \citet{allen_etal05} that produce a range of predictions for the sub-
stellar population based on various assumptions for the initial mass function, for the star formation 
history in our Galaxy, and for the luminosity and effective temperature evolution of sub-stellar objects.  
We find that within the framework of these analyses our data are most consistent with a flat mass 
function ($dN/dM\propto M^{0.0}$) in the sub-stellar regime.

The rise of number density from the L's into the T's, and especially toward the late T spectral types, 
indicates that a large number of even cooler, $>$T8, dwarfs may also exist.  These objects have 
remained undetected likely because of their very small intrinsic luminosities.  We make an 
approximate projection of the surface density of such faint and cool dwarfs in \S~\ref{sec_Ydensity}.  

\subsubsection{Upper Limit on the Surface Density of $>$T8 Dwarfs \label{sec_Ydensity}}

Over two-thirds of the number density of T0--T8 dwarfs is expected to be in T dwarfs of spectral types 
between T6 and T8 (Table~\ref{tab_space_dens}).  If we presume that the sub-stellar spectral type (or 
effective temperature) distribution function does not change functional form at spectral types $>$T8 
($T_{\rm eff}\lesssim750$~K), the space density of $>$T8 dwarfs should be at least comparable to that 
of T6--T8 dwarfs.  Thus, potential T9 dwarfs (at projected $M_z\sim20.5$~mag, $M_J\sim17.0$~mag) 
should be detectable out to approximately 8--10~pc at SDSS $z$ and 2MASS $J$, and their 
anticipated number in that volume is 10--21.  Only a twentieth of these (i.e., $\sim$0.5--1) is expected 
to be detectable in the 2099~deg$^2$ area of SDSS DR1, so the lack of an identification of a $>$T8 
dwarf in the present 2MASS/SDSS-DR1 cross-match is not surprising.  We can only put an ($\approx
$95\%) upper limit of 0.003~deg$^{-2}$, i.e., 3 per 1000 deg$^2$, on the surface density of T9 dwarfs 
in SDSS and 2MASS.

However, in the 8000~deg$^2$ area of the complete SDSS survey we would expect 2--4 T9 dwarfs at 
$S/N\geq8.3$ at $z$ band that should also be detectable at $S/N\geq5$ in at least one 2MASS band.  
These may be missing from current compilations of T dwarfs in SDSS because of the relatively recent 
release date of SDSS DR5 (June 2006), because of being below the $S/N\geq5$ cut-off in the other 
two 2MASS bands, or because of the various reasons (\S~\ref{sec_missed_Ts}) that may have led to 
the omission of the two newly-identified SDSS DR1 T dwarfs presented here.  A systematic search for 
such cool T dwarfs by cross-correlating SDSS and 2MASS should recover these T objects.  Given at 
least 2 expected T9 dwarfs in the SDSS DR5 footprint, the probability of finding at least one is $1-{\rm 
e}^{-2}=86\%$.

Even cooler objects, potential Y dwarfs, may also be recovered in an expanded 2MASS/SDSS-DR5 
cross-match.  Given that Y dwarfs are expected to be significantly fainter than the coolest known T 
dwarfs, they will be detectable to much smaller heliocentric distances and, hence, will have higher 
proper motions.  The radius of the present 2MASS/SDSS-DR1 cross-match was chosen conservatively 
to avoid large numbers of spurious alignments between artifacts in 2MASS and SDSS.  However, 
having developed a highly-automated false candidate rejection algorithm (\S\S~\ref{sec_cull1}--\ref
{sec_cull2}), the cross-match radius can be safely enlarged in a future re-iteration to include very high 
proper motion objects and to allow for the larger epoch separation between 2MASS and SDSS DR5.

\section{DISCUSSION}
 \label{sec_discussion}

\subsection{The Two Previously Overlooked T Dwarfs} 

\subsubsection{Reasons for Omission in Previous 2MASS Searches}

The most extensive and complete search for T dwarfs in 2MASS remains that of \citet{burgasser02}.
The two newly identified T dwarfs are at relatively high galactic latitudes ($b\gtrsim50\degr$), in areas 
of the sky that were included in Burgasser's search.  However, their near-IR colors fall outside of 
Burgasser's color-color search box.  To reduce contamination from interloping main sequence stars 
and L dwarfs,  \citeauthor{burgasser02} focused his search on T dwarfs with blue near-IR colors ($J-
H<0.3$~mag and $H-K_S<0.3$~mag) only, corresponding to spectral type $\geq$T5.  Both of the 
newly identified T dwarfs, on the other hand, have early-T spectral types.  Their near-IR colors ($J-
H=0.8$~mag, $J-K_S=0.9$~mag, and $J-H=1.0$~mag, $J-K_S=1.5$~mag, respectively) blend with 
those of the vastly more numerous main sequence stars and L dwarfs.   Therefore, the omission of the 
two new T dwarfs from Burgasser's sample is due to their early T spectral types. 

\subsubsection{Reasons for Omission in Previous SDSS Searches \label{sec_missed_Ts}}

\begin{figure}
\plotone{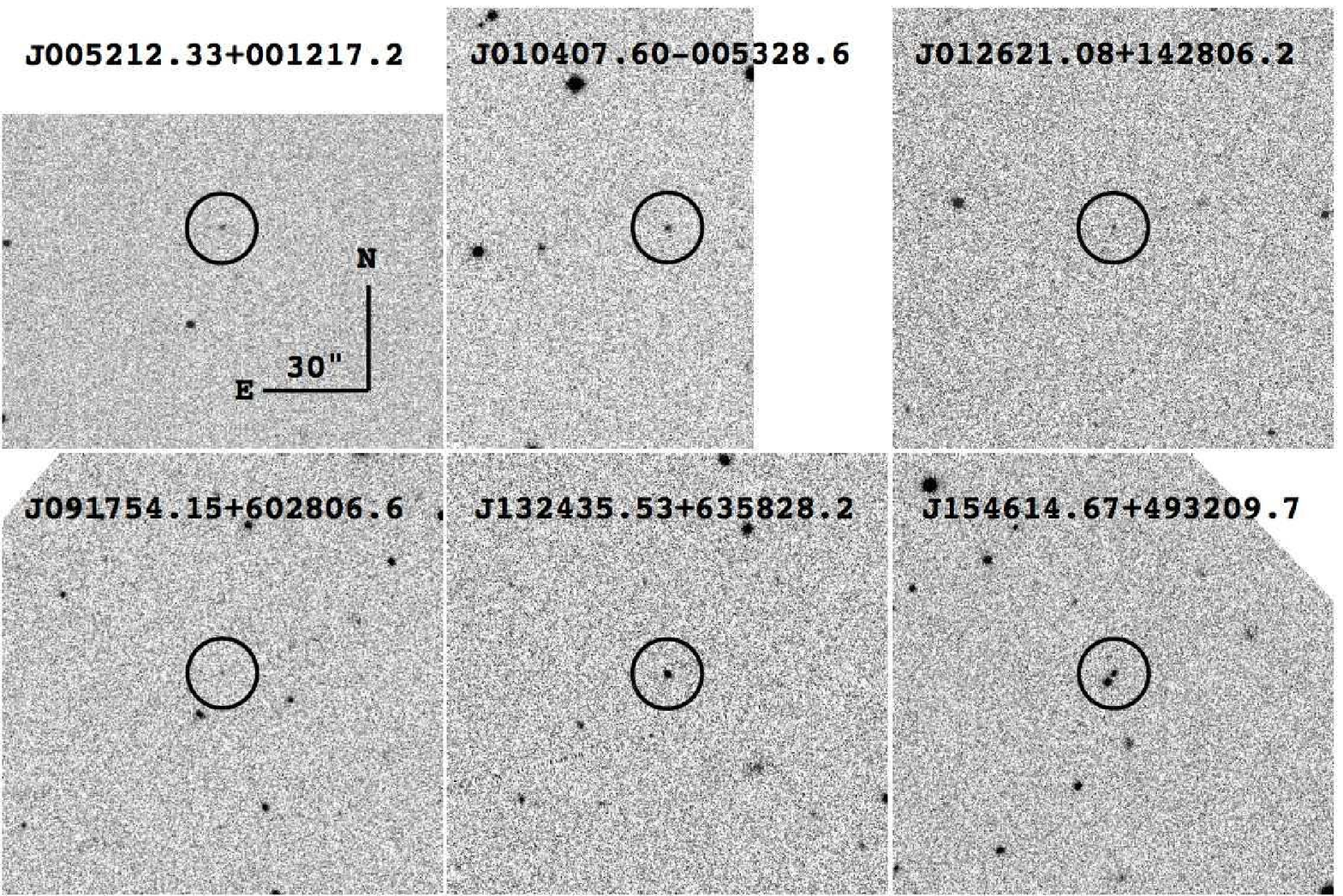}
\figcaption{SDSS $z$-band finding charts for the new ultra-cool dwarfs presented in this work.  The 
coordinate identifiers here follow the SDSS nomenclature and are similar to the 2MASS identifiers 
used throughout the paper.
\label{fig_findcharts}}
\end{figure}

\begin{figure}
\plotone{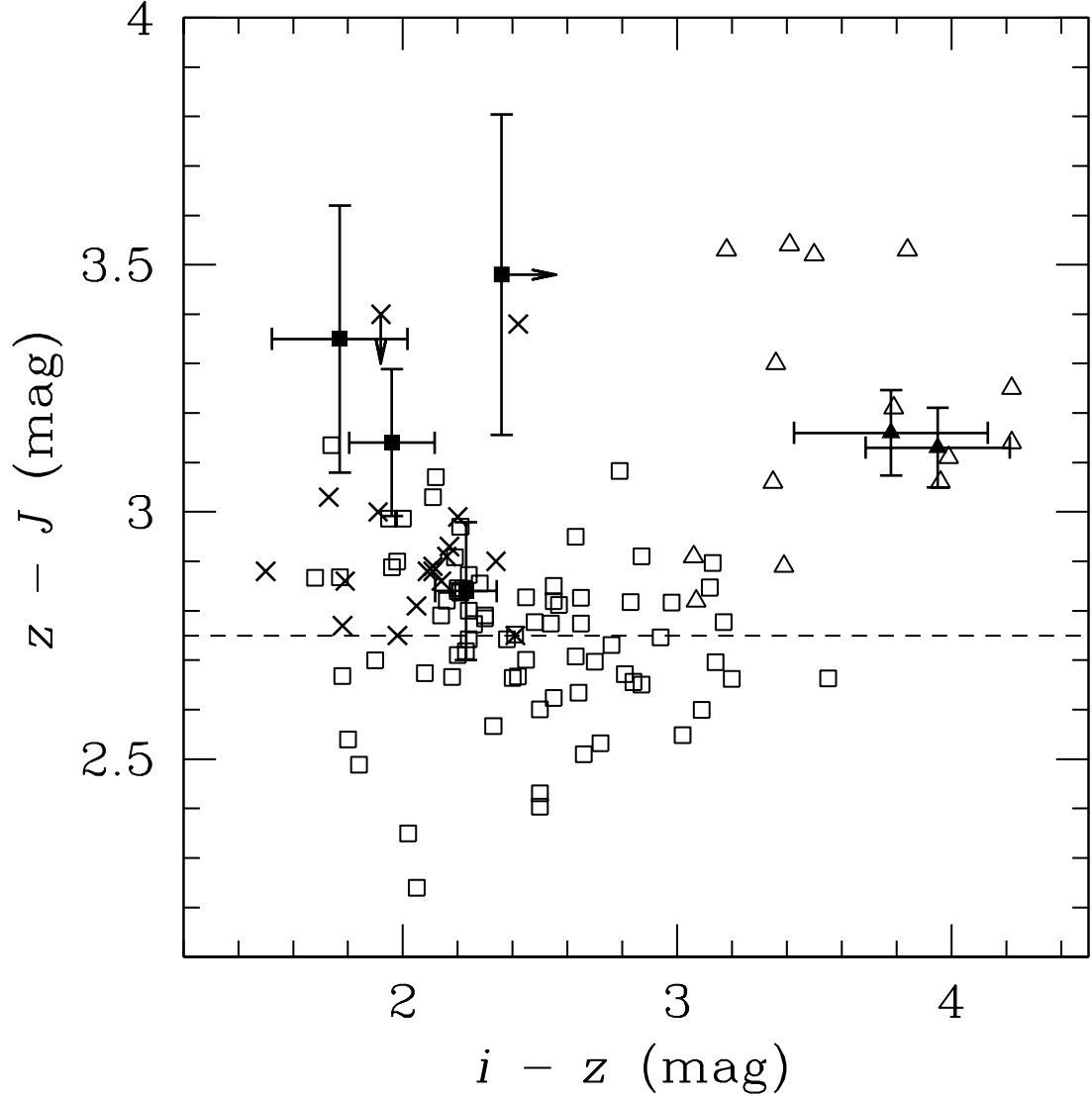}
\figcaption{An SDSS/2MASS $z-J$ versus $i-z$ diagram of known L (open squares) and T (open 
triangles) dwarfs \citep[data from][]{knapp_etal04, chiu_etal06}.  Among the known objects only those 
detected in the SDSS $i$ band ($i<23.0$~mag) are plotted.  The $z-J\geq2.75$~mag color cut used in 
out 2MASS/SDSS-DR1 cross-match is marked by the horizontal dashed line.  Symbols with error bars 
denote the six new bona-fide L (solid squares) and T dwarfs (solid triangles).  In order of increasing $i-
z$, these are: 2MASS~J01262109+1428057, 2MASS~J00521232+0012172, 
2MASS~J01040750--0053283, 2MASS~J09175418+6028065 (not detected at $i$), 
2MASS~J15461461+4932114, and 2MASS~J13243553+6358281.  Objects marked with `$\times$' 
are other candidate L dwarfs found in our cross-match.  Arrows, where present, indicate upper limits 
on the $z-J$ colors or lower limits on the $i-z$ colors.
SDSS magnitudes ($i$ and $z$) are on the AB asinh magnitude system; 2MASS magnitudes ($J$ and 
$K_S$) are on the Vega magnitude system. 
\label{fig_zJiz}}
\end{figure}

\begin{figure}
\plotone{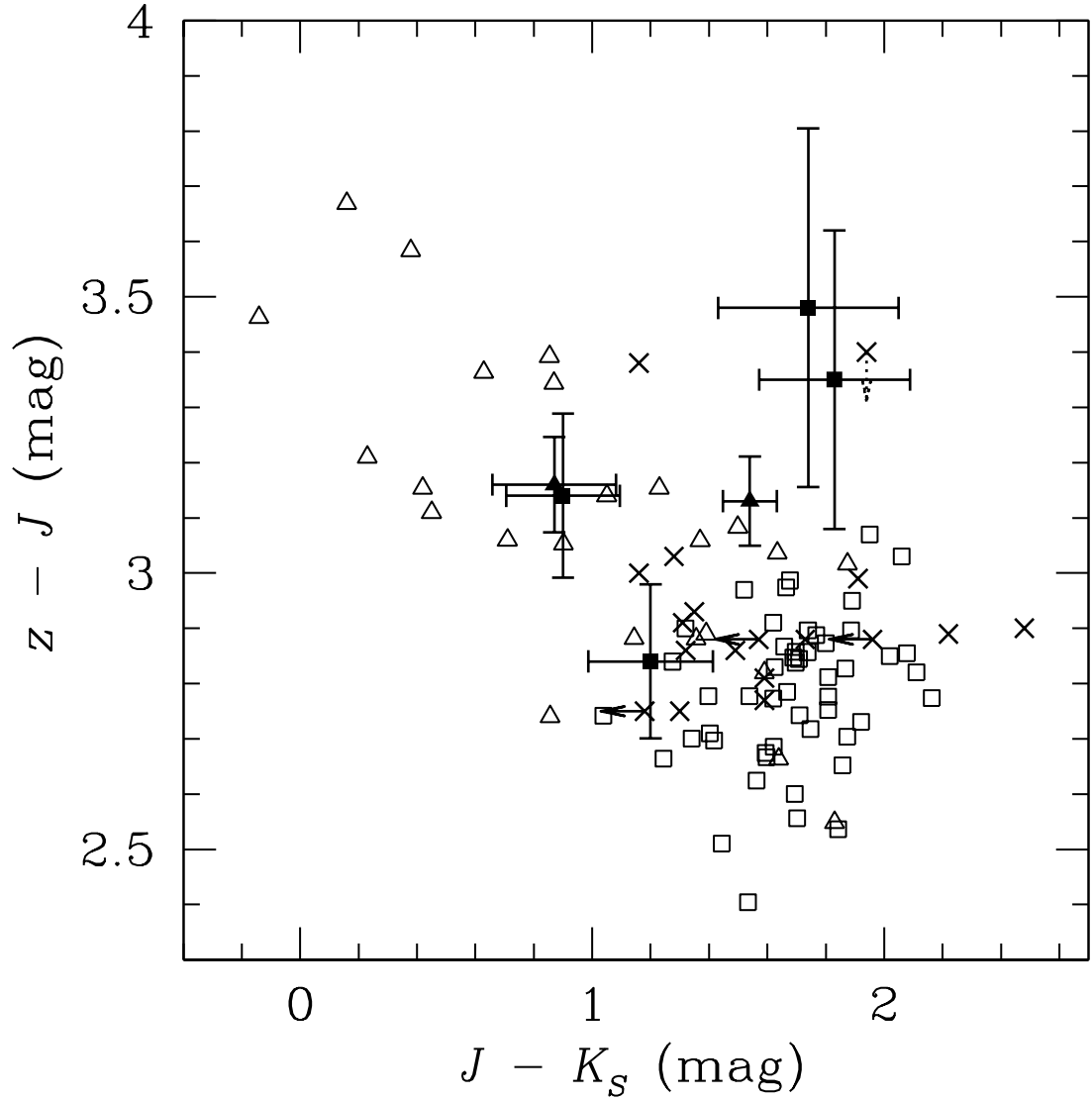}
\figcaption{As in Figure~\ref{fig_zJiz}, but for $z-J$ versus SDSS $i-z$.  In order of increasing $J-K_S$, 
the six L and T dwarfs discussed here are: 2MASS~J15461461+4932114, 2MASS~J00521232
+0012172, 2MASS~J01040750--0053283, 2MASS~J13243553+6358281, 2MASS~J09175418
+6028065, and 2MASS~J01262109+1428057.  One of the remaining candidate L dwarfs is an $H$-
band only detection in 2MASS, hence its $J-K_S$ color is unknown.  The upper limit on it $z-J$ color 
is denoted with a dotted arrow.
\label{fig_zJJK}}
\end{figure}

\begin{figure}
\plotone{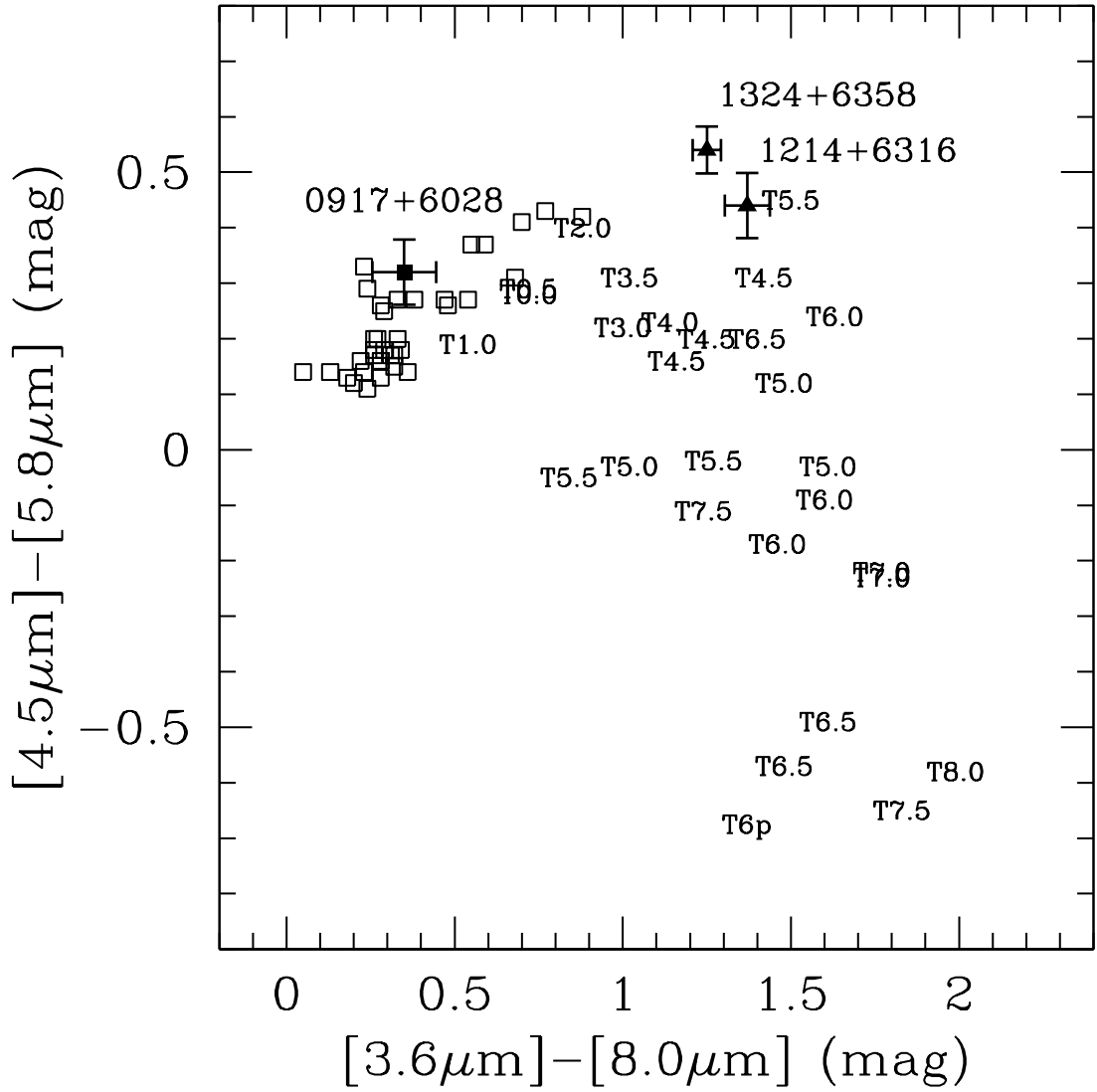}
\figcaption{A {\sl Spitzer}/IRAC color-color diagram (in Vega magnitudes) for L and T dwarfs.  The 
newly-discovered ultra-cool dwarfs 2MASS~J13243553+6358281 and 2MASS~J09175418
+6028065, and the independently announced \citep{chiu_etal06} T3.5 dwarf 2MASS~J12144089
+6316434 are plotted with solid symbols (square for the L dwarf; triangles for the T dwarfs) with error 
bars.  Known L dwarfs are shown with open squares and known T dwarfs are indicated by their 
spectral type (Tx.x).  Comparison data are from \citet{patten_etal06}.
\label{fig_irac_lt}}
\end{figure}

\begin{figure}
\plotone{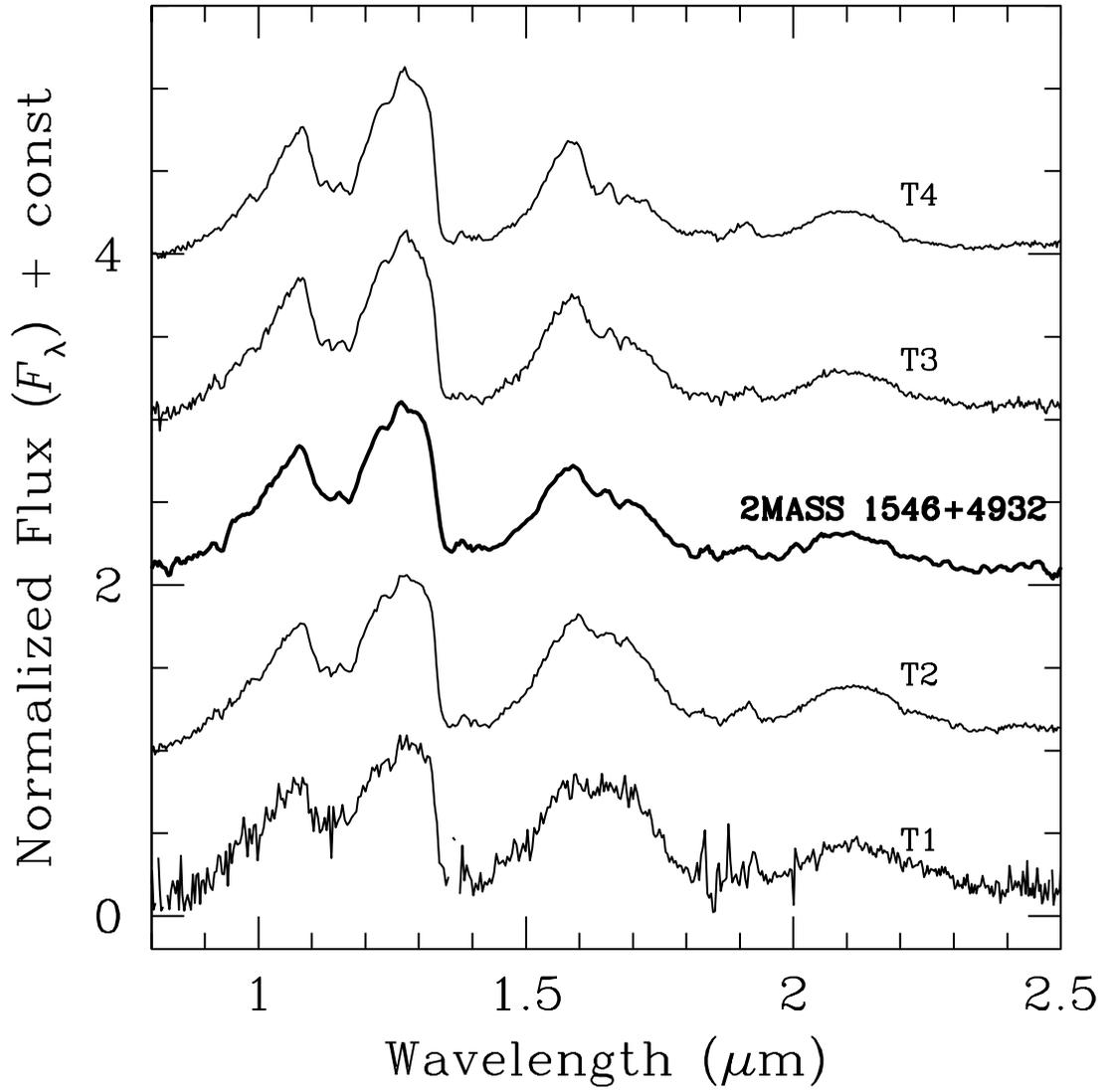}
\figcaption{A $R\approx150$ IRTF/SpeX prism spectrum (thick line) of the new T2.5 dwarf 
2MASS~J15461461+4932114.  The comparison SpeX prism spectra (thin lines) are of SDSS 
J083717.21--000018.0 (T1), SDSS~J125453.90--012247.4 (T2), 2MASS~J12095613--1004008 (T3), 
and 2MASS~J22541892+3123498 (T4), with spectroscopic classifications from \citet
{burgasser_etal06b}.
\label{fig_2mass1546_spec}}
\end{figure}

\begin{figure}
\plotone{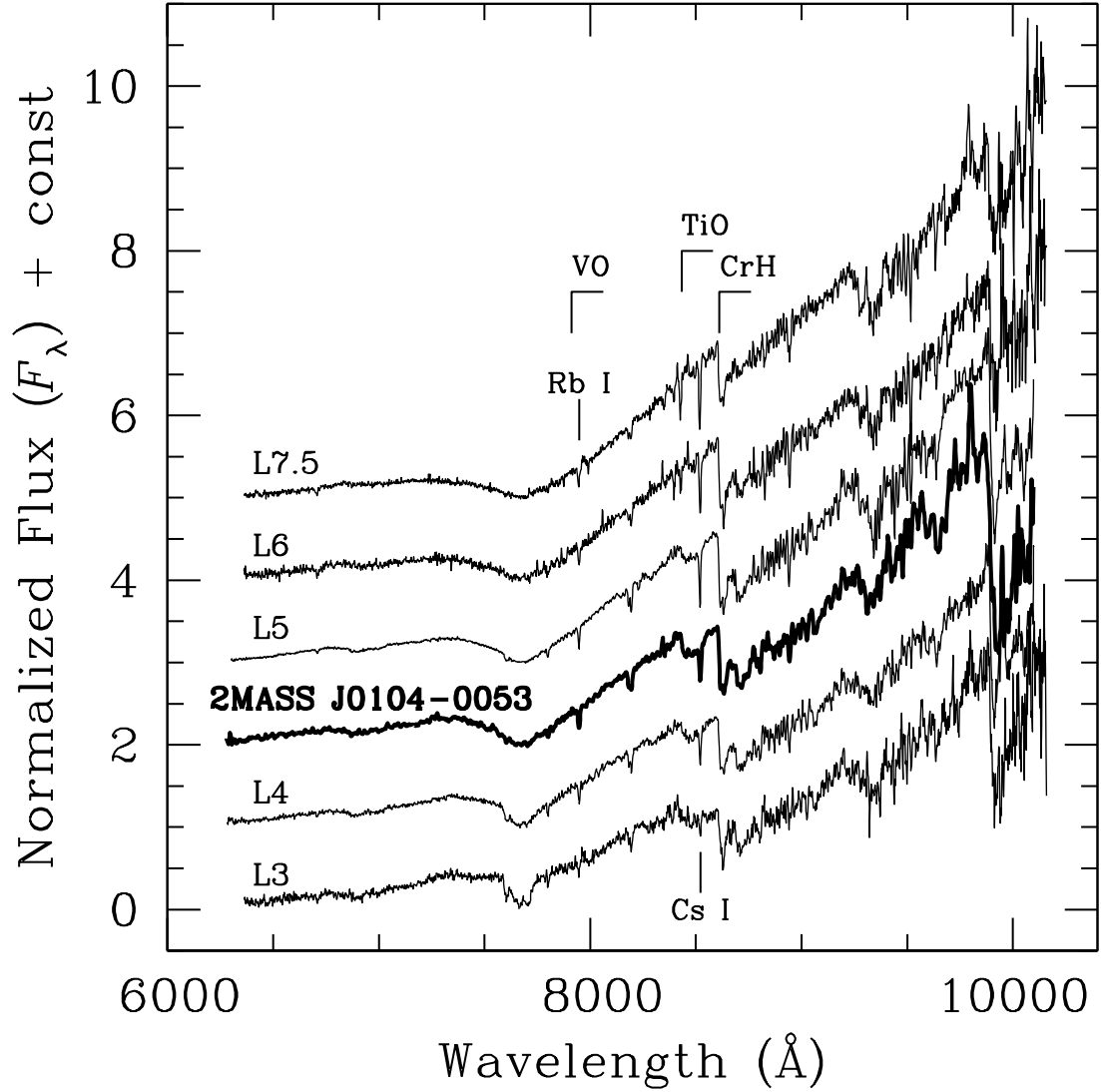}
\figcaption{A $R\approx900$ Keck/LRIS spectrum (thick line) of the L5 dwarf 
2MASS~J01040750--0053283.  The comparison spectra (thin lines) are of 2MASS~J03020122
+1358142 (L3), 2MASS~J01291221+3517580 (L4), DENIS-P~J1228.2--1547 (L5), 
2MASS~J01033203+1935361 (L6), and 2MASS~J08251968+2115521 (L7.5).  All spectral types are 
anchored to the optical classification scheme of \citet{kirkpatrick_etal99}.  Atomic and molecular 
features used in the spectral index classification are labeled.
\label{fig_2mass0104_spec}}
\end{figure}

\begin{figure}
\plottwo{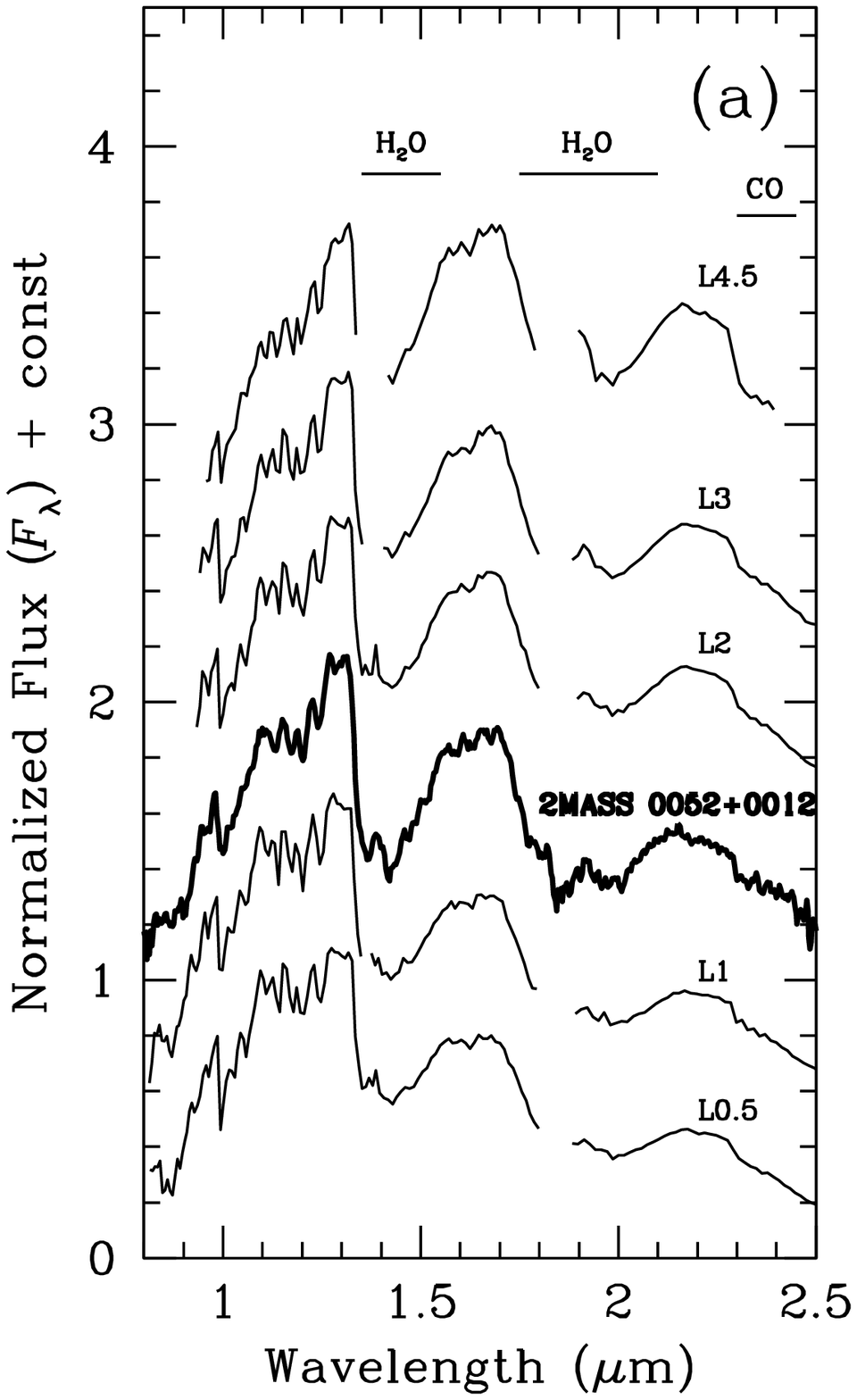}{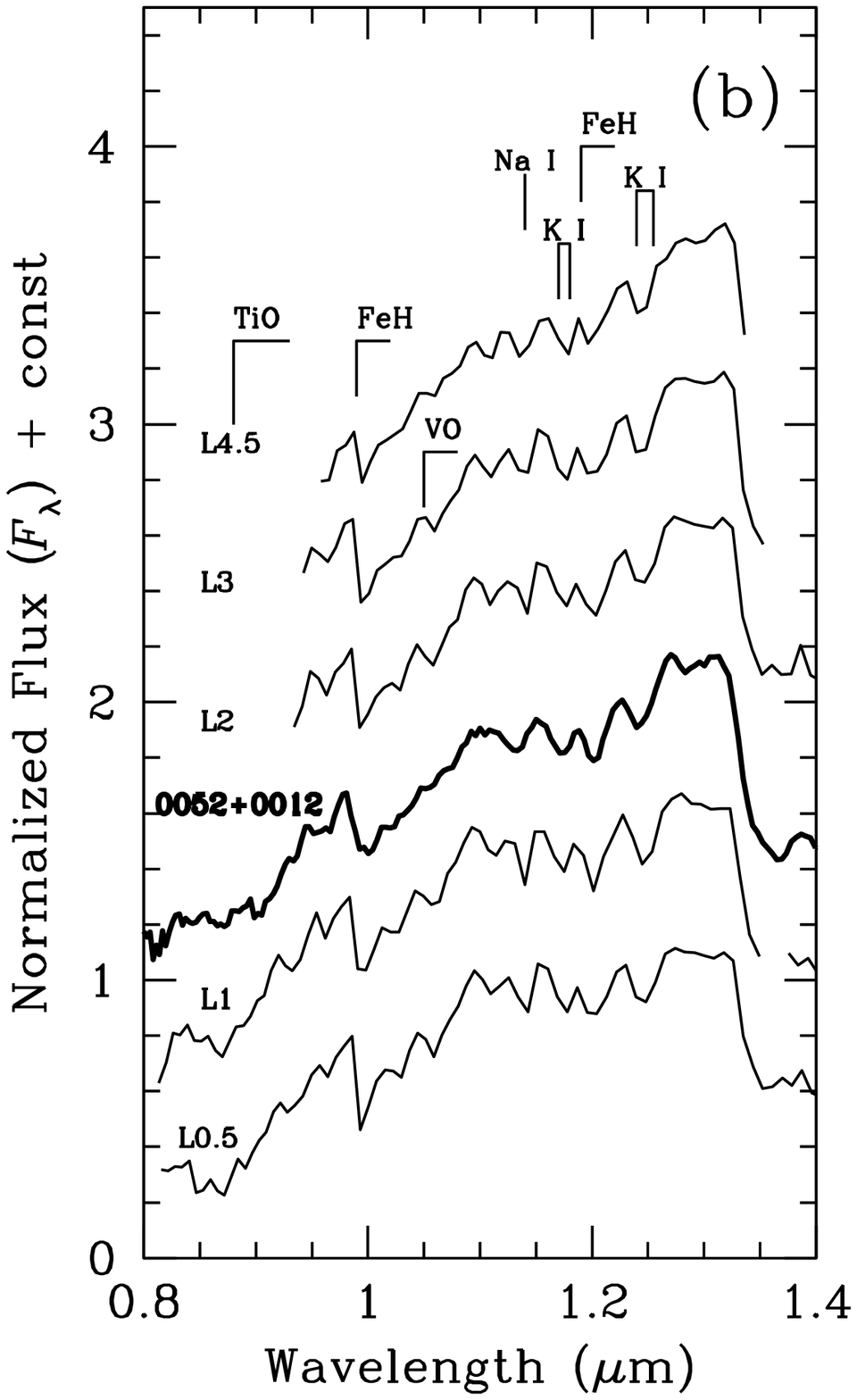}
\figcaption{A $R\approx150$ IRTF/SpeX prism spectrum (thick line) of the new L2 dwarf 
2MASS~J00521232+0012172.  The comparison spectra (thin lines) are of 2MASS~J07464256
+2000321~AB (L0.5), 2MASS~J14392836+1929149 (L1), Kelu--1~AB (L2), 2MASS~J15065441
+1321060 (L3), and 2MASS~J22244381--0158521 (L4.5) from the IRTF Spectral Library of \citet[]
[available at http://irtfweb.ifa.hawaii.edu/\textasciitilde spex/spexlibrary/IRTFlibrary.html]
{cushing_etal05}, and are smoothed to the same $R\approx150$ resolution.  All spectra are 
normalized to unity at 1.25~$\micron$.  Spectral types are anchored to the optical classification 
scheme of \citet{kirkpatrick_etal99}.  Panel (a) shows the spectra over the entire 0.8--2.5~$\micron$ 
region, and panel (b) zooms in on the 0.8--1.4~$\micron$ region.
\label{fig_2mass0052_spec_Ls}}
\end{figure}

\begin{figure}
\plottwo{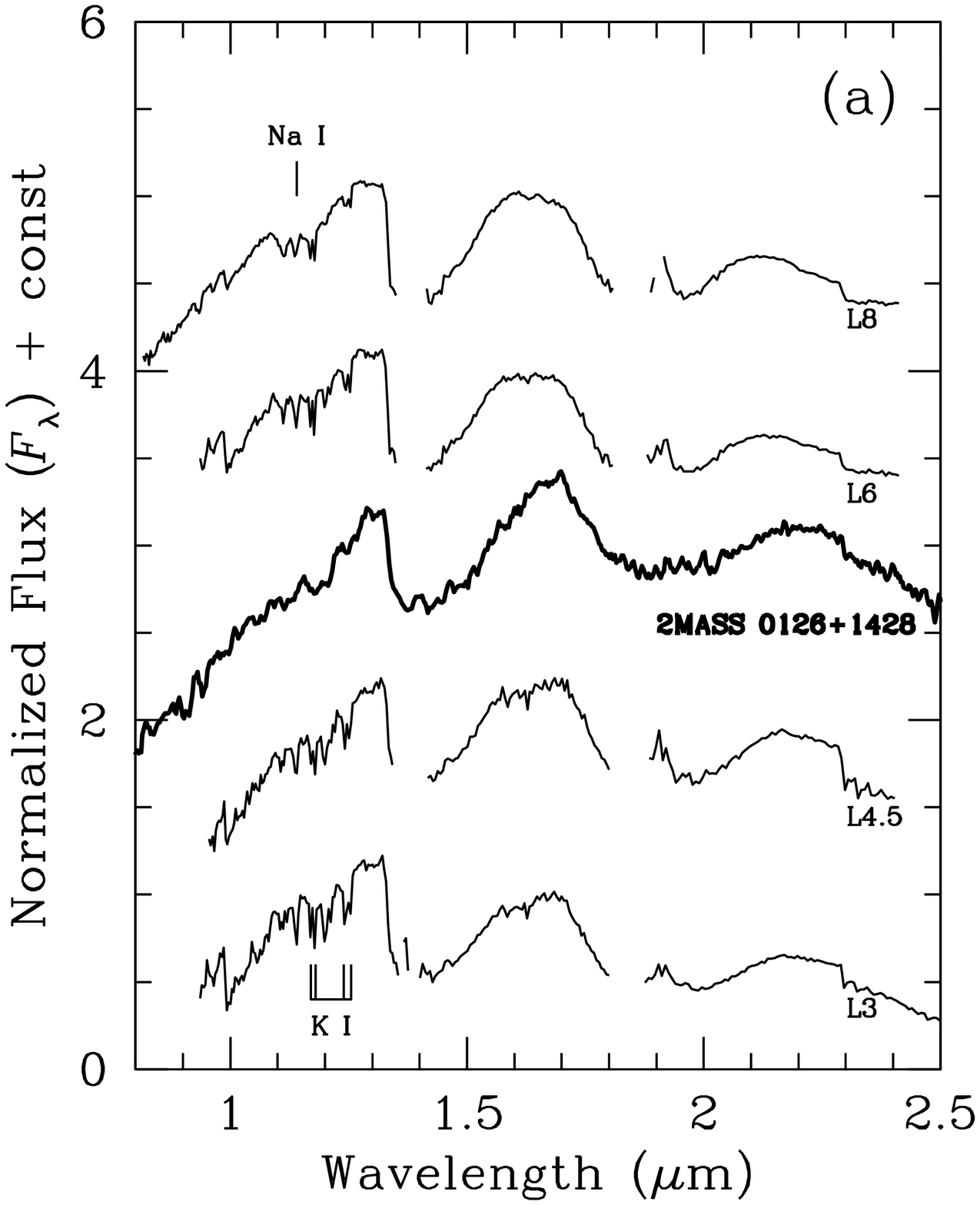}{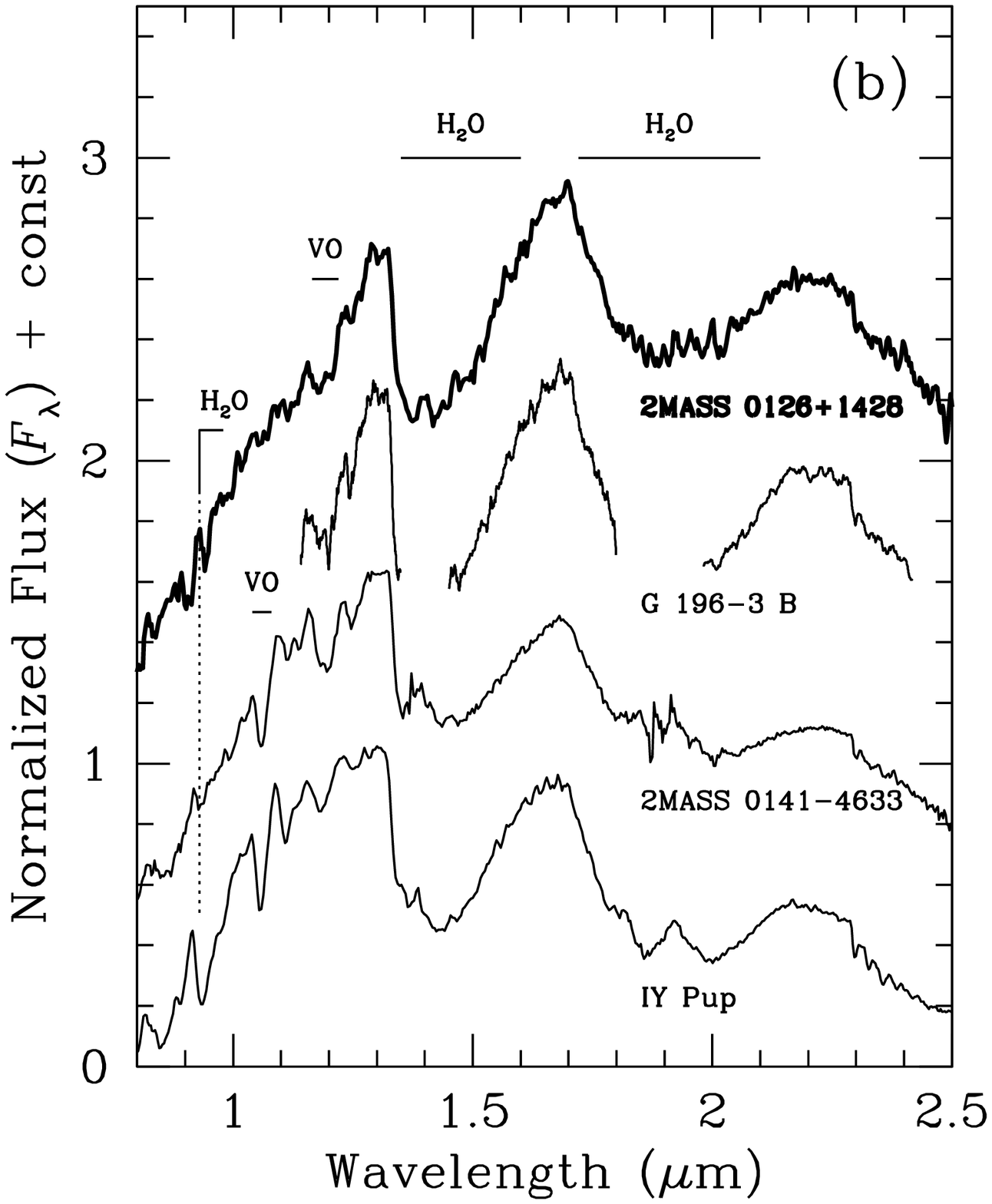}
\figcaption{$R\approx150$ IRTF/SpeX prism spectra of 2MASS~J01262109+1428057 compared to: 
(a) SpeX spectra of field L dwarfs and (b) SpeX spectra of the late-M giant IY~Pup, the 1--50~Myr L0 
dwarf 2MASS~J01415823--4633574 \citep{kirkpatrick_etal06}, and the 60--300~Myr L2 dwarf 
G~196--3B \citep{allers_etal07}.  All spectra are normalized to unity at 1.25~$\micron$ and are 
smoothed to the same resolution.  The spectrum of 2MASS~J01262109+1428057 does not show the 
strong $J$-band alkali absorption lines characteristic of old L dwarfs in the field (a), and shares 
characteristics (a peaked $H$-band continuum, enhanced VO and H$_2$ absorption) with the low 
surface gravity young L dwarfs and M giant (b).
\label{fig_2mass0126_spec}}
\end{figure}

\begin{figure}
\plotone{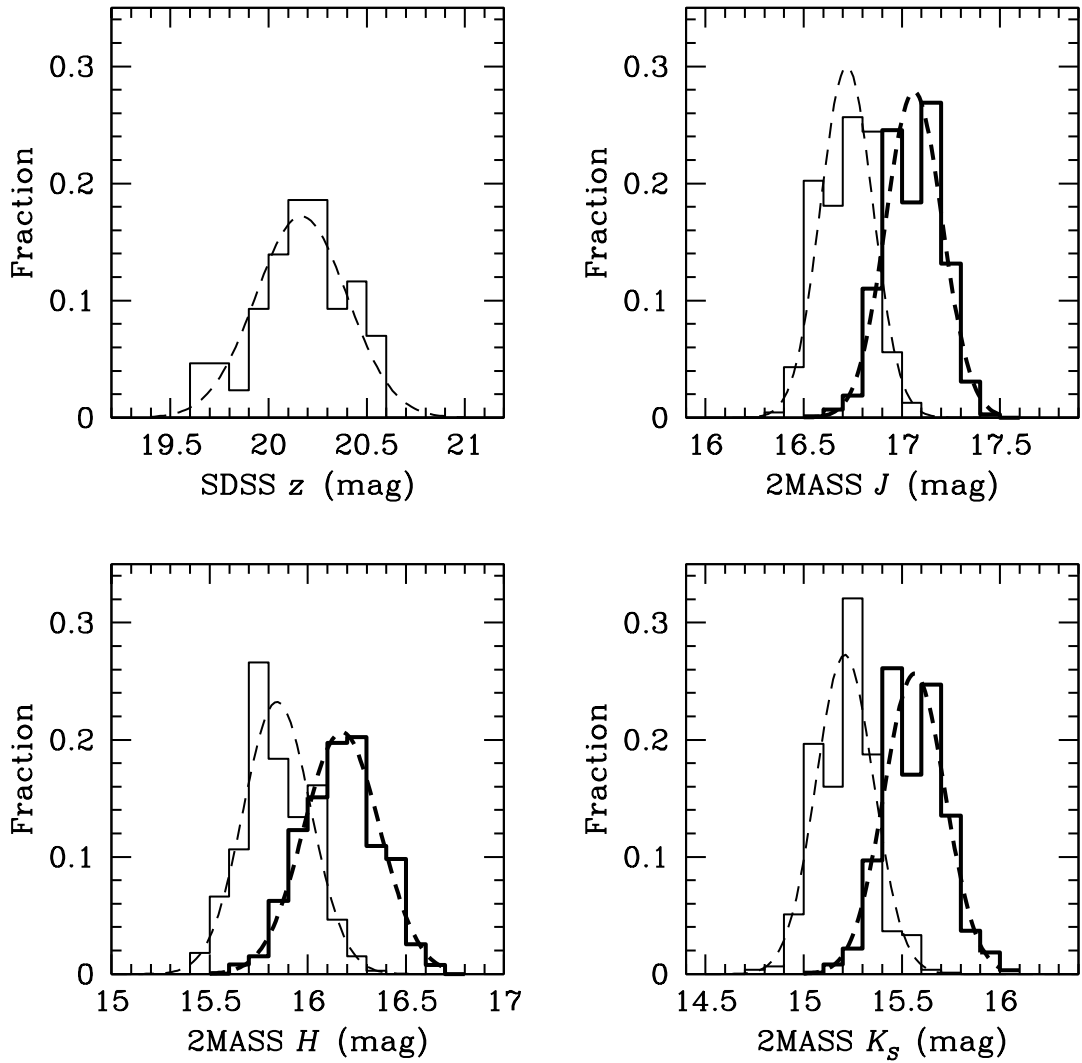}
\figcaption{Apparent magnitude distributions (histograms) of hypothetical $S/N=8.3$ T dwarfs in SDSS 
$z$ and of actual $S/N=5$ (thick lines) and $S/N=7$ (thin lines) point sources in 2MASS.  The means 
of the empirical distributions are adopted as the mean flux limits of the SDSS and 2MASS surveys.  
The dashed lines in each of the four panels are Gaussians with means and standard deviations 
corresponding to those of the histogram data.
\label{fig_limmags}}
\end{figure}

\begin{figure}
\plotone{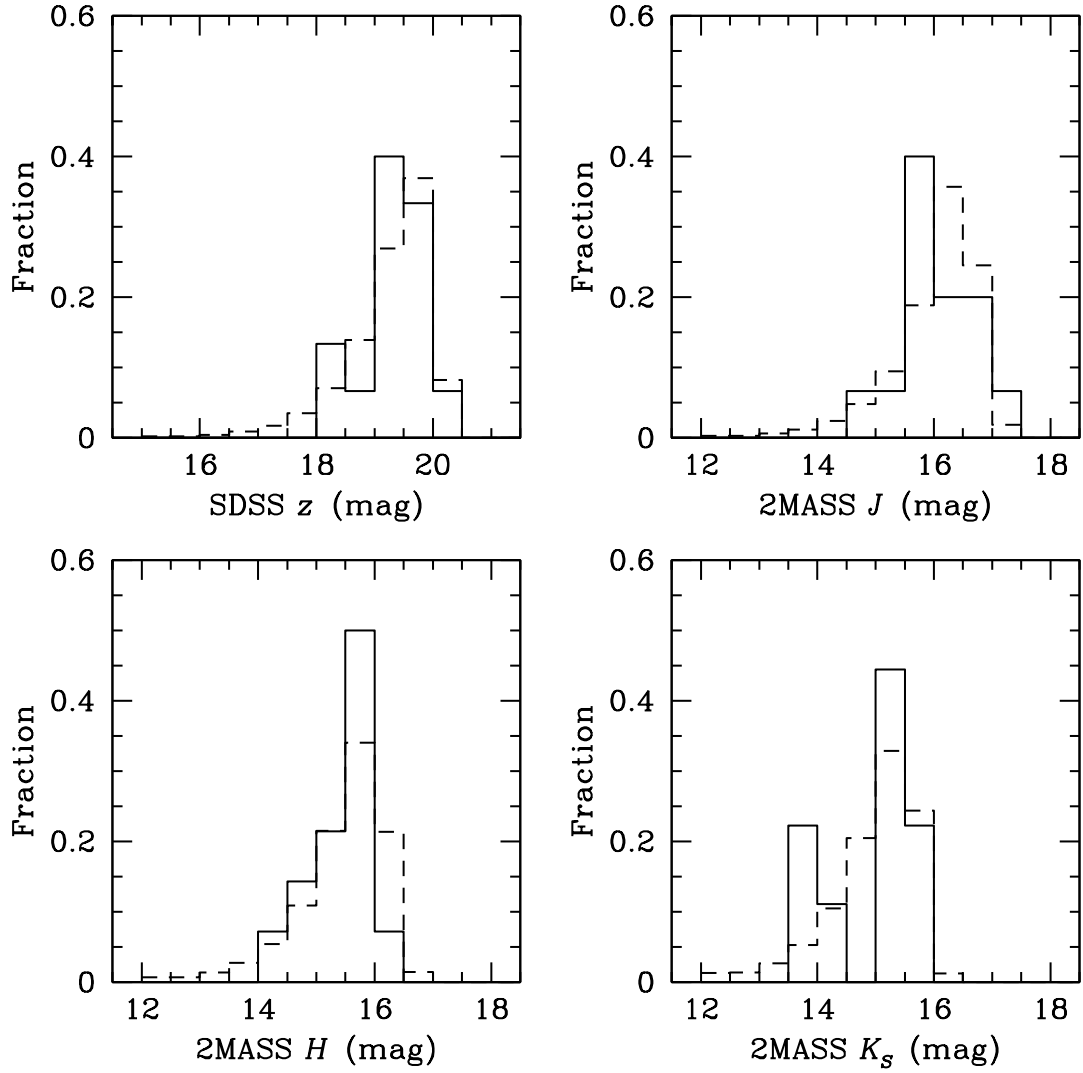}
\figcaption{Apparent magnitude distributions of the 15 known T dwarfs in
SDSS DR1 and and 2MASS (solid histograms) and of the $\approx$160000 detected T dwarfs in our 
Monte Carlo simulations (dashed histograms).  The K-S probabilities that the observed and simulated 
distributions are obtained from the same parent distribution of apparent magnitudes are 47\%, 57\%, 
76\%, and 72\% at $z$, $J$, $H$, and $K_S$ bands, respectively.
\label{fig_mag_hist}}
\end{figure}

The most comprehensive searches for T dwarfs in SDSS to date are those of \citet[][focusing mostly on 
areas contained in SDSS DR1]{knapp_etal04} and \citet[][focusing on more recent SDSS data]
{chiu_etal06}.  Both employ the $i$-dropout technique initially designed by \citet{fan_etal01} to search 
for high-redshift quasars in SDSS.  The most relaxed version of the $i$-dropout criteria are those 
employed in the most recent search by \citet{chiu_etal06}:
\begin{equation}
z<20.4, \qquad \sigma(z) < 0.12, \qquad i-z>2.2.
\end{equation}
Given SDSS $z$-band magnitudes of $\approx$19, errors $\sigma(z)<0.1$~mag, and colors $i-z>3.5
$~mag, both of the new T dwarfs satisfy these criteria.  We already pointed out (\S~\ref
{sec_2mass1546}) that 2MASS~J15461461+4932114 may have remained unidentified because of 
confusion with a nearby point source.  However, 2MASS~J13243553+6358281 is well separated from 
other points sources in SDSS (Fig.~\ref{fig_findcharts}) and is detected at a signal-to-noise ratio of at 
least 20.  The lack of prior identification of either of these two T dwarfs from SDSS requires closer 
scrutiny.

Upon an investigation the object flags for all T dwarfs known in SDSS DR1, we note that both of the 
new T dwarfs have a larger than usual number of flags set by the SDSS photometric pipeline when 
compared to other T dwarfs in SDSS (Table~\ref{tab_sdss_flags}).  Most notable among these is the 
PSF\_FLUX\_INTERP flag, which is present only for the two new T dwarfs, and is not set for the 
previously known T dwarfs.  This flag means that during PSF photometry more than 20\% of the PSF 
flux was from interpolated pixels (due to bad columns or bleed trails), which may make the photometry 
suspect \citep{stoughton_etal02}.  The recommendation on the SDSS website\footnote{See {\url http://
www.sdss.org/dr5/products/catalogs/flags.html} .} is that, when seeking a clean sample of point 
sources, this flag (among others) should be screened against.  This is the adopted procedure in at 
least one paper \citep{finkbeiner_etal04} from the SDSS collaboration.  It is therefore conceivable that, 
to the degree to which these documented examples are correct representations of the adopted 
practice, SDSS-only searches for T dwarfs may have also screened against the presence of the PSF
\_FLUX\_INTERP flag, thus explaining the omission of the two new T dwarfs presented here.   Another 
flag that is set only in the case of 2MASS~J13243553+6358281, but not for the other T dwarfs is the 
DEBLEND\_NOPEAK flag.  This flag indicates that after deblending the remnant (``child'') source in 
question did not have a peak.  The SDSS documentation\footnote{\url{http://www.sdss.org/dr5/
products/catalogs/flags\_detail.html} .} states that ``objects with [this flag] set (especially nominal point 
sources in a nominally high $S/N$ band) should be treated with suspicion.''  Given that 
2MASS~J13243553+6358281 appears single (Fig~\ref{fig_findcharts}), it seems that the decision by 
the photometric pipeline to target it for deblending may have been misguided.

The discovery of these two new early T dwarfs in an already well scrutinized part of the SDSS survey 
gives a clear demonstration of the higher sensitivity to ultra-cool objects that can be attained by cross-
correlating near-IR and optical databases.  In particular, by allowing us to impose less stringent criteria 
for object detection in either database, namely fewer object flag checks, the power of cross-correlation 
has enabled us to identify previously overlooked T dwarfs.

\subsection{Lessons Learned from Cross-Correlating Large Imaging Databases \label{sec_lessons}}

The method and research described here were conceived as a demonstration project for the NVO to 
explore the feasibility and the utility of cross-comparing large astronomical imaging databases, all of 
which have unique structures and distinct characteristics.  Our experience with SDSS and 2MASS has 
led us to conclude that a team attempting such a task needs to combine the necessary technological 
and science expertise, and to have intimate knowledge of the organization of each database.  From a 
technological point of view, our experience with cross-matching SDSS and 2MASS has lead us to 
conclude that cross-correlation of large imaging astronomical databases requires fast access to the 
data in each database and dedicated expertise in database management.  We found that by far the 
fastest way to run the cross-match was locally at IRSA (which houses the 2MASS database) in 
Pasadena, with the 462 gigabytes of SDSS DR1 catalog data contained on a personal computer 
shipped to us from Johns Hopkins University and cross-mounted on the local area network.  We also 
greatly benefitted from having a dedicated computer programmer (Mr.\ Serge Monkewitz) to create 
and run the computer code that performed the database cross-correlation.  From a scientific point of 
view, our combined expertise on the subject matter of T dwarfs helped us design simple and efficient 
cross-matching criteria (\S~\ref{sec_cull0}).  Finally, J.D.K.'s intimate knowledge of the various 2MASS 
object flags helped us eliminate spurious candidates based on their 2MASS descriptors early-on.  
However, none of us possessed the necessary close understanding of the SDSS database and of the 
tools available for its exploration.  As a result, we spent a significant amount of time getting acquainted 
with SDSS.  Here we list some of the lessons extracted from this learning process.

Unlike 2MASS, SDSS does not list objects as non-detections at any band, but reports flux 
measurements in all 5 bands for all objects that are detected in at least one of the bands.  Taking such 
measurements at face value, without proper consideration of the detection limits of SDSS, would 
greatly skew the inferred optical colors of faint objects of interest.  Therefore, listed SDSS magnitudes 
always have to be considered in the context of the adopted completeness limits of the survey.  In a 
similar vein, we were greatly confused in the beginning when our initial position/color cross-match 
(\S~\ref{sec_cull0}) found extremely red ($z-J>10$~mag) candidates, consisting of easily identifiable 
2MASS sources with no apparent SDSS counterparts.  These were later found to be results of the 
near-alignment (within 6$\farcs$0) of blank-sky SDSS pointings (object \verb|type = 8|) with quoted 
magnitudes in the 27--30~mag range, with 2MASS point sources.  Clearly, this is not an issue once 
one knows to take into account the value of the SDSS object \verb|type| flag.  To reject such spurious 
candidates we implemented an upper limit on the SDSS $z$-band magnitude and a check of the 
object \verb|type| flag in our secondary selection criteria (\S~\ref{sec_cull1}).

In a separate instance, we found that the default $ugriz$ SDSS magnitudes (object flags 
\verb|u, g, r, i, z|) are always based on extended-source fits to the point-spread function (PSF) profile.
This is the 
case even for objects classified as point sources.  In addition, all SDSS object magnitudes have 
associated extinction corrections (\verb|extinction_u, extinction_g|, etc) based on maps of the Galactic 
100~$\micron$ emission \citep{schlegel_etal98} from the {\sl COBE}/DIRBE and {\sl IRAS}/ISSA maps.  
In the context of unavoidable inaccuracies in the galaxy/star separation algorithm at faint flux levels, 
such a uniform approach is certainly justified.  However, from the perspective of an accustomed user 
of 2MASS, which contains separate point- and extended-source catalogs, the 
presence of extended-source descriptors for point sources may be misleading.  All of these SDSS 
extended source flags are irrelevant for our science, since T dwarfs are point sources, and the ones 
detectable in SDSS reside within $\sim$100~pc from the Sun, i.e., within the Local Bubble, where the 
interstellar extinction is $A_V=0.0$~mag.  Instead, we considered the non-extinction-corrected PSF 
magnitudes (SDSS object flags \verb|psfMag_u, psfMag_g|, etc) for each candidate and used these to 
calculate optical/near-IR colors.

The issue of galaxy/star separation at faint flux levels in both SDSS and 2MASS deserves special 
attention.  In the current project we focused only on objects explicitly identified as point sources 
(criteria \ref{crit_type} and \ref{crit_ext} in \S~\ref{sec_cull1}).  However, automated galaxy/star 
separation algorithms are unreliable at very low signal levels.  As we noted in \S~\ref
{sec_completeness}, erroneous morphological typing in SDSS DR1 was the reason for which we 
failed to recover one of the previously known T dwarfs in the SDSS DR1 footprint.   Although that T 
dwarf has been correctly re-classified as a point source by the presumably better morphological 
identification algorithm used for DR2, it is still highly probable that fainter T dwarfs may remain 
classified as galaxies, especially near the $z=21.0$~mag ($S/N\sim3$) cut-off of our cross-match.  Our 
T dwarf surface and space density analysis uses a more stringent $z=20.4$~mag limit ($S/N=8.3$; \S~
\ref{sec_considerations}), corresponding to that employed in previous SDSS-only T dwarf searches.  
Therefore, we believe that our results for the local population of T dwarfs are largely unaffected by the 
inefficiency of star/galaxy separation algorithms at faint flux levels.  Nevertheless, a more careful 
treatment of the problem will be necessary in our planned future iteration of the cross-match with DR5.

As an example that our unfamiliarity with SDSS also inadvertently helped, we note that our lack of 
knowledge of the various SDSS object flags may have been the very reason for the discoveries of the 
two new T dwarfs reported here, since none of us knew to screen against flags commonly regarded as 
suspect (\S~\ref{sec_missed_Ts}).  Nonetheless, as a counter-example, our inadequate 
understanding of the suite of SDSS object flags that govern galaxy/star separation in the different 
filters probably prevented us from successfully recovering the one T dwarf that was missed because of 
being erroneously classified as an extended object (\verb|type = 3|) in SDSS DR1 (\S~\ref
{sec_completeness}).

Looking beyond the automated selection processes and our familiarization with SDSS, a somewhat 
cumbersome stage of our program was the visual inspection of the 1160 T-dwarf candidates that 
survived all of the automated culls.  All of these were inspected individually on both the SDSS and the 
2MASS survey images.  A fraction of these turned out to be 2MASS persistence or line--128 artifacts 
(\S~\ref{sec_cull3}).  Another set of the candidates had very uncertain photometry because of being 
embedded in the bright halos of saturated stars.  We could not find an a priori reason to exclude these 
without at least a visual inspection of the survey images.  Even after the visual inspection, such 
candidates contributed to $\sim$25\% of the final ``good'' candidates that required follow-up.  In the 
expanded cross-match planned for 2MASS and SDSS DR5, the number of ``good'' candidates is 
expected to be at least an order of magnitude larger than in the present cross-match.  More stringent 
automated candidate culling may thus be necessary before the visual inspection and observational 
follow-up stages.  At the same time however, care will need to be taken to keep the cross-match 
constraints relaxed in comparison to the constraints that would otherwise be applied in each database 
individually, in order to maintain the superior completeness of the combined search.

Finally, because of the large number of remaining good candidates even after the visual inspection, a 
significant fraction (20--30\%) of which may still turn out to be artifacts (in both databases), fast imaging 
follow-up is necessary to confirm the existence of any objects before more time-consuming 
spectroscopy is attempted.  We found that a 2--3 meter class telescope with a simple optical or near-IR 
camera is well suited for the task.

\section{CONCLUSIONS}
 \label{sec_conclusion}

Our pilot project to search for previously overlooked T dwarfs in 2MASS and SDSS DR1 demonstrates 
the feasibility and utility of large database cross-correlation in discovering rare interesting objects.
Our simultaneous positional and color cross-match of the 2MASS and SDSS DR1 databases 
uncovered 2 more T dwarfs in addition to the 13 already known in the SDSS DR1 footprint.  Despite 
the great scrutiny with which this area has already been explored for T dwarfs, both of the new T 
dwarfs had previously been overlooked, probably because of suspect photometry flags in SDSS.  

The discovery of the two new T dwarfs demonstrates the superior sensitivity to ultra-cool dwarfs that 
can be attained by simultaneously cross-correlating large optical and near-IR databases, compared to 
searches based on individual optical or near-IR databases alone.  As a by-product of our search, 
which focused on objects with very red optical minus near-IR colors, we also report the discovery of 
two new peculiar L dwarfs: an L2 dwarf with unusually blue near-IR colors, potentially linked to mildly 
sub-solar metallicity, and another young L2 dwarf.

We took advantage of the high degree of completeness attained through our approach to obtain a flux-
limited estimate of the local T dwarf space density.  We used Monte Carlo analysis to reproduce the 
observed T dwarf population in the overlap area of SDSS DR1 and 2MASS, and found that the local 
space density of T dwarfs is $0.0070_{-0.0030}^{+0.0032}$~pc$^{-3}$ (95\% confidence interval), i.e., 
about one per 140 pc$^3$.  This empirical result is the first empirical estimate of the number density of 
T dwarfs over the entire range of T0--T8 spectral type range and extends earlier work by \citet
{burgasser02} that focused on T5--T8 dwarfs.  In the context of various predictions for the local sub-
stellar population \citep{burgasser04, burgasser07, allen_etal05}, we find that our result is most 
consistent with model-dependent estimates that assume a flat sub-stellar mass function, $d N/d M 
\propto M^{0.0}$.

Given the success of the 2MASS/SDSS-DR1 cross-match, we expect that the approach will be 
instrumental for the identification of brown dwarfs cooler than the coolest ones presently known, with 
spectral types $>$T8.  While no such brown dwarfs were identified in the present cross-match 
covering the 2099~deg$^2$ area of SDSS DR1, we anticipate with a 86\% probability that at least one 
T9 dwarf will be detectable in a similar cross-comparison of the entire 8000~deg$^2$ SDSS DR5 
footprint with 2MASS.

\acknowledgments

{\bf Acknowledgments.}  The SDSS-DR1/2MASS database cross-match was funded as a 
demonstration project by the NSF National Partnership for Advanced Computational Infrastructure, 
and by the National Virtual Observatory, sponsored by the NSF.  We thank Mr.\ Serge Monkewitz for 
developing the cross-match engine.  We thank Gillian Knapp for assistance with SDSS, and Katelyn 
Allers for providing us with her IRTF/SpeX spectrum of G~196--3B.  We acknowledge Tiffany Meshkat 
for help with the acquisition and reduction of the Lick PFCam data.  This research has made use of the 
NASA/IPAC Infrared Science Archive (IRSA), which is operated by the Jet Propulsion Laboratory, 
California Institute of Technology, under contract with the National Aeronautics and Space 
Administration.  This research has also benefitted from the M, L, and T dwarf compendium housed at 
DwarfArchives.org and maintained by Chris Gelino, Davy Kirkpatrick, and Adam Burgasser.  This 
publication makes use of data products from the Two Micron All Sky Survey, which is a joint project of 
the University of Massachusetts and the Infrared Processing and Analysis Center/California Institute of 
Technology, funded by the National Aeronautics and Space Administration and the National Science 
Foundation.  The Sloan Digital Sky Survey is managed by ARC for the Participating Institutions: the 
University of Chicago, Fermilab, the Institute for Advanced Study, the Japan Participation Group, The 
Johns Hopkins University, the Korean Scientist Group, Los Alamos National Laboratory, the Max 
Planck Institute for Astronomy, the Max Planck Institute for Astrophysics, New Mexico State University, 
the University of Pittsburgh, the University of Portsmouth, Princeton University, the United States Naval 
Observatory, and the University of Washington. Funding for SDSS has been provided by the Alfred P. 
Sloan Foundation, the Participating Institutions, NASA, the NSF, the US Department of Energy, the 
Japanese Monbukagakusho, and the Max Planck Society.  Support for S.A.M.\ was provided by NASA 
through the {\sl Spitzer} Fellowship Program, under award 1273192.

\facility{{\it Facilities:} Keck I Telescope, Palomar Observatory's 1.5~meter Telescope, University of 
Hawaii 2.2~meter Telescope, Infrared Telescope Facility, Shane Telescope, Spitzer Space Telescope 
satellite} 

\appendix 

\section{BAYESIAN INFERENCE OF THE SURFACE DENSITY OF T DWARFS}
 \label{sec_appendix}

To estimate the mean space density of T dwarfs in any given spectral type bin, we need to take into 
account the fact that the observed number of T dwarfs per bin is small, and is likely derived from a 
Poisson distribution.  That is, if the mean number of T dwarfs per spectral type bin in any 2099 deg$^2
$ area of SDSS and 2MASS is $\eta$, the probability of detecting $k$ dwarfs belonging to the same 
spectral type bin in the 2099~deg$^2$ area of SDSS DR1 is:
\begin{eqnarray}
P(k \vert \eta) = \frac{e^{-\eta} \eta^{k}}{k!}.
\label{eqn_poisson}
\end{eqnarray}
Given observed numbers of T dwarfs $k$, we would like to find $\eta$, which is a simple exercise in 
Bayesian inference:
\begin{eqnarray}
P(\eta \vert k) = \frac{P(k \vert \eta) P (\eta)}{\int P(k \vert \eta^\prime) P (\eta^\prime) d\eta^\prime}.	
\label{eqn_bayes}
\end{eqnarray}
$P(\eta)$ above summarizes our prior guess for the probability distribution of $\eta$.  We expect that 
$P(\eta \vert k)$ will follow the same functional form as $P(k \vert \eta)$, which would be the case if, for 
example, we set the prior $P (\eta)$ to a constant.  Adopting a ``uniform prior'' is common practice in 
Bayesian analysis in the lack of an educated guess \cite[e.g.,][]{kraft_etal91}.  However, our data 
demonstrate that large numbers of T dwarfs in any given bin are unlikely, so we can improve our initial 
guess by adopting
\begin{eqnarray}
P(\eta) \equiv P(k \vert \eta) = \frac{e^{-\eta} \eta^{k}}{k!}.
\label{eqn_erlang}
\end{eqnarray}
$P(\eta)$ is the ``conjugate prior'' of $P(\eta \vert k)$.  Conjugate priors are also a popular choice in 
Bayesian analysis \citep[e.g.,][]{raiffa_schlaifer61}.  As we shall see below, the choice of the conjugate, 
as opposed to a flat uniform prior decrease the expectation value of $\eta$ and narrow its confidence 
interval.

\begin{figure}
\plottwo{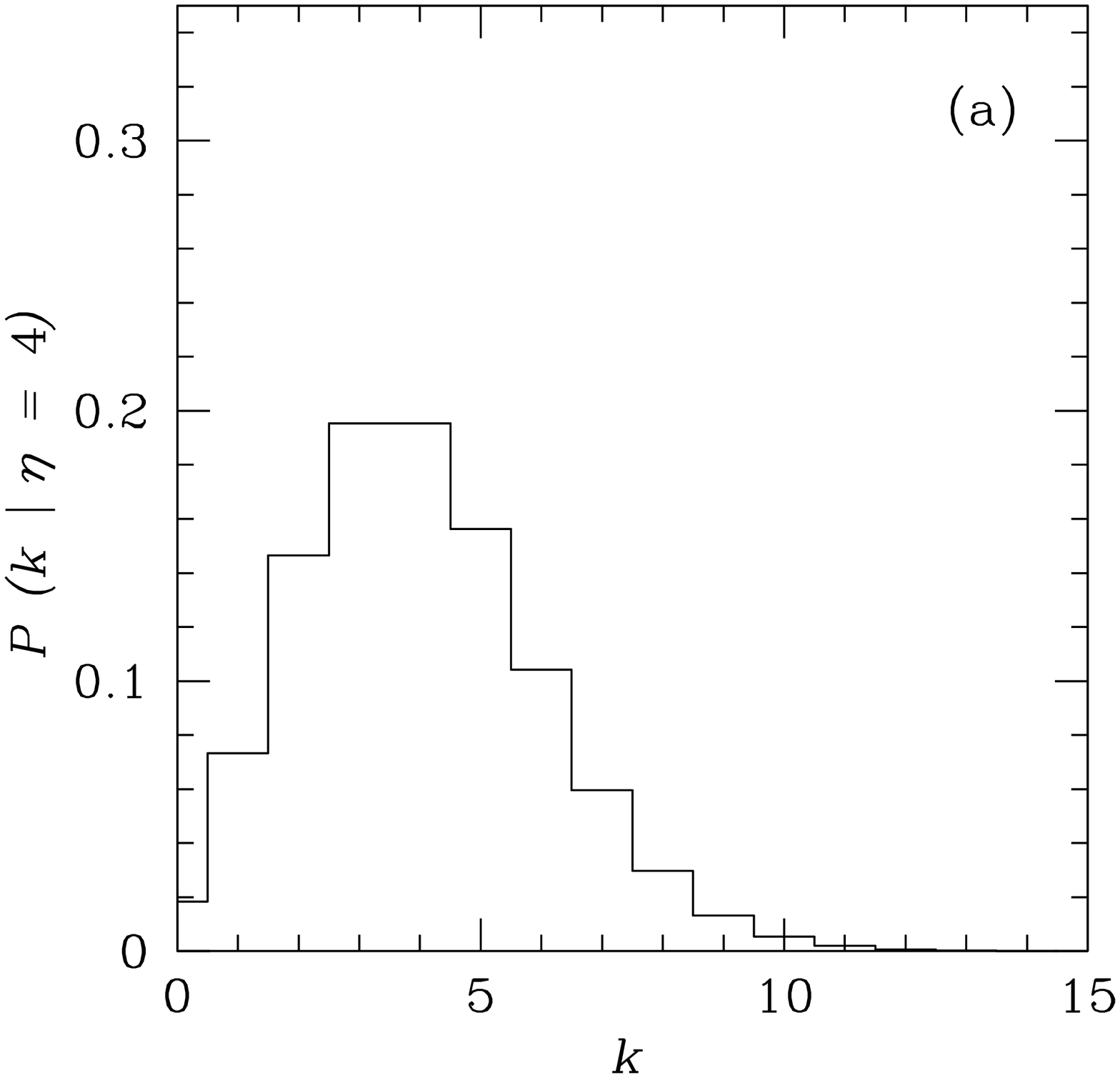}{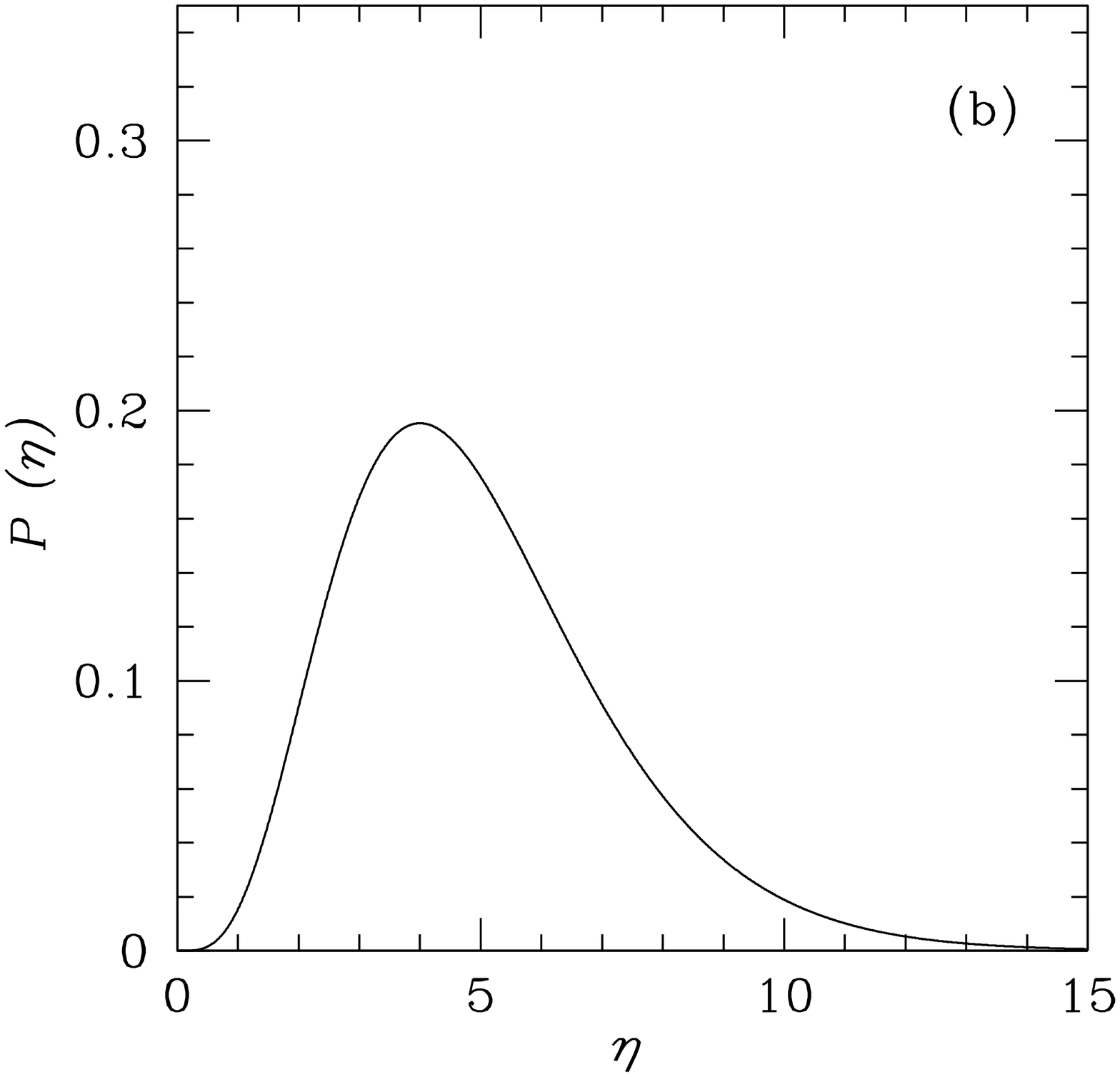}
\figcaption{(a) The discrete Poisson distribution $P(k\vert\eta)$ with a
population mean value of $\eta=4$.  (b)  The continuous Gamma
distribution $P(\eta)$ for $k=4$ (see Equation \ref{eqn_erlang}).
Although the distributions appear similar, we note, for example, that
$P(k=0 \vert \eta=4)>0$, while $P(\eta=0)=0$.  $P(\eta)$ is the
conjugate prior of the Poisson distribution in panel (a), and is the
Bayesian prior that we have adopted for our inference for the probability distribution $P(\eta\vert k)$ of 
the population mean $\eta$ (Equation \ref{eqn_posterior}).  
\label{fig_erlang}}
\end{figure}

We note that although $P(\eta)$ and $P(k \vert \eta)$ are identical, one is a function of $\eta$ (at a 
constant $k$), while the other is a function of $k$ (at a constant $\eta$), so their functional forms are 
different: $P(k \vert \eta)$ is a discrete Poisson distribution and $P(\eta)$ is a continuous Gamma 
distribution (Fig.~\ref{fig_erlang}).  We also note that our choice for $P(\eta)$ peaks at the observed 
value $k$, indicating that our prior guess for $\eta$ is that its most likely value is the observed one, $k
$.  We now substitute the expressions from equations (\ref{eqn_poisson}) and (\ref{eqn_erlang}) in 
Equation (\ref{eqn_bayes}) and, after performing the integration, we find the Bayesian posterior 
distribution
\begin{equation}
P(\eta \vert k) = \frac{2 e^{-2\eta} (2\eta)^{2k}}{(2k)!} = \frac{2 e^{-2\eta} (2\eta)^{2k}}{\Gamma(2k+1)},
\label{eqn_posterior}
\end{equation}
where we have used the fact that the complete Gamma function $\Gamma(a)\equiv \int_0^\infty t^{a-1} 
e^{-t} dt$ evaluates to $(a-1)!$ when $a$ is a positive integer.  $P(\eta \vert k)$ gives the probability 
density distribution that describes how likely different values for the mean number of T dwarfs per 
2099~deg$^2$ area $\eta$ are given an observed number of $k$.  We find the mean value of $\eta$ 
from
\begin{eqnarray}
\langle\eta\rangle & = & 
\frac{\int_0^\infty \eta^\prime P(\eta^\prime \vert k) d\eta^\prime}{\int_0^\infty P(\eta^\prime \vert k) d
\eta^\prime} \nonumber \\
& = & \frac{(2k+1) \int_0^\infty e^{-2\eta^\prime} (2\eta^\prime)^{2k+1} d(2\eta^\prime) / (2k+1)!}{2 
\int_0^\infty e^{-2\eta^\prime} (2\eta^\prime)^{2k} d(2\eta^\prime) / (2k)!} \nonumber \\
& = & \frac{(2k+1) \Gamma(2k+2) / \Gamma(2k+2)}{2 \Gamma(2k+1)/\Gamma(2k+1)} 
\nonumber \\
& = & k + 0.5.
\label{eqn_mean}
\end{eqnarray}

\begin{figure}
\plotone{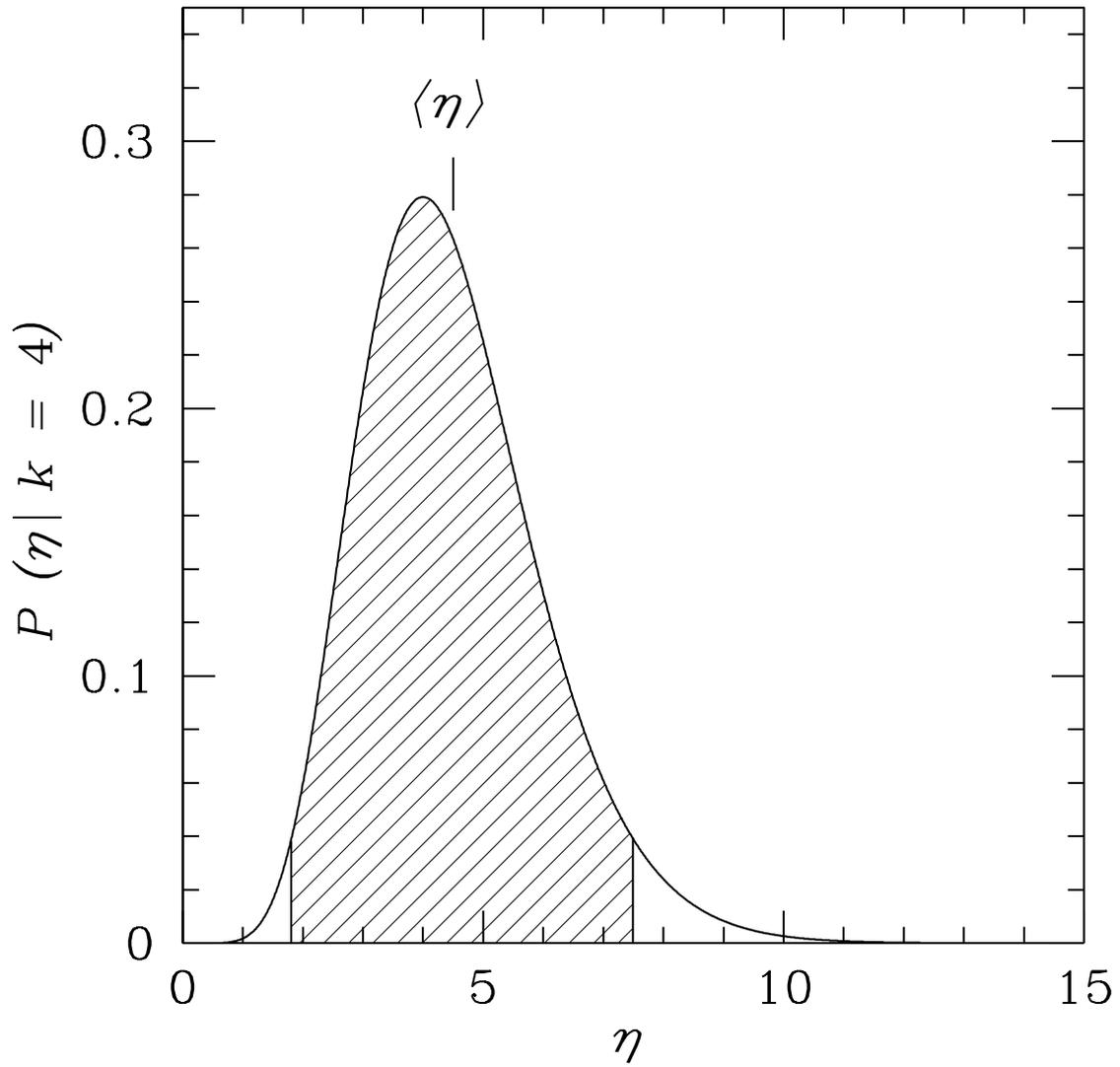}
\figcaption{The Bayesian posterior probability distribution $P(\eta\vert k)$ given $k=4$ detections.  The 
mean number of detections (e.g., T5--T8 dwarfs in any 2099~deg$^2$ area of SDSS) is expected to 
be $\langle\eta\rangle=4.5$, while the most likely number of detections is 4, as observed.  The shaded 
area represents the CL~= 0.95 confidence interval.
\label{fig_posterior}}
\end{figure}

Having found the expectation value $\langle\eta\rangle$, we would also like to determine a 
confidence interval $[\eta_l, \eta_u]$, such that $\eta_l\leq\eta\leq\eta_u$ at a desired confidence level 
CL.  We choose the lower and upper bounds $\eta_l$ and $\eta_u$ of the confidence interval CL, 
such that
\begin{equation}
\int_{\eta_l}^{\eta_u} P(\eta^\prime \vert k) d\eta^\prime = {\rm CL}
\label{eqn_cl}
\end{equation}
and
\begin{equation}
P(\eta_l \vert k) = P(\eta_u \vert k).
\label{eqn_cl_limits}
\end{equation}
Equations (\ref{eqn_cl}) and (\ref{eqn_cl_limits}) define the minimum size confidence interval $[\eta_l, 
\eta_u]$ at confidence level CL \citep{kraft_etal91}.  The system of equations can not be inverted 
analytically, and has to be solved for $\eta_l$ and $\eta_u$ numerically.  We do so for the CL~= 0.95 
confidence level and accordingly quote the 95\% confidence limits on the space density of T dwarfs in 
each spectral type bin in Table~\ref{tab_space_dens}.  Figure~\ref{fig_posterior} shows an example of 
the posterior Bayesian probability distribution, and of the 0.95 confidence interval, $[\eta_l, \eta_u]=
[1.81,7.49]$, for $k=4$ detections (corresponding to the number of SDSS DR1 T dwarfs in our T6--T8 
bin).  We note that the areas under the $P(\eta \vert k)$ curve for $\eta<\eta_l=1.81$ and $\eta>
\eta_u=7.49$ are not equal: a result of the requirement to minimize the confidence interval $[\eta_l, 
\eta_u]$ \citep{kraft_etal91}.  Also, the expectation value of $\eta$ is not in the middle of the 
confidence interval.

Had we chosen a uniform prior, $P(\eta)={\rm const}$, instead of the expression in Equation (\ref
{eqn_erlang}), the expectation value of $\eta$ would have been $\langle\eta\rangle=k+1=5$, as 
opposed to $k+0.5=4.5$ (Equation \ref{eqn_mean}), and the 1$\sigma$ confidence limits on $\eta$ 
would have been [1.21, 9.43].  That is, our educated guess that not all T dwarf surface densities in a 
given spectral type bin are equally probable, based on the observed counts of T dwarfs in the three 
spectral type bins, has allowed us to constrain the confidence interval of $\eta$.  Finally, we note that 
the widths of our 1$\sigma$ confidence intervals for either prior are narrower than what would have 
been inferred from a frequentist, rather than a Bayesian, point of view.  The 1$\sigma$ confidence 
interval derived in frequentist manner would have been [1.09, 10.24] \citep{gehrels86}.  This justifies 
our choice of Bayesian inference to determine the narrowest confidence interval $[\eta_l, \eta_u]$ for 
any chosen confidence level CL.


\input{ms.bbl}
\clearpage

\clearpage

\begin{deluxetable}{lccccccc}
\tablewidth{0pt}
\tabletypesize{\scriptsize}
\tablecaption{Optical and Near-IR Photometry of Candidate and Known Ultra-Cool Dwarfs \label
{tab_izJHK}}
\tablehead{\colhead{2MASS ID} & \colhead{SDSS $i$\tablenotemark{a}} & 
	\colhead{SDSS $z$} &
	\colhead{2MASS $J$} & \colhead{2MASS $H$} & \colhead{2MASS $K_S$} &
	\colhead{Sp.T.\tablenotemark{a}} & \colhead{Ref.} \\
	\colhead{(J2000.0)} & \colhead{(mag)} & \colhead{(mag)} & 
	\colhead{(mag)} & \colhead{(mag)} & \colhead{(mag)} & \colhead{(mag)}}
\startdata
\input{tab1.tex}
\enddata
\tablenotetext{a}{SDSS $i$-band magnitudes are listed if they are brighter than the $i\approx23.0
$~mag 3$\sigma$ detection limit.  Otherwise, 23.00~mag is listed as the lower magnitude limit.}
\tablenotetext{b}{The spectral types of all previously known T dwarfs have been updated to conform to 
the uniform near-IR T dwarf classification scheme of \citet{burgasser_etal06b} and are as listed on the 
\url{DwarfArchives.org} website \citep{kirkpatrick03, gelino_etal04}. }
\tablenotetext{c}{Below the $S/N=5$ limit in 2MASS.  The photometry was obtained by fitting a PSF to 
the signal at the known location of the object from the other two 2MASS bands.}
\tablenotetext{d}{The de-blending of the source from a nearby star in SDSS was re-done to obtain 
more reliable photometry.}
\tablerefs{1.\ \citet{kirkpatrick_etal00}, 2.\ \citet{geballe_etal02}, 3.\ \citet{strauss_etal99}, 4.\ \citet
{knapp_etal04}, 5.\ \citet{hawley_etal02}, 6.\ \citet{chiu_etal06}, 7.\ \citet{burgasser_etal99}, 8.\ \citet
{leggett_etal02}, 9.\ \citet{fan_etal00}, 10.\ \citet{tsvetanov_etal00}.}
\end{deluxetable}

\begin{deluxetable}{l}
\tablewidth{0pt}
\tablecaption{Additional Candidates Near Bright Stars \label{tab_bright_star_candidates}}
\tablehead{\colhead{2MASS ID (J2000.0)}}
\startdata
07302933+2709051 \\
08201812+5101519 \\
08460641+4606208 \\
15350377+0219239 \\
16442092+4615156 \\
16581425+3147372 \\
17080715+6109134
\enddata
\end{deluxetable}

\begin{deluxetable}{cl}
\tablewidth{0pt}
\tablecaption{Discarded Candidates \label{tab_junk_candidates}}
\tablehead{\colhead{2MASS ID (J2000.0)} & \colhead{Notes}}
\startdata
\input{tab3.tex}
\enddata
\end{deluxetable}

\begin{deluxetable}{lcccccc}
\tabletypesize{\scriptsize}
\tablewidth{0pt}
\tablecaption{Spectroscopic Observations of Candidate L and T Dwarfs \label{tab_spectr_obs}}
\tablehead{\colhead{Object} & \colhead{Date} & \colhead{Telescope/Instrument} &
	\colhead{Wavelength} & \colhead{Resolution} & \colhead{$J$} & \\colhead{Exposure} \\
	\colhead{(2MASS ID)} & \colhead{(UT)} & & \colhead{($\micron$)} & 
	\colhead{($\lambda/\Delta\lambda$)} & \colhead{(mag)} & \colhead{(min)}}
\startdata
00521232$+$0012172 & 2006 Dec 20 & IRTF/SpeX & 0.8--2.5 & 150 & 16.36 & 24 \\
01040750$-$0053283 & 2006 Jan 3 & Keck/LRIS & 0.63--1.01 & 900 & 16.53 & 25  \\
01075242$+$0041563 & 2005 Oct 20 & IRTF/SpeX & 0.8--2.5 & 150 & 15.82 & 16 \\
01262109$+$1428057 & 2006 Dec 8 & IRTF/SpeX & 0.8--2.5 & 150 & 17.11 & 32 \\
15461461$+$4932114 & 2005 Sep 9 & IRTF/SpeX & 0.8--2.5 & 150 & 15.90 & 16 
\enddata
\end{deluxetable}	 

\begin{deluxetable}{llcccccc}
\tablewidth{0pt}
\tabletypesize{\scriptsize}
\tablecaption{{\sl Spitzer}/IRAC Observations and Photometry of Ultra-Cool Dwarfs \label
{tab_spitzer_phot}}
\tablehead{\colhead{2MASS ID} & \colhead{Obs. Date} & \colhead{AOR key} & 
	\colhead{Exposure} & \colhead{[3.6$\micron$]} & \colhead{[4.5$\micron$]} & 
	\colhead{[5.8$\micron$]} & \colhead{[8.0$\micron$]}\\
	\colhead{(J2000.0)} & \colhead{(UT)} & & \colhead{(s)} & \colhead{(mag)} & 
	\colhead{(mag)} & \colhead{(mag)} & \colhead{(mag)}}
\startdata	
09175418+6028065 & 2005 Oct 26 & 13778176 & 134 & $14.17\pm0.03$ & 
	$14.12\pm0.03$ & $13.80\pm0.05$ & $13.82\pm0.09$\\
12144089+6316434 & 2005 Nov 24 & 13778688 & 134 & $14.14\pm0.03$ & 
	$13.74\pm0.03$ & $13.30\pm0.05$ & $12.77\pm0.06$\\
13243553+6358281 & 2005 Jun 13 & 13777920 & 52 & $12.56\pm0.03$ & 
	$12.33\pm0.03$ & $11.79\pm0.03$ & $11.31\pm0.03$
\enddata
\end{deluxetable}

\begin{deluxetable}{lccccccc}
\tablewidth{0pt}
\tabletypesize{\scriptsize}
\tablecaption{Spectral Types and Colors of Confirmed and Candidate Ultra-Cool Dwarfs \label
{tab_izJHK_colors}}
\tablehead{\colhead{2MASS ID} & \colhead{Sp.T.} & \colhead{$i-z$} & 
	\colhead{$z-J$} & \colhead{$J-H$} & \colhead{$H-K_S$} &
	\colhead{$J-K_S$} & \colhead{$J$} \\
	\colhead{(J2000.0)} & & \colhead{(mag)} & \colhead{(mag)} & 
	\colhead{(mag)} & \colhead{(mag)} & \colhead{(mag)} & \colhead{(mag)}}
\startdata
\input{tab6_astroph.tex}
\enddata
\tablenotetext{a}{$H$-only detection in 2MASS.}
\end{deluxetable}	

\begin{deluxetable}{lllc}
\tablewidth{0pt}
\tablecaption{All Known T Dwarfs in SDSS DR1 \label{tab_sdss_tdwarfs}}
\tablehead{\colhead{SDSS ID} & \colhead{2MASS ID} & \colhead{SpT} & 
	\colhead{Ref.} \\
	\colhead{(J2000.0)} & \colhead{(J2000.0)}}
\startdata
\multicolumn{4}{c}{Previously Known T Dwarfs} \\
015141.69+124429.6 & 01514155+1244300 & T1 & 1 \\
020742.48+000056.2 & 02074284+0000564 & T4.5 & 1 \\
083048.80+012831.1 & 08304878+0128311 & T4.5 & 2 \\
092615.38+584720.9 & 09261537+5847212 & T4.5 & 1 \\
111010.01+011613.1 & 11101001+0116130 & T5.5 & 1 \\
120747.17+024424.8 & 12074717+0244249 & T0 & 3 \\
121440.95+631643.4 & 12144089+6316434 & T4 & 4 \\
121711.19$-$031113.3 & 12171110$-$0311131 & T7.5 & 5 \\
123739.35+652613.6 & 12373919+6526148 & T6.5 & 5 \\
125453.90$-$012247.4 & 12545393$-$0122474 & T2 & 6 \\
134646.45$-$003150.4 & 13464634$-$0031501 & T6.5 & 7 \\
151603.03+025928.9 & 15160303+0259292 & T0 & 2 \\
162414.37+002915.6 & 16241436+0029158 & T6 & 8 \\
\\
\multicolumn{4}{c}{New T Dwarfs} \\
132435.53+635828.2 & 13243553+6358281 & T2.5: & 9\\
154614.67+493209.7 & 15461461+4932114 & T2.5$\pm$1.0 & 9 \\
\enddata
\tablerefs{1.\ \citet{geballe_etal02}; 2.\ \citet{knapp_etal04}; 3.\ \citet{hawley_etal02}; 4.\ \citet
{chiu_etal06}; 5.\ \citet{burgasser_etal99}; 6.\ \citet{leggett_etal00}; 7.\ \citet{tsvetanov_etal00}; 8.\ \citet
{strauss_etal99}; 9.\ this paper.}
\end{deluxetable}

\begin{deluxetable}{lcccccccccc}
\tabletypesize{\scriptsize}
\tablewidth{0pt}
\tablecaption{Spectral Type Distribution, Absolute Magnitudes, Colors and Multiplicity of T Dwarfs 
\label{tab_absmags}}
\tablehead{\colhead{Sp.T.} & \colhead{$N_{\rm SDSS,DR1}$} &
	\colhead{$z-J$} & \colhead{sample} &
	\colhead{$M_J$} & \colhead{sample} & \colhead{$J-H$} &
	\colhead{sample} & \colhead{$J-K_S$} & \colhead{sample} & 
	\colhead{$N_{\rm bin}/N_{\rm tot}$\tablenotemark{a}} \\
	 & & \colhead{(mag)} & & \colhead{(mag)} & & \colhead{(mag)} & & \colhead{(mag)}}
\startdata
\input{tab8a_astroph.tex}
\\
\input{tab8b.tex}
\enddata
\tablenotetext{a}{From Table 1 in \citet{burgasser07}.}
\tablenotetext{b}{Assumed.  No T8 dwarfs are known in SDSS yet.}
\end{deluxetable}	

\begin{deluxetable}{lccc}
\tablewidth{0pt}
\tablecaption{Limiting Magnitudes for SDSS and 2MASS \label{tab_limmags}}
\tablehead{\colhead{Filter} & \multicolumn{3}{c}{Limiting Magnitude} \\
	\cline{2-4}
	& \colhead{$S/N=8.3$} & \colhead{$S/N=7$} & \colhead{$S/N=5$}}
\startdata
SDSS $z$\tablenotemark{a} & $20.17\pm0.23$ & $20.35\pm0.23$ & 
	$20.72\pm0.23$ \\
2MASS $J$ & \nodata & $16.69\pm0.15$ & $17.06\pm0.14$ \\
2MASS $H$ & \nodata & $15.80\pm0.17$ & $16.17\pm0.19$ \\
2MASS $K_S$ & \nodata & $15.19\pm0.15$ & $15.56\pm0.16$
\enddata
\tablenotetext{a}{Estimated for T dwarfs only.}
\end{deluxetable}

\begin{deluxetable}{lrrrr}
\tablewidth{0pt}
\tablecaption{Mean Imaging Depth (in Parsecs) for Single T Dwarfs in the SDSS $z$ Band (at $S/
N=8.3$) and in the 2MASS $J$, $H$, and $K_S$ Bands (at $S/N=5$) \label{tab_survey_depth}}
\tablehead{\colhead{Sp.T.}  & {$z$} & {$J$} & {$H$} & \colhead{$K_S$}}
\startdata
\input{tab10_astroph.tex}
\enddata
\tablenotetext{a}{Based on an assumed $z$-band absolute magnitude $M_z=20.0$~mag (see 
Table~\ref{tab_absmags}).}
\end{deluxetable}

\begin{deluxetable}{lrrrccrrr}
\tablewidth{0pt}
\tablecaption{Monte Carlo Simulations of T Dwarfs Detectable in SDSS DR1 and 2MASS \label
{tab_montecarlo}}
\tablehead{\colhead{Sp.T.} & \colhead{$r_{\rm sim}$} & 
	\colhead{$N_{\rm sim}$} & \colhead{$N_{\rm det}$} & 
	\colhead{$N_{\rm exp}$} & 
	\colhead{$N_{\rm SDSS,DR1}$} & \colhead{$f_{\rm bin, sim}$} & 
	\colhead{$f_{\rm bin, det}$} & \colhead{$f_{\rm bin, obs}$} \\
	& \colhead{(pc)} & & & & & \colhead{(\%)} & \colhead{(\%)} & 
	\colhead{(\%)}}
\startdata
\input{tab11.tex}
\enddata
\tablecomments{Results are based on 10000 simulations of $N_{\rm sim}$ brown dwarfs in a volume 
of radius $r_{\rm sim}$ at each spectral type bin.  Only dwarfs that fall within an area of 2099 deg$^2$, 
equivalent to the footprint of the SDSS DR1 imaging survey, have been considered as detected.  The 
number of detected dwarfs per spectral type bin is in column $N_{\rm det}$, while the number of 
expected T dwarfs of the same sub-type per 2099~deg$^2$ sky area is in column $N_{\rm exp}$.  $N_
{\rm exp}$ is determined to be 0.5 higher than the observed number of brown dwarfs in each bin, $N_
{\rm SDSS,DR1}$ (see Appendix).  The errors on the expectation value $N_{\rm exp}$ denote its 95\% 
confidence interval.  $f_{\rm bin, sim}$, $f_{\rm bin, det}$, and $f_{\rm bin, obs}$ are the fractions of 
binary systems that are input into the simulations, detected from the simulations, and observed in high-
resolution imaging, respectively.} 
\end{deluxetable}

\begin{deluxetable}{lccccccc}
\tablewidth{0pt}
\tablecaption{$H$ and $K_S$ Band Drop-outs among the Observed and the Simulated T Dwarfs \label
{tab_dropouts}}
\tablehead{\colhead{Data Set} & \colhead{Sp.T.} & \colhead{Population Size} & 
	\multicolumn{2}{c}{$K_S$ Drop-outs} & &
	\multicolumn{2}{c}{$H,K_S$ Drop-outs} \\
	\cline{4-5} \cline{7-8}
	& & & \colhead{Number} & \colhead{Fraction} & & \colhead{Number} & 
	\colhead{Fraction}}
\startdata
SDSS DR1 & T0--T8 & 15 & 5 & 33\% & & 1 & 7\% \\
	& T6--T8 & 4 & 3 & 75\% & & 0 & 0\% \\
SDSS DR5 & T0--T8 & 58 & 19 & 33\% & & 6 & 10\% \\
	& T6--T8 & 7 & 3 & 43\% & & 1 & 14\% \\
Simulated & T0--T8 & 164,451 & 46,607 & 28\% & & 10,250 & 6\% \\
	& T6--T8 & 45,282 & 22,845 & 50\% & & 6478 & 14\%
\enddata
\end{deluxetable}

\begin{deluxetable}{lccc}
\tablewidth{0pt}
\tablecaption{Space Density of T Dwarfs \label{tab_space_dens}}
\tablehead{\colhead{Sp.T.} & \colhead{$\rho$} & 
	\colhead{$\rho_{2\times f_{\rm bin,obs}}$} & \colhead{$\rho_0$} \\
	& \colhead{(10$^{-3}$ pc$^{-3}$)} & \colhead{(10$^{-3}$ pc$^{-3}$)}
	& \colhead{(10$^{-3}$ pc$^{-3}$)}}
\startdata
T0--T2.5 & $0.86_{-0.44}^{+0.48}$ & $0.45_{-0.23}^{+0.26}$ & 
	$1.3_{-0.7}^{+0.7}$ \\
T3--T5.5 & $1.4_{-0.8}^{+0.8}$ & $1.2_{-0.6}^{+0.7}$ & 
	$1.6_{-0.9}^{+1.0}$ \\
T6--T8    & $4.7_{-2.8}^{+3.1}$ & $4.3_{-2.6}^{+2.9}$ & 
	$5.1_{-3.0}^{+3.4}$ \\
\\
T0--T8 & $7.0_{-3.0}^{+3.2}$ & $6.0_{-2.7}^{+2.9}$ & $8.0_{-3.3}^{+3.6}$ \\
\enddata
\tablecomments{$\rho$ is the space density of T dwarfs per spectral type bin for the observed T dwarf 
binarity rate $f_{\rm bin, obs}$ from direct imaging.  $\rho_{2\times f_{\rm bin,obs}}$ is the 
corresponding space density for twice the observed binarity rate (e.g., including potential unresolved 
spectroscopic binaries).  $\rho_0$ is the space density in the hypothetical case when all T dwarfs are 
single.  The errors denote 95\% confidence limits based on the number of SDSS-DR1/2MASS T 
dwarfs detected in each spectral type bin, and are obtained as described in the Appendix.}
\end{deluxetable}	

\begin{deluxetable}{ll}
\tablewidth{0pt}
\tabletypesize{\scriptsize}
\tablecaption{SDSS Flags of All Known T Dwarfs in DR1 \label{tab_sdss_flags}}
\tablehead{\colhead{SDSS ID} & \colhead{SDSS Flags} \\
	\colhead{(J2000.0)}}
\startdata
\multicolumn{2}{c}{Previously Known T Dwarfs} \\
015141.69+124429.6 & TOO\_FEW\_GOOD\_DETECTIONS 
	BINNED1 NOPETRO \\
020742.48+000056.2\tablenotemark{a} & TOO\_FEW\_GOOD\_DETECTIONS
	BINNED1 INTERP NOPETRO \\
083048.80+012831.1 & \nodata\tablenotemark{b} \\ 
092615.38+584720.9 & TOO\_FEW\_GOOD\_DETECTIONS BINNED1 NOPETRO \\
111010.01+011613.1 & TOO\_FEW\_GOOD\_DETECTIONS BINNED1 INTERP
	MANYPETRO NOPETRO \\ 
120747.17+024424.8 & TOO\_FEW\_GOOD\_DETECTIONS BINNED1 NOPETRO \\
121440.95+631643.4 & TOO\_FEW\_GOOD\_DETECTIONS STATIONARY BINNED1
	NOPETRO \\ 
121711.19$-$031113.3 & TOO\_FEW\_GOOD\_DETECTIONS BINNED1 
	DEBLENDED\_AS\_PSF INTERP COSMIC\_RAY \\
	& NOPETRO CHILD \\ 
123739.35+652613.6 & TOO\_FEW\_GOOD\_DETECTIONS BINNED1
	DEBLENDED\_AS\_PSF INTERP COSMIC\_RAY \\
	& MANYPETRO NOPETRO CHILD \\ 
125453.90$-$012247.4 & TOO\_FEW\_GOOD\_DETECTIONS BINNED1 NOPETRO \\
134646.45$-$003150.4 & TOO\_FEW\_GOOD\_DETECTIONS STATIONARY BINNED1
	INTERP NOPETRO \\
151603.03+025928.9\tablenotemark{a,c} & TOO\_FEW\_GOOD\_DETECTIONS 
	BINNED1 MANYPETRO NOPETRO \\
162414.37+002915.6 & TOO\_FEW\_GOOD\_DETECTIONS BINNED1 MANYPETRO 
	NOPETRO \\ 
\\
\multicolumn{2}{c}{New T Dwarfs} \\
132435.53+635828.2 & TOO\_FEW\_GOOD\_DETECTIONS PSF\_FLUX\_INTERP 
	DEBLEND\_NOPEAK \\
	& STATIONARY MOVED BINNED1 INTERP NOPETRO CHILD \\
154614.67+493209.7 & TOO\_FEW\_GOOD\_DETECTIONS PSF\_FLUX\_INTERP 
	STATIONARY BINNED1 \\
	& DEBLENDED\_AS\_PSF INTERP NOPETRO CHILD \\ 
\enddata
\tablenotetext{a}{Not recovered in the present cross-match (see \S~\ref{sec_completeness}).}
\tablenotetext{b}{The field containing this object (run, rerun, camcol, field = 2125, 40, 1, 49) is included 
in the SDSS Data Archive Server (DAS), but data for the object are unavailable through the Catalog 
Archive Server (CAS).}
\tablenotetext{c}{The SDSS ID is given instead of the 2MASS ID.}
\end{deluxetable}

\end{document}

%% file: tab1.tex
00283943$+$1501418 & $21.70\pm0.13$ & $19.58\pm0.09$ & $16.51\pm0.11$ & $15.26\pm0.09$ & $14.56\pm0.07$ &           L4.5 & 1 \\ 
00521232$+$0012172 & $21.46\pm0.12$ & $19.50\pm0.10$ & $16.36\pm0.11$ & $15.56\pm0.13$ & $15.46\pm0.16$ &                &  \\ 
01040750$-$0053283 & $21.60\pm0.10$ & $19.37\pm0.05$ & $16.53\pm0.13$ & $15.64\pm0.14$ & $15.33\pm0.17$ &                &  \\ 
01075242$+$0041563 & $21.19\pm0.08$ & $18.64\pm0.03$ & $15.82\pm0.06$ & $14.51\pm0.04$ & $13.71\pm0.04$ &             L8 & 2 \\ 
01262109$+$1428057 & $22.23\pm0.18$ & $20.46\pm0.17$ & $17.11\pm0.21$ & $16.17\pm0.22$ & $15.28\pm0.15$ &                &  \\ 
01514155$+$1244300 & $22.85\pm0.35$ & $19.46\pm0.08$ & $16.57\pm0.13$ & $15.60\pm0.11$ & $15.18\pm0.19$ &             T1 & 2 \\ 
02292794$-$0053282 & $21.56\pm0.11$ & $19.40\pm0.06$ & $16.49\pm0.10$ & $15.75\pm0.10$ & $15.18\pm0.14$ &                &  \\ 
07354882$+$2720167 & $21.70\pm0.11$ & $19.97\pm0.11$ & $16.94\pm0.13$ & $16.11\pm0.12$ & $15.66\pm0.17$ &                &  \\ 
08095903$+$4434216 & $21.84\pm0.16$ & $19.29\pm0.06$ & $16.44\pm0.11$ & $15.18\pm0.10$ & $14.42\pm0.06$ &             L6 & 4 \\ 
08202996$+$4500315 & $21.42\pm0.09$ & $19.31\pm0.05$ & $16.28\pm0.11$ & $15.00\pm0.09$ & $14.22\pm0.07$ &             L5 & 1 \\ 
08304878$+$0128311 &        $>23.0$ & $19.82\pm0.10$ & $16.29\pm0.11$ & $16.14\pm0.21$ &       $>16.36$ &           T4.5 & 4 \\ 
09175418$+$6028065 &        $>23.0$ & $20.64\pm0.18$ & $17.16\pm0.27$\tablenotemark{c} & $15.96\pm0.13$ & $15.42\pm0.15$ &                 & \\ 
09261537$+$5847212 &        $>23.0$ & $19.01\pm0.06$ & $15.90\pm0.07$ & $15.31\pm0.09$ & $15.45\pm0.19$ &           T4.5 & 4 \\ 
09264992$+$5230435 & $21.15\pm0.13$\tablenotemark{d} & $19.65\pm0.24$ & $16.77\pm0.14$ &       $>15.58$ &       $>15.20$ &                 & \\ 
10440942$+$0429376 & $21.66\pm0.08$ & $18.79\pm0.03$ & $15.88\pm0.08$ & $14.95\pm0.07$ & $14.26\pm0.09$ &             L7 & 4 \\ 
11101001$+$0116130 &        $>23.0$ & $19.64\pm0.10$ & $16.34\pm0.12$ & $15.92\pm0.14$ &       $>15.13$ &           T5.5 & 2 \\ 
11191046$+$0552484 & $21.75\pm0.11$ & $19.64\pm0.06$ & $16.76\pm0.16$ & $15.48\pm0.10$ & $15.03\pm0.15$ &                &  \\ 
11571680$-$0333279 & $22.19\pm0.25$ & $20.14\pm0.14$ & $17.33\pm0.22$ & $16.30\pm0.16$ & $15.74\pm0.24$ &                &  \\ 
12074717$+$0244249 & $21.47\pm0.12$ & $18.40\pm0.04$ & $15.58\pm0.07$ & $14.56\pm0.06$ & $13.99\pm0.06$ &             T0 & 5 \\ 
12144089$+$6316434 &        $>23.0$ & $19.65\pm0.10$ & $16.59\pm0.12$ & $15.78\pm0.16$ & $15.88\pm0.23$ &             T4 & 6 \\ 
12171110$-$0311131 & $22.88\pm0.34$ & $19.38\pm0.06$ & $15.86\pm0.06$ & $15.75\pm0.12$ &       $>15.89$ &           T7.5 & 7 \\ 
12172372$-$0237369 & $22.09\pm0.18$ & $19.89\pm0.11$ & $16.90\pm0.16$ & $15.81\pm0.13$ & $14.99\pm0.13$ &                &  \\ 
12373919$+$6526148 &        $>23.0$ & $19.59\pm0.08$ & $16.05\pm0.09$ & $15.74\pm0.15$ &       $>16.06$ &           T6.5 & 7 \\ 
12545393$-$0122474 & $22.25\pm0.29$ & $18.03\pm0.03$ & $14.89\pm0.03$ & $14.09\pm0.03$ & $13.84\pm0.05$ &             T2 & 8 \\ 
13081228$+$6103486 & $21.40\pm0.12$ & $19.42\pm0.10$ & $16.67\pm0.15$ & $16.16\pm0.21$ &       $>15.49$ &                &  \\ 
13141551$-$0008480 & $21.50\pm0.10$ & $19.52\pm0.07$ & $16.62\pm0.14$ & $16.17\pm0.17$ & $15.30\pm0.16$ &           L3.5 & 2 \\ 
13243553$+$6358281 & $22.68\pm0.26$ & $18.73\pm0.04$ & $15.60\pm0.07$ & $14.58\pm0.06$ & $14.06\pm0.06$ &                &  \\ 
13262981$-$0038314 & $21.68\pm0.11$ & $19.05\pm0.04$ & $16.10\pm0.07$ & $15.05\pm0.06$ & $14.21\pm0.07$ &             L8 & 9 \\ 
13464634$-$0031501 &        $>23.0$ & $19.21\pm0.06$ & $16.00\pm0.10$ & $15.46\pm0.12$ & $15.77\pm0.27$ &           T6.5 & 10 \\ 
14140586$+$0107102 & $21.74\pm0.14$ & $19.60\pm0.09$ & $16.74\pm0.20$ & $15.73\pm0.19$ & $15.25\pm0.20$ &                &  \\ 
14232186$+$6154005 & $21.73\pm1.24$ & $19.56\pm0.12$ & $16.63\pm0.15$\tablenotemark{c} & $15.96\pm0.15$ & $15.28\pm0.13$ &                & \\ 
15341068$+$0426410 & $21.57\pm0.08$ & $19.78\pm0.07$ & $16.92\pm0.17$ & $16.42\pm0.23$ & $15.60\pm0.22$ &                &  \\ 
15422494$+$5522451 & $22.45\pm0.25$ & $20.53\pm0.17$ &       $>17.13$ & $15.95\pm0.15$ &       $>15.19$ &                &  \\ 
15423630$-$0045452 & $21.87\pm0.14$ & $19.46\pm0.06$ & $16.71\pm0.13$ & $15.98\pm0.14$ & $15.41\pm0.20$ &                &  \\ 
15461461$+$4932114 & $22.84\pm0.35$ & $19.06\pm0.05$ & $15.90\pm0.07$ & $15.14\pm0.09$ & $15.03\pm0.20$\tablenotemark{c} &                & \\ 
15513546$+$0151129 & $21.40\pm0.11$ & $19.62\pm0.10$ & $16.85\pm0.15$ & $16.63\pm0.24$ & $15.26\pm0.17$ &                &  \\ 
16154255$+$4953211 & $22.03\pm0.14$ & $19.69\pm0.07$ & $16.79\pm0.14$ & $15.33\pm0.10$ & $14.31\pm0.07$ &                &  \\ 
16241436$+$0029158 & $22.86\pm0.28$ & $19.02\pm0.04$ & $15.49\pm0.05$ & $15.52\pm0.10$ &       $>15.52$ &             T6 & 3 \\ 
17164260$+$2945536 & $21.97\pm0.12$ & $20.06\pm0.10$ & $17.06\pm0.20$ & $16.47\pm0.22$ & $15.90\pm0.27$ &                &  \\ 
17310140$+$5310476 & $21.55\pm0.14$ & $19.34\pm0.07$ & $16.37\pm0.11$ & $15.48\pm0.11$ & $14.85\pm0.14$ &             L6 & 6 \\ 
17373467$+$5953434 & $22.68\pm0.36$ & $20.26\pm0.14$ & $16.88\pm0.16$ & $16.44\pm0.24$ & $15.72\pm0.26$ &                &  \\ 
21163374$-$0729200 & $22.20\pm0.17$ & $20.09\pm0.13$ & $17.20\pm0.21$ & $16.21\pm0.21$ & $14.98\pm0.13$ &                &  \\ 
21203387$-$0747208 & $21.79\pm0.71$ & $19.70\pm0.11$ & $16.82\pm0.15$ &       $>15.77$ &       $>14.86$ &                &  \\ 

%% file: tab3.tex
00530603$-$0920330 &	artifact \\
01174188$-$0929305 &	nearby bright star artifact \\
01505720$-$0038177 &	nearby bright star artifact \\
03383405$-$0103222 &	artifact \\
07583541+4118142 &	artifact \\
08414399+0212593 &	artifact \\
08465686+4503341 &	artifact \\
08540505+0408554 &	artifact \\
08593179+0024568 &	artifact \\
09002850+4833141 &	nearby bright star; background M dwarf\\
09021214+5240568 &	artifact \\
09044567+5305476 &	artifact \\
09181838+0116413 &	artifact \\
09185052+0135107 &	2MASS asteroid? \\
09305216+5246191 &	artifact \\
12301772+0429075 &	nearby bright star artifact \\
12440895+0048101 &	artifact \\
13222708+0443076 &	artifact \\
13334829+6015313 &	nearby bright star; background M dwarf \\
13435828+6034197 &	artifact \\
13493774+0339254 &	nearby bright star artifact \\
14381450$-$0055409 &	artifact \\
14520086+5540300 &	artifact \\
14521363+6141509 &	artifact \\
15064154+0356529 &	nearby bright star; background M dwarf\\
15323835+0209166 &	nearby bright star; background M dwarf\\
15350377+0219239 &	nearby bright star; background M dwarf\\
15573023+5223194 &	nearby bright star artifact \\
16200993+0015135 &	nearby bright star; background M dwarf\\
16330761+4152025 &	artifact \\
16443142+4246142 &	artifact \\
21435405$-$0633498 &	artifact \\
23163032$-$0033131 &	artifact

%% file: tab6_astroph.tex
00521232$+$0012172 &      L2p$\pm$1 &  $1.96\pm0.16$ &  $3.14\pm0.15$ &  $0.80\pm0.17$ &  $0.10\pm0.21$ &  $0.90\pm0.19$ & $16.36\pm0.11$ \\
01040750$-$0053283 &   L5.0$\pm$0.5 &  $2.23\pm0.11$ &  $2.84\pm0.14$ &  $0.89\pm0.19$ &  $0.31\pm0.22$ &  $1.20\pm0.21$ & $16.53\pm0.13$ \\
01262109$+$1428057 & young L2$\pm$2 &  $1.77\pm0.25$ &  $3.35\pm0.27$ &  $0.94\pm0.30$ &  $0.89\pm0.27$ &  $1.83\pm0.26$ & $17.11\pm0.21$ \\
02292794$-$0053282 &             L? &  $2.16\pm0.13$ &  $2.91\pm0.12$ &  $0.74\pm0.14$ &  $0.57\pm0.17$ &  $1.31\pm0.17$ & $16.49\pm0.10$ \\
07354882$+$2720167 &             L? &  $1.73\pm0.16$ &  $3.03\pm0.17$ &  $0.83\pm0.18$ &  $0.45\pm0.21$ &  $1.28\pm0.21$ & $16.94\pm0.13$ \\
09175418$+$6028065 &          mid$-$L &        $>2.36$ &  $3.48\pm0.32$ &  $1.20\pm0.30$ &  $0.54\pm0.20$ &  $1.74\pm0.31$ & $17.16\pm0.27$ \\
09264992$+$5230435 &             L? &  $1.50\pm0.27$ &  $2.88\pm0.28$ &        $<1.19$ &        \nodata &        $<1.57$ & $16.77\pm0.14$ \\
11191046$+$0552484 &             L? &  $2.11\pm0.13$ &  $2.88\pm0.17$ &  $1.28\pm0.19$ &  $0.45\pm0.18$ &  $1.73\pm0.22$ & $16.76\pm0.16$ \\
11571680$-$0333279 &             L? &  $2.05\pm0.29$ &  $2.81\pm0.26$ &  $1.03\pm0.27$ &  $0.56\pm0.29$ &  $1.59\pm0.33$ & $17.33\pm0.22$ \\
12172372$-$0237369 &             L? &  $2.20\pm0.21$ &  $2.99\pm0.19$ &  $1.09\pm0.21$ &  $0.82\pm0.18$ &  $1.91\pm0.21$ & $16.90\pm0.16$ \\
13081228$+$6103486 &             L? &  $2.00\pm0.14$ &  $2.75\pm0.19$ &  $0.51\pm0.26$ &        $<0.67$ &        $<1.18$ & $16.67\pm0.15$ \\
13243553$+$6358281 &          T2.5: &  $3.95\pm0.26$ &  $3.13\pm0.08$ &  $1.02\pm0.09$ &  $0.52\pm0.08$ &  $1.54\pm0.09$ & $15.60\pm0.07$ \\
14140586$+$0107102 &             L? &  $2.14\pm0.17$ &  $2.86\pm0.22$ &  $1.01\pm0.28$ &  $0.48\pm0.28$ &  $1.49\pm0.28$ & $16.74\pm0.20$ \\
14232186$+$6154005 &             L? &  $2.17\pm1.25$ &  $2.93\pm0.19$ &  $0.67\pm0.21$ &  $0.68\pm0.20$ &  $1.35\pm0.20$ & $16.63\pm0.15$ \\
15341068$+$0426410 &             L? &  $1.79\pm0.11$ &  $2.86\pm0.18$ &  $0.50\pm0.29$ &  $0.82\pm0.32$ &  $1.32\pm0.28$ & $16.92\pm0.17$ \\
15422494$+$5522451 &             L? &  $1.92\pm0.30$ &        $<3.40$ &        $>1.18$ &        $<0.76$ &        \nodata &       $>17.13$\tablenotemark{a} \\
15423630$-$0045452 &             L? &  $2.41\pm0.15$ &  $2.75\pm0.14$ &  $0.73\pm0.19$ &  $0.57\pm0.24$ &  $1.30\pm0.24$ & $16.71\pm0.13$ \\
15461461$+$4932114 &   T2.5$\pm$1.0 &  $3.78\pm0.35$ &  $3.16\pm0.09$ &  $0.76\pm0.11$ &  $0.11\pm0.22$ &  $0.87\pm0.21$ & $15.90\pm0.07$ \\
15513546$+$0151129 &             L? &  $1.78\pm0.15$ &  $2.77\pm0.18$ &  $0.22\pm0.28$ &  $1.37\pm0.29$ &  $1.59\pm0.23$ & $16.85\pm0.15$ \\
16154255$+$4953211 &             L? &  $2.34\pm0.16$ &  $2.90\pm0.16$ &  $1.46\pm0.17$ &  $1.02\pm0.12$ &  $2.48\pm0.16$ & $16.79\pm0.14$ \\
17164260$+$2945536 &             L? &  $1.91\pm0.16$ &  $3.00\pm0.22$ &  $0.59\pm0.30$ &  $0.57\pm0.35$ &  $1.16\pm0.34$ & $17.06\pm0.20$ \\
17373467$+$5953434 &             L? &  $2.42\pm0.39$ &  $3.38\pm0.21$ &  $0.44\pm0.29$ &  $0.72\pm0.35$ &  $1.16\pm0.31$ & $16.88\pm0.16$ \\
21163374$-$0729200 &             L? &  $2.11\pm0.21$ &  $2.89\pm0.25$ &  $0.99\pm0.30$ &  $1.23\pm0.25$ &  $2.22\pm0.25$ & $17.20\pm0.21$ \\
21203387$-$0747208 &             L? &  $2.09\pm0.72$ &  $2.88\pm0.19$ &        $<1.05$ &        \nodata &        $<1.96$ & $16.82\pm0.15$ \\

%% file: tab8a_astroph.tex
T0--T0.5 & 2 & 2.85$\pm$0.19 &  9 & 13.76$\pm$0.28 &  2 &  0.99$\pm$0.11 & 11 &  1.58$\pm$0.14 &  9 & 2/2 \\
T1--T1.5 & 1 & 2.93$\pm$0.19 & 11 & 14.72$\pm$0.21 &  3 &  0.96$\pm$0.12 & 14 &  1.32$\pm$0.32 & 11 & 1/3 \\
T2--T2.5 & 3 & 3.10$\pm$0.16 & 11 & 14.54$\pm$0.00 &  2 &  0.83$\pm$0.18 & 13 &  1.03$\pm$0.29 & 11 & 0/1 \\
T3--T3.5 & 0 & 3.25$\pm$0.09 &  6 & 14.41$\pm$0.39 &  2 &  0.68$\pm$0.18 &  6 &  0.89$\pm$0.18 &  4 & 1/2 \\
T4--T4.5 & 4 & 3.41$\pm$0.13 &  7 & 14.51$\pm$0.00 &  2 &  0.33$\pm$0.17 &  8 &  0.34$\pm$0.11 &  6 & 1/4 \\
T5--T5.5 & 1 & 3.48$\pm$0.13 &  8 & 15.12$\pm$0.20 &  3 &  0.19$\pm$0.16 & 17 &  0.14$\pm$0.16 & 12 & 1/8 \\
T6--T6.5 & 3 & 3.51$\pm$0.02 &  3 & 15.49$\pm$0.28 &  8 &  0.13$\pm$0.20 & 11 &  0.10$\pm$0.43 &  7 & 1/8 \\
T7--T7.5 & 1 & 3.40$\pm$0.19 &  5 & 15.98$\pm$0.43 &  3 &  0.03$\pm$0.19 &  8 &  0.24$\pm$0.37 &  6 & 1/6 \\
T8       & 0 & 3.50$\pm$0.20\tablenotemark{b} &  0 & 16.49$\pm$0.58 &  2 &  0.01$\pm$0.28 &  3 & -0.15$\pm$0.59 &  2 & 0/1\\

%% file: tab8b.tex
T0--T2.5 & 6 & 2.97$\pm$0.20 & 31 & 14.61$\pm$0.20 &  7 &  0.92$\pm$0.15 & 38 &  1.29$\pm$0.34 & 31 \\
T3--T5.5 & 5 & 3.39$\pm$0.15 & 21 & 14.74$\pm$0.41 &  7 &  0.32$\pm$0.25 & 31 &  0.33$\pm$0.32 & 22 \\
T6--T8   & 4 & 3.44$\pm$0.16 &  8 & 15.75$\pm$0.50 & 13 &  0.08$\pm$0.20 & 22 &  0.12$\pm$0.41 & 15 \\

%% file: tab10_astroph.tex
T0--0.5 & 51.6 & 45.8 & 47.0 & 45.8 \\
T1--1.5 & 32.0 & 29.4 & 30.3 & 25.3 \\
T2--2.5 & 32.1 & 32.0 & 28.4 & 24.2 \\
T3--3.5 & 31.8 & 33.8 & 30.3 & 26.8 \\
T4--4.5 & 28.3 & 32.4 & 22.8 & 18.0 \\
T5--5.5 & 20.6 & 24.4 & 17.3 & 12.7 \\
T6--6.5 & 17.2 & 20.6 & 14.6 & 10.5 \\
T7--7.5 & 14.4 & 16.5 & 11.0 &  7.9 \\
T8      & 10.9\tablenotemark{a} & 13.0 &  8.3 &  6.1 \\

%% file: tab11.tex
T0--T2.5 &  91.3 & 2750 & 6.47$\pm$0.03 & $6.5_{-3.3}^{+3.6}$ & 6 & 26.1 & 50.1$\pm$0.2 & 50.0 \\
T3--T5.5 &  89.8 & 4250 & 5.49$\pm$0.02 & $5.5_{-3.0}^{+3.3}$ & 5 &  8.8 & 21.5$\pm$0.2 & 21.4 \\
T6--T8   &  63.8 & 5090 & 4.48$\pm$0.02 & $4.5_{-2.7}^{+3.0}$ & 4 &  5.1 & 13.2$\pm$0.2 & 13.3 \\